\documentclass{jpp}
\pdfoutput=1
\usepackage{graphicx}
\usepackage{dcolumn}% Align table columns on decimal point
\usepackage{bm}% bold math
\usepackage{bigints}
\usepackage{appendix}
\usepackage{xcolor}
\usepackage{subcaption}
\usepackage{verbatim}

\usepackage[utf8]{inputenc}
\usepackage[T1]{fontenc}
\usepackage{amsmath}

\usepackage[english]{babel}

\usepackage{tikz}

\DeclareUnicodeCharacter{2009}{\,} 

\shorttitle{Temperature separation in moderately-coupled plasma}
\shortauthor{H. Fetsch, T. E. Foster, and N. J. Fisch}

\title{Temperature separation under compression of moderately-coupled plasma}

\author{H. Fetsch\aff{1}
  \corresp{\email{hfetsch@princeton.edu}},
  T. E. Foster\aff{1}
 \and N. J. Fisch\aff{1}}

\affiliation{\aff{1}Department of Astrophysical Sciences, Princeton University, Princeton, NJ 08540, USA}

\begin{document}

\maketitle

\date{12 March, 2023}

%%total={6.5in,8.75in}, top=1.2in, left=0.9in, includefoot,
%%height=10in,a5paper,hmargin={3cm,0.8in},
%]{geometry}

\newcommand{\citethomas}{\citep{Foster_et}}%{(Foster, Fetsch, and Fisch 2023)}

\newcommand{\bs}[1]{\boldsymbol{#1}}
\newcommand{\mc}[1]{\mathcal{#1}}
\newcommand{\todo}[1]{\textcolor{blue}{#1}}

\newcommand{\half}{\frac{1}{2}}
\newcommand{\Lambdainv}{\epsilon}
\newcommand{\energynorm}{\mc E_0}

\newcommand{\sety}{\{Y_i\}}
\newcommand{\setf}{\{F_i\}}
\newcommand{\nisq}{{|\widetilde n_i|^2}}
\newcommand{\nisqk}{{|\widetilde n_i(\bs k)|^2}}
\newcommand{\tni}{{\widetilde n_i}}
\newcommand{\tnik}{{\widetilde n_i(k)}}
\newcommand{\trho}{{\widetilde \rho}}
\newcommand{\trhok}{{\widetilde \rho(\bs k)}}
\newcommand{\ci}{S_{ii}}
\newcommand{\savg}{\langle S_{ii} \rangle}
\newcommand{\civ}{S_{ii}}
\newcommand{\ciavg}{\langle S_{ii} \rangle}
\newcommand{\fem}{\mc F_e}
\newcommand{\fim}{\mc F_i}
\newcommand{\vhat}{\hat{V}}
\newcommand{\fiv}{(\vhat \mc F_i)}
\newcommand{\fev}{(\vhat \mc F_e)}

\newcommand{\im}{\text{i}}
\newcommand{\e}{\text{e}}

\newcommand{\consfrac}[3]{\left(\frac{\partial #1}{\partial #2}\right)_{#3}}
\newcommand{\consfracfunc}[3]{\left(\frac{\delta #1}{\delta #2}\right)_{#3}}
\newcommand{\avg}[1]{\langle #1 \rangle}

\newcommand{\sone}{a}
\newcommand{\stwo}{b}

\newcommand{\Gii}{C_{ii}}
\newcommand{\Gei}{C_{ei}}
\newcommand{\Gee}{C_{ee}}
\newcommand{\Gonetwo}{C_{\sone\stwo}}

\newcommand{\nuee}{{\nu_{ee}}} %electron-electron collisions
\newcommand{\nuii}{{\nu_{ii}}} %ion-ion collisions
\newcommand{\nuie}{{\nu_{ie}}} %electron-ion energy exchange rate

\begin{abstract}

%In moderately-coupled plasma, a significant fraction of the internal energy resides in electric fields; as these plasmas are heated or compressed, the shifting partition of energy between particles and fields leads to surprising effects. In this work, a thermodynamic description of moderately-coupled two-temperature plasma is developed analytically from first principles and used to derive coupled quasi-Equations of State for the electrons and ions. Adiabatic compression under these conditions generates a separation between electron and ion temperatures. A derivation is also presented for "correlation heating," an increase in the ion temperature following heating of the electrons even in the absence of interspecies collisions. These results may be relevant in the design of inertial confinement fusion (ICF) experiments.

In moderately-coupled plasmas, a significant fraction of the internal energy resides in electric fields. As these plasmas are heated or compressed, the shifting partition of energy between particles and fields leads to surprising effects, particularly when ions and electrons have different temperatures. In this work, quasi-equations of state (quasi-EOS) are derived for two-temperature moderately-coupled plasma in a thermodynamic framework and expressed in a simple form. These quasi-EOS readily yield expressions for correlation heating, in which heating of the electrons causes a rapid increase in ion temperature even in the absence of collisional energy exchange between species. It is also shown that, remarkably, compression of moderately-coupled plasma drives a temperature difference between electrons and ions, even when the species start at equal temperature. These additional channels for ion heating may be relevant in designing ignition schemes for inertial confinement fusion (ICF).

\end{abstract}

\maketitle

%\tableofcontents

\section{Introduction}

Plasmas in which electrons and ions have different temperatures are a subject of intense interest, in part because this regime is relevant for both magnetic confinement and inertial confinement fusion (ICF) \citep{Strachan_Bitter_Ramsey_Zarnstorff_Arunasalam_Bell_Bretz_Budny_Bush_Davis_et,Han_Park_Sung_Kang_Lee_Chung_Hahm_Kim_Park_Bak_et,Edwards_Patel_Lindl_Atherton_Glenzer_Haan_Kilkenny_Landen_Moses_Nikroo_et, Eliezer_Henis_Nissim_Pinhasi_Val_2015}. Temperature separation can have a substantial impact on plasma properties including pressure, heat capacity, fast particle stopping power, and driver-target coupling, which significantly affect the fusion yield in a system near ignition \citep{Fan_Liu_Liu_Yu_He_2016,Rinderknecht_Rosenberg_Li_Hoffman_Kagan_Zylstra_Sio_Frenje_Gatu}.  In a plasma consisting of electrons and ions at comparable temperatures with masses $m_e, m_i$, the electron-ion energy exchange rate is smaller than the electron-electron energy exchange rate by a factor of $m_e/m_i$ and smaller than the ion-ion energy exchange rate by a factor of $\sqrt{m_e/m_i}$ \citep{Braginskii_1958,Scullard_Serna_Benedict_Ellison_Graziani_2018}. The temperature equilibration between species can therefore be much slower than the thermalization of a single species, and so it is useful to treat the two species as equilibrium systems in quasi-steady state at different temperatures \citep{Boercker_More_1986,Dharma-wardana_Perrot_1998}. 

This work studies a pair of fundamental questions relevant to two-temperature plasma. First: how is energy partitioned between electrons, ions, and electric fields? Second: if the system is perturbed, e.g. by heating or compression, how do the temperatures of both species evolve? We answer these questions in a regime where first-principles analytical treatment is possible, and we find that the intrinsic asymmetry between ions and electrons leads to intriguing effects.

\subsection{Quasi-Equations of State}

It is natural to approach these questions using an equation of state (EOS), which allows prediction of the system's internal properties as well as the evolution of these properties under various constraints \citep{Craxton_Anderson_Boehly_Goncharov_Harding_Knauer_McCrory_McKenty_Meyerhofer_Myatt_et, Lindl_Amendt_Berger_Glendinning_Glenzer_Haan_Kauffman_Landen_Suter_2004,Fortov_Ilkaev_Arinin_Burtzev_Golubev_Iosilevskiy_Khrustalev_Mikhailov_Mochalov_Ternovoi_et, Tully_Hawker_Ventikos_2016}.
When quantum effects are negligible and the number of particles in a Debye sphere ($\Lambda = n\lambda_D^3$) is large, plasma can often be treated using the EOS of ideal gas. However, non-ideal corrections to the EOS are important in a vast array of applications, including modelling solar oscillations \citep{ChristensenDalsgaard_Dappen_1992,Stix_Skaley_1990}, predicting nuclear reaction rates \citep{Heckler_1994}, and generating corrections to plasma opacity \citep{Hummer_Mihalas_1988}. Such EOS generally take a single temperature as a parameter and so are not generally valid in two-temperature plasma because standard thermodynamics does not apply to systems out of thermodynamic equilibrium.

A physical system that exhibits near-steady-state behavior despite being far out of thermal equilibrium can be described by an effective equation of state, or `quasi equation of state' (quasi-EOS). Like a classical EOS, a quasi-EOS gives relationships between a system's macroscopic properties, such as internal energy, pressure, and heat capacity, but holds even away from true thermal equilibrium, at least on the timescales of interest. Such quasi-EOS have been formulated for systems as diverse as metals \citep{Petrov_Migdal_Inogamov_Zhakhovsky_2015}, warm dense matter \citep{Harbour_Forster_Dharma-wardana_Lewis_2018}, turbulent fluids and plasmas \citep{Volkov_1999,Davidovits_Fisch_2019}, and the intergalactic medium \citep{Ricotti_Gnedin_Shull_2000}. Other examples of quasi-EOS include the compression of plasmas laden with either linear waves \citep{Schmit_Dodin_Fisch_2010} or nonlinear waves \citep{Schmit_Dodin_Rocks_Fisch_2013}, as well as rotating plasmas \citep{Geyko_Fisch_2017}. In these cases, there is excess heat capacity associated with collective behavior, whether turbulence, waves, or rotation. 

In the case at hand, excess heat capacity appears as a result of internal electrostatic potential energy. The importance of potential energy in a thermal system is captured by the coupling parameter, which is heuristically given by $\Gamma = \langle U_\phi \rangle / T$, the characteristic inter-particle potential energy divided by the thermal energy. As the density increases or the temperature decreases, the magnitude of the potential energy $U_\phi$ can rise to become comparable to the temperature. A weakly-coupled plasma is one in which $\Gamma \ll 1$, meaning that thermal kinetic energy is dominant. In this work, we focus on cases in which the potential energy, while still smaller than the kinetic energy, is not so small as to be negligible. We refer to plasma in this regime as `moderately coupled.' At the high densities achieved in ICF implosions, especially in the dense stage immediately before ignition, moderate coupling effects often become important \citep{Hu_Militzer_Goncharov_Skupsky_2011}. Many of the effects discussed in this work have also been observed in the context of ultracold neutral plasma (UCP) \citep{Bergeson_Baalrud_Ellison_Grant_Graziani_Killian_Murillo_Roberts_Stanton_2019}. 

Several models have been developed for quasi-EOS in two-temperature plasma using various combinations of computational, empirical, and semi-analytical techniques \citep{Ramshaw_Cook_2014,Triola_2022,Gleizes_Chervy_Gonzalez_1999,Chen_Han_1999,Kraeft_Schlanges_Kremp_Riemann_DeWitt_1998, Boercker_More_1986, More_Warren_Young_Zimmerman_1988}. 
First-principles analytical models are rare because the complex physics of dense plasmas, including degeneracy, ionization, and many-body correlations, usually renders such treatments intractable. However, analytical models offer valuable insight into the fundamental physics of two-temperature plasma. By manipulating parameters and evaluating limits, analytical expressions offer a powerful tool to disentangle the essential physical effects at work. Although the regime of applicability for analytical theory is more restrictive than for simulation, rigorous analytical results can serve as a benchmark and complement to computational studies. In this work, we develop a first-principles analytical quasi-EOS for two-temperature moderately-coupled plasma and use it to determine the evolution of electron and ion temperatures on timescales where collisional energy transfer between species can be neglected.

\subsection{Approach in this work}

Although it will not be assumed beforehand in this work, the natural expansion parameter for the electrostatic potential energy is $\Lambdainv = 1/4\upi\Lambda$, where as usual $\Lambda = n\lambda_D^3$ represents the number of particles in a Debye sphere \citep{Kelly_1963}\footnote{The numerical prefactors on these quantities are discussed in Appendix~\ref{sec_regimes}.}. The plasma parameter is related to the coupling strength as $\Lambda \propto \Gamma^{-3/2}$. We seek corrections to the plasma equation of state (EOS), to leading-order in $\Lambdainv$, relevant in a moderately-coupled plasma. The regime of applicability of our weak-coupling approximation is discussed further in Appendix~\ref{sec_regimes}.

This work presents a novel derivation of the EOS of a moderately-coupled plasma by statistical methods. Although some of the results could be obtained by other means \citep{Salpeter_1963, Ecker_Kroll_1964, Boercker_More_1986, Triola_2022, Foster_et}, the approach taken here yields a simple and physically transparent formula. Through its simple form, the quasi-EOS obtained in this work provides insight into the mechanisms of collisionless heating and allows the prediction of novel effects in temperature separation between species. Because the standard procedures of statistical mechanics fail in non-equilibrium systems, we take advantage of the fact that electrons equilibrate much faster than ions and therefore can be considered ergodic for any fixed set of ion positions. Using this separation of timescales, we define an ensemble in which the electron and ion subsystems are statistically independent except for the constraint that electrons equilibrate with each other rapidly enough to balance the force applied by the ions at every ion microstate. This approach is in line with that of other authors who have approached the problem \citep{Salpeter_1963, Ecker_Kroll_1964, Boercker_More_1986}, and is reminiscent of the widely-used Born-Oppenheimer approximation \citep{Essen_1977, Dharma-wardana_Perrot_1998}. 

Using this quasi-EOS, we derive intriguing effects unique to two-temperature plasma in the moderate coupling regime. We find that upon compression, it is possible to change the temperatures of the two species by different amounts, generating a greater temperature difference than would be expected by following ideal gas adiabats. We additionally study how the ion temperature responds when the electron temperature is increased or decreased. We show that heating the electrons causes the ions to heat as well, even when collisional energy exchange is disallowed; interestingly, the change in ion temperature depends on the rate at which the electron temperature is varied. This `correlation heating' or `disorder-induced heating' effect has been observed in experiments and simulations \citep{Murillo_2001, Killian_Pattard_Pohl_Rost_2007, Kuzmin_ONeil_2002, Gericke_Murillo_2003, Lyon_Bergeson_2011, Lyon_Bergeson_Murillo_2013} and studied analytically \citep{Morawetz_Bonitz_Morozov_Ropke_Kremp_2001, Foster_et}. Our approach complements this previous work by offering a simple and physically-transparent formula.

In complementary work, \citet{Foster_et} study moderately-coupled two-temperature plasma in the same regime using kinetic theory. Relying on timescale separation as in this paper, the ions are found to behave as a One-Component Plasma (OCP), interacting through a shielded potential due to electron screening. The result obtained for heating of ions following sudden electron heating exactly matches the result in this paper. However, the kinetic formalism provides additional information by capturing the evolution of the ion distribution function over time, demonstrating that this heating takes place on the timescale of the ion plasma frequency. The rigorous descriptions of the evolution toward quasi-equilibrium after sudden heating offer valuable insights. These complement the statistical formalism of this paper, which is more easily able to describe compression and multiple modes of heating, but not the transient state of the system during equilibration.

For the sake of generality, we work with arbitrary interaction potentials through most of the derivation, and then specialize to Coulomb potentials in the final steps to obtain concrete results. Extensions to physical systems interacting through other effective potentials are briefly discussed. In the final results of this work, we assume weak coupling of both electrons and ions; however, intermediate results depend only on weakly-coupled electrons and therefore could be applied to systems where the ions are strongly coupled and the ion structure factor has been determined through some other means, such as molecular dynamics simulations, solving the hypernetted chain equation, or various analytical approximations \citep{Shaffer_Tiwari_Baalrud_2017, Gregori_Ravasio_Holl_Glenzer_Rose_2007, Slattery_Doolen_DeWitt_1980}.

\subsection{Outline of the paper}

This paper is organized as follows. In \S\ref{sec_twotemp_partition}, we introduce approximations to extend standard techniques from statistical mechanics to a two-temperature system with large timescale separation. We then apply this procedure to derive partition functions for both species, and we discuss how thermal averages can be computed. In \S\ref{sec_electron_ion}, we specialize our results to a two-temperature plasma with Coulomb interactions, and obtain simple, explicit formulas for quantities such as the energy, entropy, and two-particle correlations for each species. In \S\ref{sec_results}, we use these quantities to study the response of the system to changing parameters. We derive simple analytical expressions for the temperature change of both species resulting from compression and from heat input to electrons. Finally, in \S\ref{sec_discussion}, we discuss the results for energy partition, compressional heating, and correlation heating graphically and by taking instructive limits. The physical mechanisms of energy exchange that can be identified in this way, beyond their academic interest as fundamental plasma phenomena, may remain qualitatively similar outside of the moderate coupling regime and therefore could be of interest in the design of experiments even beyond the regime treated in this work.

\section{Two-Temperature Partition Function}
\label{sec_twotemp_partition}

\subsection{Setup}
\label{sec_sub_setup}

In this section, we define the physical system under consideration as well as relevant notation. We outline the formalism that will be used to separate the electrons and ions by timescale and relate the partition functions of the subsystems. We then proceed to derive thermodynamic potentials for each subsystem in terms of general interaction potentials. Finally, we outline the procedure for deriving thermal expectation values. The central results of this section are thermodynamic potentials for each species under various constraints. These are listed for electrons with general ion structure factor \eqref{eq_Phie_Gs}, for electrons with equilibrium ions \eqref{eq_Psie_Gs}, and for ions \eqref{eq_Psii_Gs}.

We consider a homogeneous, net-neutral, non-magnetized plasma in volume $V$ consisting of $N_e$ electrons and $N_i = N_e/Z$ ions with charges $q_e = -e$ and $q_i = +Ze$, where for simplicity only a single ion charge state is considered. The mass, position, and velocity of particle $j$ of species $s$ are written respectively as $m_s, \bs r_{s,j}, \bs v_{s,j}$ and the electron and ion densities are $n_{e} = N_e/V$ and $n_{i} = N_i/V$ respectively. For now, we set the entire system to temperature $T = 1/\beta$. The electrostatic potential at location $\bs r$ generated by a particle of species $s$ at location $\bs r_{s,j}$ is denoted by $\varphi_s(\bs r - \bs r_{s,j})$ and so the total potential at location $\bs r$ due to particles of species $s$ is
\begin{equation}
    \phi_s(\bs r) = \sum_j^{N_s}\varphi_s(\bs r - \bs r_{s,j}) 
\end{equation}
and the energy associated with each species is defined as
\begin{equation}
    E_s = E_{Ks} + E_{\phi s}
\end{equation}
where the kinetic energy is, as usual,
\begin{equation}
    E_{Ks} = \sum_j^{N_s} \frac{1}{2}m_sv_{s,j}^2. 
\end{equation}

The electrostatic energy is given by
\begin{equation}
    E_{\phi s} = \sum_j^{N_s}q_s \big( \phi_s(\bs r_{s,j}) - \varphi(0)\big),
\end{equation}
where the second term inside the summation subtracts self-energy, the energy of a particle placed at the center of its own potential well.

There is additionally an inter-species interaction energy $E_\mathrm{int}$, which we can write as the energy of electrons placed in the potential generated by ions, i.e.
\begin{equation}
    E_\mathrm{int} = \sum_j^{N_e} -e\phi_i(\bs r_{e,j}) .
\end{equation}

Having defined each component of the system's microscopic energy, we can proceed with a statistical description. As we will see, partitioning the interaction energy in the two-temperature case is a nontrivial problem. The remainder of this section is devoted primarily to capturing the effect of the interaction energy in a systematic way even out of equilibrium. 

\subsection{Timescale separation}

The partition function offers a statistical description of a system from which various averaged quantities can be determined by taking the appropriate derivatives. For a single-temperature equilibrium plasma, the partition function can straightforwardly be written as

\begin{equation}
    \mathcal{Z} = \int_V dX_i dX_e \exp\left\{-\beta(E_e + E_i + E_\mathrm{int})\right\} .
\end{equation}

Here, $dX_i, dX_e$ are the phase space integration measures for the ions and electrons respectively, normalized so that $dX_s = (1/N_s!)\left(m_s/2\upi\hbar\right)^{3N_s}d^{3N_s}r_sd^{3N_s}v_s$. The subscript $V$ denotes that the spatial integrals are taken over a fixed volume $V$.

When electrons and ions have different temperatures, the standard partition function formalism does not apply. In the absence of interactions, the partition function would be separable into ion and electron parts, and we could simply write the Boltzmann factor for each species using the energy of that species divided by its temperature. To describe the partition of the interaction energy, some authors have attempted to define an effective `cross-temperature' \citep{Triola_2022, Seuferling_Vogel_Toepffer_1989}. However, there is no consensus on a procedure for choosing this temperature, and no guarantee that a particular choice of effective cross-temperature will yield appropriate physical behavior. Some models involving a cross-temperature have been shown to correspond well with simulation \citep{Shaffer_Tiwari_Baalrud_2017}; while this validation is useful, a first-principles approach is useful for analyzing processes where, for example, energy is added to the system and must be somehow partitioned between species. 

To resolve ambiguity associated with the interaction energy, we adopt an approach, similar to that used by \citet{Boercker_More_1986}, in which the electron dynamics are much faster than the ion dynamics. We allow the electrons to come to equilibrium at temperature $T_e$ within a fixed potential established by the ions. The ions then equilibrate with each other at temperature $T_i$ under constraints imposed by the rapidly-established electron shielding. Boercker \& More make an ansatz for the way that electron screening affects the ion partition function, which is similar to the Born-Oppenheimer approximation \citep{Essen_1977, Dharma-wardana_Perrot_1998}. Denoting the electron free energy as $A_e$, they apply an additional factor of $\exp\left\{-A_e/T_i\right\}$ weighting every ion configuration in the ion partition function \citep{Boercker_More_1986}. In this way, the electrons provide an effective potential that modifies the $\exp\left\{-E_i/T_i\right\}$ term corresponding to bare ion-ion interactions.

In our approach, we incorporate electron response into the ion dynamics using the similar constraint that electrons are in equilibrium at all ion configurations. In other words, the electrons respond rapidly compared to the timescale on which ion configurations change, equilibrating among themselves such that they generate an effective force, which balances exactly the effective force applied by the ions. These effective forces are discussed in more detail below; in brief, the ion subsystem is an equilibrium system that generates a (generalized) force acting on the ions themselves and driving the system toward a thermodynamically favorable state, analogously to the pressure in standard thermodynamics. Positioned in screening clouds around the ions, the electrons also generate a force that acts on the ions, but this force in general drives them toward a different configuration. Our approach imposes that, since the electrons have sufficient time to equilibrate following every change in ion configuration, these forces must be in balance.

We describe our system in a generalized Gibbs ensemble, with generalized displacement $\ci$ (defined below) parameterizing the ion two-point correlation function, and generalized entropic forces $\mc F_e, \mc F_i$ acting on this displacement from each subsystem. It is convenient to approach the problem using entropic potentials \citep{Planes_Vives_2002}, in part because the resulting expressions make fewer references to species' temperatures and generalize more readily. 

\subsection{Definitions}

The statistical formalism used in this work is worked out in more detail in Appendix~\ref{sec_statistical_ensembles}, starting from basics. Here, we briefly summarize the approach and define quantities that are used in subsequent sections.

Formally, we consider the electrons and ions to be independent subsystems in contact with heat reservoirs of temperatures $T_e $ and $ T_i$ respectively. As in many constructions of the canonical ensemble, these baths need not be external to the system, but rather each subsystem can be some subvolume of one species, which is allowed to exchange heat with the rest of that species.

We begin in the canonical ensemble, describing a subsystem of ions in volume $V$ at temperature $T_i = 1/\beta_i$. We define the normalized Fourier-space ion distribution to be
\begin{equation}
    \trho_i(\bs k) = \frac{1}{\sqrt{N_i}}\sum_j^{N_i} \e^{-\im\bs k\cdot \bs r_{i,j}}.
\end{equation} 
Now we use the assumption that statistical quantities in our system are translationally invariant, so that the correlation functions depend only on spatial separation, i.e. if $g_{\sone\stwo}$ is the two-particle position-space correlation function, then $g_{\sone\stwo}(\bs r_2,\bs r_2) = g_{\sone\stwo}(\bs r_2 - \bs r_1)$. Then the ion-ion structure factor (a normalized Fourier-space two-particle correlation function) depends on only a single $\bs k$ argument and is defined as 
\begin{equation}
\label{eq_Sii_defn}
    \ci(\bs k) = |\trho_i(\bs k)|^2 .
\end{equation}

%TODO: reword arguments about translational invariance
We will eventually choose $\ci$ as the thermodynamic variable\footnote{This choice differs subtly from the formalism reviewed in Appendix~\ref{sec_statistical_ensembles} in which we begin in the canonical ensemble and all extensive variables except entropy are held constant. Here $\ci$ is an intensive variable and its conjugate variable $\fiv$ introduced below is extensive. We nevertheless begin by fixing $\ci$ because is a more conventional variable and more physically meaningful.} 
parameterizing the ion positions, even though the integral form of the partition function \eqref{eq_Zi_canonical} appears to require knowledge of $\trho_i$, which is not uniquely determined by $\ci$. Translational invariance ensures that $\trho_i$ will not appear linearly in the thermodynamic potentials. Higher-order correlations (e.g. $\trho_i(-\bs k)\trho_i(-\bs q)\trho_i(\bs k + \bs q)$) could appear, but to the order of coupling strength considered in this theory, we will not encounter them.

The ion partition function in the fixed-$\ci$ ensemble is
\begin{equation}
\label{eq_Zi_canonical}
    \mc Z_i(\beta_i, V, \ci) = \int_{V,\ci} dX_i \e^{-\beta_i E_i(X_i)} ,
\end{equation}
where the subscripts $V, \ci$ indicate that we only integrate over regions of ion phase space that are consistent with the known volume $V$ and known structure factor $\ci$. We note that the ion phase space is represented by the positions of the $N_i$ discrete particles and so, in general, different values of $\ci$ occupy different phase space volumes in the integral over $dX_i$. The configurational entropy associated with fixing $\ci$ therefore varies with the choice of $\ci$.

We now consider a subsystem of electrons in the same volume at inverse temperature $\beta_e$. Because we have grouped the interaction energy $E_\mathrm{int}$ into the electron subsystem, the total energy of the electrons depends on the configurations of both ions and electrons. However, we will choose a form for the interaction energy in which the dependence on ion configuration enters only through $\trho_i$. Therefore, we can write the electron partition function $\mc Z_e$ as
\begin{equation}
    \mc Z_e(\beta_e, V, \civ) = \int_{V} dX_e \e^{-\beta_e (E_e(X_e) + E_\mathrm{int}(X_e, \trho_i))} ,
\end{equation}
where the $\trho_i$ used as a parameter of $E_\mathrm{int}$ is any Fourier-space distribution $\trho_i$ compatible with the fixed $\ci$. By the argument above, the phase on $\trho_i$ is irrelevant as long as we are considering only two-particle correlations, as we do here. We can then write the corresponding entropic Massieu potential as
\begin{equation}
    \Phi_e(\beta_e, V, \civ) = \ln \mc Z_e .
\end{equation}

See Appendix~\ref{sec_statistical_ensembles} for more detailed discussion of the thermodynamic potentials used in this work. In short, the Massieu potential is an entropic analogue to the Helmholtz free energy, $A_e = -T_e \Phi_e$, which is convenient in this two-temperature system because its use means that fewer species-dependent temperature factors appear in our thermodynamic relations and that entropic quantities for the subsystems add without relative weighting factors.

With these expressions, we could describe the thermodynamics of electrons and ions in some volume if we knew the ion distribution in that volume. In general, however, we do not know $\ci$ \textit{a priori}; rather, we want to solve for its thermal average under the constraint of some known extensive quantity. The natural quantity is the generalized force applied to the electrons by the ions, which we can find by requiring that the electrons come to equilibrium separately for every ion microstate. This assumption is discussed in detail through examples in Appendix~\ref{sec_strings_model}. We transform to an ensemble in which we apply a constraint on this force and allow $\ci$ to fluctuate.

For the ion subsystem, we consider $\civ(\bs k)$ at all $\bs k$, and define the entropic force as the conjugate variable $\fiv$, where $\vhat = V/(2\upi)^3$ is the volume scaled by a factor that is convenient with the present Fourier transform convention. The quantity $\fim$ (without the factor of volume) is the one that we will most often use. In the canonical ensemble, we would impose that the ion subsystem be in contact with some external reservoir, enforcing that the reservoir shares the same $\ci$ as the ions but allowing $\mc F_i$ to fluctuate.  
%\todo{Expand explanation?}

If instead $\mc F_i$ is known in the reservoir (and therefore $\civ$ is of course allowed to fluctuate), then we can write the partition function by the same procedure as in \eqref{eq_zeta_Gibbs_general}, giving
\begin{equation}
\label{eq_zetai}
    \zeta_i(\beta_i, V, \fiv) = \int_{V} dX_i \e^{-\beta_i E_i(X_i) - \mc F_i \cdot \ci} ,
\end{equation}
where as shorthand, we have defined a dot product over functions of $\bs k$ such that\begin{equation}
\label{eq_dot_defn}
    g \cdot h \doteq V\int \frac{d^3k}{(2\upi)^3} g^*(\bs k) h(\bs k) .
\end{equation}

In order to write this change of ensembles in the form of a Legendre transform, we define the average $\langle q\rangle$ over ion configurations of some quantity $q$ in the standard way as
\begin{equation}
    \langle q \rangle = \int_V dX_i q(X_i) p_i(X_i)
\end{equation}
where the ion probability density is
\begin{equation}
    p_i(X_i) = \zeta_i^{-1} e^{-\beta_i E_i(X_i) - \mc F_i\cdot\ci(X_i)} .
\end{equation}
The entropic potential $\Psi_i = \ln \zeta_i$ in the fixed $\mc F_i$ ensemble is then
\begin{equation}
    \Psi_i (\beta_i, V, \fiv) = \Phi_i(\beta_i, V, \ciavg) - \mc F_i \cdot \savg .
\end{equation}

We can do the same for the electron subsystem to write an electron partition function  in terms of the entropic force $\mc F_e$ acting on the electrons\footnote{The prefactor $\vhat = V/(2\upi)^3$, included to make the dot product unitless, causes us to write $\fem$ rather than $\fev$ in this exponent.}, which gives
\begin{equation}
\label{eq_zetae}
    \zeta_e(\beta_e, V, \fev) = \int_{V} dX_e \e^{-\beta_e (E_e(X_e) + E_x(X_e)) - \mc F_e \cdot \ci}
\end{equation}
while the electron Planck potential is
\begin{equation}
\label{eq_Psie_singleq}
    \Psi_e (\beta_e, V, \fev) = \Phi_e(\beta_e, V, \ciavg) - \mc F_e \cdot \ciavg .    
\end{equation}

Now in order to evaluate the ion partition function, we need a prescription for fixing $\mc F_i$. We start by finding the electron response to a given ion distribution, which is the entropic force $\mc F_e$, found from the electron Massieu potential using
\begin{equation}
    \fev = \consfracfunc{\Phi_e}{\civ}{\beta_e, V} .
\end{equation}

We now say that the reservoir with which the ions are interacting is the electron subsystem. The force from the reservoir can therefore be identified with the force $F_e$ due to the electrons. We apply this condition at all wavevectors $\bs q$, i.e. $F_e(\bs q) + F_i(\bs q) = 0$. This constraint is discussed in more detail through a toy model in Appendix~\ref{sec_strings_model}. Defining the temperature ratio $\tau = T_e/T_i$ for convenience and noting that the force $F_R$ applied by a reservoir of temperature $T_R$ is related to the entropic forces by $F_R = T_R\mathcal{F}_R$, the equilibrium condition becomes
\begin{equation}
\label{eq_force_balance}
\begin{split}
    0 &=T_e\mathcal{F}_e + T_i\mathcal{F}_i,
    \\
    \mathcal{F}_i &= - \tau \frac{1}{\vhat}\consfracfunc{\Phi_e}{\civ}{\beta_e, V} .
\end{split}
\end{equation}

It is understood that everywhere $\mc F_i$ appears in future expressions, we insert the entropic force as determined by the equilibrium constraint \eqref{eq_force_balance}. We note that this constraint relies only on the other parameters of the two subsystems, which in this case are $\{\beta_i, V, \beta_e\}$. Therefore, we can write a form for $\Psi_i$ that depends only on volume and on the temperatures of the two species, where electron equilibrium for all ion configurations is implicitly assumed. We write this version of the ion Planck potential as
\begin{equation}
    \hat \Psi_i(\beta_e, \beta_i, V) = \ln \zeta_i
\end{equation}
where the hat on $\hat\Psi$ is meant as a reminder that the arguments $\beta_e, \beta_i, V$ are no longer the natural variables of the Planck potential, and therefore the expected thermodynamic relations may no longer hold when taking partial derivatives of $\hat\Psi_i$. For instance, if $\mc F_i$ depends on $\beta_i$ (which we will find later in the case of interest), we have in general that
\begin{equation}
    \consfrac{\hat\Psi_i(\beta_e, \beta_i, V)}{\beta_i}{\beta_e, V} \neq \consfrac{\Psi_i(\beta_i, V, \fiv)}{\beta_i}{\fim, V}
\end{equation}
where the derivative of $\Psi_i(\beta_i, V, \mc F_i)$ is taken at constant $\fiv$, and then the (generally $\beta_i$-dependent) expression for $\fiv$ is substituted in the final step.

\subsection{Electron subsystem}

\label{sec_sub_electron_subsystem}

With this formalism in place, we can apply it to describe a plasma of electrons and ions with a large separation of dynamical timescales.

We first evaluate the electron energy and partition function. We work in Fourier space with the electrostatic potential
\begin{equation}
    \widetilde\phi(\bs k) = \int d^3r \phi(\bs r)\e^{-i\bs k\cdot \bs r}
\end{equation}
and normalized Fourier-space electron distribution
\begin{equation}
    \trho_e(\bs k) = \frac{1}{\sqrt{N_e}}\sum_j^{N_e} \e^{-i\bs k \cdot \bs r_{e,j}} .
\end{equation}
For generality, and notational simplicity, we will describe the electrostatic potential in terms of Green's functions $\Gonetwo(\bs k)$ acting on the particle distributions; these functions can be specified at the end of the calculation, allowing the derivation to apply for a variety of effective potentials. Gathering temperature, volume, and charge factors in the Green's function definitions for convenience, the potential seen by the electrons is defined to be

\begin{equation}
\label{eq_phi_greensfunctions}
    \widetilde\phi(\bs k) = -V\frac{T_e}{e} \frac{1}{\sqrt{N_e}} \left[\Gee(\bs k)\trho_e(\bs k) + \Gei(\bs k)\trho_i(\bs k)\right] .
\end{equation}

This definition allows the potential energy to be written in simple form as
\begin{equation}
     E_{\phi e} = T_e\Gee\trho_e\cdot\trho_e + T_e\Gei\trho_i\cdot\trho_e - T_e\Gee \cdot 1.
\end{equation}
The final term in this equation, included to remove the contribution from self-interactions, has a possibly counterintuitive form but is discussed further in Appendix~\ref{sec_sub_energy_defns}.

The calculation of the electron partition function using this expression is presented in detail in Appendix~\ref{sec_sub_partition_calculation}. Here, we present only the results that will be used elsewhere in the work.

It follows from \eqref{eq_Ze_solved} that the Massieu potential for the electrons is
\begin{equation}
\label{eq_Phie_Gs}
\begin{split}
    \Phi_e =&  -\frac{V}{2} \int \frac{d^3k}{(2\upi)^3} \left[ \ln\left(2\Gee + 1 \right) - \frac{\Gei^2}{2 \Gee+1}\ci - 2\Gee \right] \\
    &- \ln N_e! + N_e\ln\left(\left(\frac{m_eT_e}{2\upi \hbar^2}\right)^{3/2}V\right) .
\end{split}
\end{equation}

The entropic force applied by the electrons to systems with which they are in force balance is then:

\begin{equation}
\label{eq_Fe}
    \mathcal{F}_e(\bs k) = \frac{\frac{1}{2}\Gei^2(\bs k)}{2\Gee(\bs k) + 1} .
\end{equation}

The condition of force balance \ref{eq_force_balance} requires that

\begin{equation}
   \mathcal{F}_i(\bs k) = -\tau\frac{\frac{1}{2} \Gei^2(\bs k)}{2 \Gee(\bs k) + 1} .
\end{equation}

We can now use this expression for $\mc F_i$ to write the ion partition function in the generalized Gibbs ensemble, in which we impose electron equilibrium for all ion configurations.

\subsection{Ion subsystem}

Now using the definitions of $\zeta_e, \zeta_i$ from \eqref{eq_zetae} and \eqref{eq_zetai}, we can write the partition functions for each subsystem. The electrons are straightforward because the additional force term can be taken outside of the integral over electron configurations, meaning $\zeta_e(\beta_e, V, \fev) = e^{-\mathcal{F}_e\cdot \avg{\ci}}\mathcal{Z}_e(\beta_e, V, \ciavg)$. The ion partition function that we are interested in finding is 
\begin{equation}
    \zeta_i = \int dX_i \e^{-E_i/T_i - \mathcal{F}_i\cdot \ci} .
\end{equation}

As we did for the electrons, we will write the ion-ion interaction potential in a general form using Green's functions, so that the ion potential energy is
\begin{equation}
    E_{\phi i} = T_i\Gii\trho_i \cdot \trho_i - T_i\Gii \cdot 1.
\end{equation}

Then the ion partition function in the generalized Gibbs ensemble is
\begin{equation}
\begin{split}
    \zeta_i &= \mathcal{Z}_{Ki}\mc Z_{\phi i}^\mathrm{self}\int d^{3N_i}r_{i,j} \exp\left\{ -\Gii\cdot\ci - \mc F_i \cdot \ci\right\}
\end{split}
\end{equation}
where the self-energy correction has been factored out into $\mc Z_{\phi i}^\mathrm{self}$. This partition function can also be written as
\begin{equation}
\label{eq_zeta_Gstar}
\begin{split}
    \zeta_i &= \mathcal{Z}_{Ki}\mc Z_{\phi i}^\mathrm{self}\int d^{3N_i}r_{i,j} \exp\left\{ -\Gii^\prime\trho_i\cdot\trho_i\right\}
\end{split}
\end{equation}
where $\Gii^\prime = \Gii + \mc F_i$ is an effective ion-ion interaction Green's function. This integral has the same form as \eqref{eq_Z_phi_e}, if we make the symbolic substitutions $\Gii \rightarrow \Gii^\prime, \Gei \rightarrow 0$, meaning that we can immediately write the result of the integral over ion positions (cf. Appendix~\ref{sec_sub_partition_calculation}) as 
\begin{equation}
    \zeta_{\phi i} = \exp\left\{-\frac{V}{2} \int \frac{d^3k}{(2\upi)^3} \left[\ln\left(2\Gii^\prime + 1 \right)- 2\Gii \right]\right\}
\end{equation}
and so the full ion Planck potential becomes
\begin{equation}
\label{eq_Psii_Gs}
\begin{split}
    \Psi_{i}(\beta_i, V, \fim) =& -\frac{V}{2}\int \frac{d^3k}{(2\upi)^3} \Bigg[\ln(2\Gii + 2\mc F_i + 1) - 2\Gii\Bigg]\\
    & - \ln N_i! + N_i\ln\left(\left(\frac{m_iT_i}{2\upi\hbar^2}\right)^{3/2}V\right)    .
\end{split}
\end{equation}
Physically, this means that the ions interact as a one-component plasma under some effective potential, which is a combination of the bare Coulomb potential and a term due to electron screening.

For completeness, we can also transform the electron Massieu potential to the new ensemble, giving the electron Planck potential explicitly as
\begin{equation}
\label{eq_Psie_Gs}
\begin{split}
    \Psi_e(\beta_e, V, \fem) =& - \frac{V}{2}\int \frac{d^3k}{(2\upi)^3} \Bigg[ \ln\left(2\Gee + 1 \right) - 2\Gee - \left(\frac{\Gei^2}{2 \Gee+1} - 2\mc F_e\right)\ciavg \Bigg]
    \\
    & - \ln N_e! + N_e\ln\left(\left(\frac{m_eT_e}{2\upi\hbar^2}\right)^{3/2}V\right).
\end{split}
\end{equation}

In the system of interest, kinetic degrees of freedom are fully separable and so the kinetic part of the partition function always yields the ideal gas result in our classical model. We will therefore sometimes only work with the electrostatic components of the thermodynamic potentials, $\Phi_{\phi e}, \Psi_{\phi e}, \Psi_{\phi i}$, corresponding to only the electrostatic energy. These are defined explicitly in Appendix~\ref{sec_sub_energy_defns}.

We could in principle now substitute the electron force given by \eqref{eq_Fe} and get explicit expressions for the Planck potentials $\hat\Psi_e, \hat\Psi_i$ in terms of only the externally-imposed variables $\beta_e, \beta_i, V$. However, in the generalized Gibbs ensemble that we have set up, differentiating the partition function should be done while holding the force fixed; it is therefore most useful to keep the potential in this expanded form for the following steps, and apply our knowledge of the force only in the final step.

The equations for the Planck potentials $\Psi_e$ \eqref{eq_Psie_Gs} and $\Psi_i$ \eqref{eq_Psii_Gs} serve as equations of state for the subsystems of a two-temperature plasma. Together, they comprise a quasi-EOS. It bears repeating that, within the regime of applicability of our assumptions, these expressions could be applied to arbitrary two-particle interactions. In \S\ref{sec_electron_ion} we will specialize to a specific physically-relevant case, but first we discuss the procedure for generating thermal averages from these equations of state.

\subsection{Thermal expectation values}

Using the Planck potentials for each species, we can calculate various quantities at thermal equilibrium. It will be of interest to calculate some quantities when the electrons are at equilibrium but the ion structure factor is held fixed, as well as when both subsystems are at equilibrium. We have already defined $\ci$ in \eqref{eq_Sii_defn}. We can define electron-electron and electron-ion structure factors as well; since in this work we are only interested in timescales longer than the electron equilibration time, quantities will always be averaged over electron configurations. The new structure factors are therefore defined as
\begin{equation}
\begin{split}
    S_{ee}(\bs q) &= \int dX_e p_e(X_e) \trho_e^*(\bs q)\trho_e(\bs q)
    \\
    S_{ie}(\bs q) &= \int dX_e p_e(X_e) \trho_e^*(\bs q)\trho_i(\bs q)
\end{split}
\end{equation}
where the electron probability density is
\begin{equation}
    p_e = \mc Z_e^{-1} e^{-\beta_e(E_e + E_\mathrm{int})}
\end{equation}
in the canonical ensemble. Since $p_e$ depends on $\ci$, so do $S_{ee}$ and $S_{ie}$. When the ions are also at equilibrium, we can substitute the equilibrium ion-ion structure factor $\avg{\ci}$ in order to find the other equilibrium structure factors $\avg{S_{ee}}, \avg{S_{ie}}$.

We will start with the ion structure factor, which can be found by differentiating the ion Planck potential with respect to $\Gii$ to be
\begin{equation}
    \avg{S_{ii}(\bs q)} = 1 - \frac{1}{\vhat} \consfracfunc{\Psi_i}{\Gii(\bs q)}{V,\fim}
\end{equation}
where we denote by the subscript that the generalized force should be held constant because $\fiv$ constitutes an independent thermodynamic variable, which should be unaffected by the differentiation even if the final formulas for this quantity depend on $\Gii$. The factor $1/\vhat = (2\upi)^3/V$ simply removes the prefactor on the integral that appears in the expression for $\Psi_i$. Some manipulation yields
\begin{equation}
\label{eq_Sii_Gs}
\begin{split}
    \avg{S_{ii}(\bs q)} &= \frac{1}{1 + 2 \Gii(\bs q) - \tau \frac{\Gei^2(\bs q)}{2 \Gee(\bs q)+1}} .
\end{split}
\end{equation}

Here we have used $\fem = \frac{\frac{1}{2} \Gei^2}{2 \Gee+1}$ from the derivation of the entropic force in \eqref{eq_Fe}. The ion-electron structure factor is similarly given by
\begin{equation}
    S_{ie}(\bs q) = -\frac{1}{\vhat}\consfracfunc{\Phi_e}{\Gei(\bs q)}{\ci, V} ,
\end{equation}
which simplifies to 
\begin{equation}
\label{eq_Sie_Gs}
\begin{split}
    S_{ie}(\bs q) &= -\frac{\Gei(\bs q)}{2\Gee(\bs q)+1} S_{ii}(\bs q) .
\end{split}
\end{equation}

Finally, for the electrons, the structure factor is given by
\begin{equation}
\begin{split}
    S_{ee}(\bs q) &= 1 - \frac{1}{\vhat} \consfracfunc{\Phi_e}{\Gee(\bs q)}{\ci, V}
\end{split}
\end{equation}
and therefore
\begin{equation}
\label{eq_See_Gs}
    S_{ee}(\bs q) = \frac{1}{2\Gee(\bs q) + 1} + \frac{\Gei^2(\bs q)}{(2\Gee(\bs q)+1)^2} S_{ii}(\bs q) .
\end{equation}

Using these structure factors, we could obtain the potential energy directly by integrating, but it is informative to derive it instead from the partition function by differentiating the Planck potential with respect to the temperature. The energy of each species is given by

\begin{equation}
    U_\sone = -\left(\frac{\partial\Psi_\sone}{\partial \beta_\sone}\right)_{V, \mc F_\sone} .
\end{equation}

The kinetic part gives the usual ideal gas result. For the electrostatic part, using the fact that we absorbed the temperatures into the definitions of the Green's functions, and so each temperature appears only as $\beta_\sone$ in the corresponding $\Gonetwo$, we can differentiate with respect to each Green's function. The potential components of each species' energy are then

\begin{equation}
\begin{split}
    U_{\phi i} &= -T_i\frac{1}{\vhat}\Gii\cdot\left(\frac{\delta\Psi_{i}}{\delta \Gii}\right)_{V, \fim}
    \\
    U_{\phi e} &= - T_e\frac{1}{\vhat} \Gee\cdot\left(\frac{\delta\Psi_{e}}{\delta \Gee} \right)_{V, \fem} - T_e\frac{1}{\vhat}\Gei\cdot\left(\frac{\delta\Psi_e}{\delta \Gei} \right)_{V, \fem} .
\end{split}
\end{equation}

These derivatives are the same ones that yielded the structure factors above, meaning that the energy can also be written as.

\begin{equation}
\label{eq_U_Ss}
\begin{split}
    U &= \frac{3}{2}N_eT_e + \frac{3}{2}N_iT_i + T_e \Gee\cdot (S_{ee} - 1) + T_e \Gei\cdot S_{ie} + T_i\Gii\cdot (S_{ii} - 1) .
\end{split}
\end{equation}

This is exactly the prescription for integrating the structure factor that we would otherwise have used to calculate the energy directly from known two-particle correlations. This is an useful check on the formalism adopted here, in that the two-timescale equilibrium procedure hasn't left artifacts in the thermodynamic state functions that we get from $\zeta_e$ and $\zeta_i$. The partition function approach adopted here also includes a prescription for dividing the potential energy between electrons and ions, which would not be uniquely specified when simply integrating the structure factor and potential to find the total energy. The electron electrostatic energy, given as a function of the structure factor for greater generality, is

\begin{equation}
\label{eq_U_phie_Gs}
\begin{split}
    U_{\phi e}(T_e; S_{ii}) =& T_e V\int \frac{d^3k}{(2\upi)^3} \Gee \left( \frac{1}{2 \Gee + 1} - 1 + \frac{\Gei^2}{(2\Gee+1)^2}S_{ii} \right)
    \\ &- T_e V\int \frac{d^3k}{(2\upi)^3} \Gei \frac{\Gei}{2\Gee+1}S_{ii} .
\end{split}
\end{equation}

When the ions are at equilibrium, we can substitute the $S_{ii}$ derived above, and the electron electrostatic energy reduces to

\begin{equation}
\begin{split}
    U_{\phi e}(T_e, T_i) =& T_e V\int \frac{d^3k}{(2\upi)^3} \left( - \frac{2\Gee^2 }{2 \Gee + 1} + \frac{\Gee}{(2\Gee+1)}\frac{\Gei^2}{(2\Gee+1)(1 + 2 \Gii) - \tau \Gei^2} \right)
    \\ &- T_eV\int \frac{d^3k}{(2\upi)^3} \frac{\Gei^2}{(2\Gee+1)(1 + 2 \Gii) - \tau \Gei^2} .
\end{split}
\end{equation}

Finally, the ion electrostatic energy is

\begin{equation}
\label{eq_U_phii_Gs}
    U_{\phi i}(T_e, T_i) = T_i  V\int \frac{d^3k}{(2\upi)^3} \Gii \left(\frac{1}{1 + 2  \Gii - \tau \frac{\Gei^2}{2 \Gee+1}} - 1\right).
\end{equation}

We define the entropy in this work using the Gibbs definition $\mathcal{S} = -\int dX_s p_s\ln p_s$ where $p_s$ is the phase space probability density of species $s$. The ion entropy, described by the generalized Gibbs ensemble, can then be written as

\begin{equation}
\label{eq_S_species_general}
\begin{split}
    \mathcal{S}_i &= \zeta_i^{-1}\int dX_i \left(\ln\zeta_i + \beta_iE_i + \mc F_i\cdot \ci \right) \e^{-\beta_iE_i - \mc F_i \cdot \ci}
    \\
    \mathcal{S}_i &= \Psi_i + \beta_i U_i + \mc F_i\cdot\ciavg .
\end{split}
\end{equation}

If system is constrained by known $\civ$ rather than by the condition of force balance, then we can still compute an entropy for the electrons, but we should use the canonical ensemble instead of the generalized Gibbs ensemble. In this case the electron entropy is given by
\begin{equation}
\label{eq_S_e_massieu}
    \mc S_e = \Phi_e + \beta_e (U_e + U_x).
\end{equation}

\begin{sloppypar}
This is equivalent, when the ions are at equilibrium, to the expression ${\mc S_e = \Psi_e + \beta_e(U_e + U_x) + \mc F_e\cdot\ciavg}$ found from the generalized Gibbs ensemble.
\end{sloppypar}

\section{Two-Component Coulomb Plasma}
\label{sec_electron_ion}

\subsection{Coulomb potentials}

Having obtained general expressions for the thermodynamic potentials of electrons and ions, we can specialize to a case of interest. This requires specifying the functional form of the electron-electron, ion-ion, and electron-ion interactions. In this section, we specify such forms and calculate explicit expressions for the resulting thermal averages.

To describe a classical plasma consisting of electrons and ions, we choose Green's functions corresponding to the Coulomb potential. It is necessary to include the factors of temperature, charge, and particle number that we grouped into the Green's functions to simplify notation in \ref{eq_phi_greensfunctions}. Including a factor $\frac{1}{2}$ to avoid overcounting within the same species, the Coulomb Green's functions are
\begin{equation}
\label{eq_G_defns}
\begin{split}
    \Gee(\bs k) &= \frac{1}{2}N_e\frac{4\upi e^2}{VT_e}\frac{1}{k^2} = \frac{1}{2}\frac{\kappa^2}{k^2} ,
    \\
    \Gei(\bs k)  &= -\sqrt{N_eN_i}\frac{4\upi Z e^2}{VT_e}\frac{1}{k^2} = -\frac{\sqrt{Z}\kappa^2}{k^2} ,
    \\
    \Gii(\bs k)  &= \frac{1}{2}N_e\frac{4\upi Z^2 e^2}{VT_i}\frac{1}{k^2} = \frac{1}{2}\frac{\chi^2}{k^2},
\end{split}
\end{equation}
where $k = |\bs k|$ and  we have defined the inverse screening lengths for electrons and ions respectively as $\kappa = \sqrt{4\upi e^2 n_{e}/T_e}$ and $\chi = \sqrt{4\upi Z^2e^2 n_{i}/T_i}$. These quantities are inversely proportional to single-species Debye lengths for electrons and ions. We can likewise define the total inverse screening length in the usual way as $k_D^2 = \kappa^2 + \chi^2$. In these variables, the entropic force from the electron subsystem given by \eqref{eq_Fe} is then

\begin{equation}
\label{eq_f_coulomb}
    \mc F_e = \frac{1}{2} \frac{1}{k^2}\frac{Z\kappa^4}{\kappa^2 + k^2} .
\end{equation}

Notably, the effective Green's function $\Gii^\prime = \Gii + \mc F_i$ for ion-ion interactions then simplifies to
\begin{equation}
    \Gii^\prime = \half \frac{\chi^2}{k^2 + \kappa^2}
\end{equation}
which is a Yukawa potential of screening parameter $\kappa$. The ions can therefore be treated as an isolated gas interacting through a Yukawa potential in a uniform neutralizing background. This is consistent with other theoretical predictions and with observations of ultracold neutral plasmas \citep{Chen_Simien_Laha_Gupta_Martinez_Mickelson_Nagel_Killian_2004,Gericke_Murillo_2003, Foster_et}.

With the potentials defined in \eqref{eq_G_defns}, the electrostatic part of the electron Massieu potential, which is a function of $\ci$, is 
\begin{equation}
\label{eq_Psie_coulomb}
\begin{split}
    \Phi_{\phi e}(\beta_e,  V, \ci) &= - \frac{V}{2}\int \frac{d^3k}{(2\upi)^3} \left[ \ln \left(\frac{\kappa^2}{k^2} + 1 \right)  - \frac{\kappa^2}{k^2} - \frac{1}{k^2}\frac{ Z \kappa^4}{\kappa^2 + k^2} \ci \right] .
\end{split}
\end{equation}

The electrostatic part of the ion Planck potential is

\begin{equation}
\label{eq_Psii_coulomb}
\begin{split}
    \Psi_{\phi i}(\beta_i, V, \fim) &= - \frac{V}{2}\int \frac{d^3k}{(2\upi)^3} \left[\ln \left(\frac{\chi^2}{k^2}  - 2\tau \mc F_e + 1 \right) - \frac{\chi^2}{k^2} \right].
\end{split}
\end{equation}

Finally, when the ions are at equilibrium, the electron Planck potential can be written in terms of the equilibrium ion-ion structure factor as
\begin{equation}
\label{eq_Xie_avg_coulomb}
\begin{split}
    \Psi_{\phi e}(\beta_e,V, \fem) &= - \frac{V}{2}\int \frac{d^3k}{(2\upi)^3} \left[ \ln \left(\frac{\kappa^2}{k^2} + 1 \right) - \frac{\kappa^2}{k^2} - \left(\frac{1}{k^2}\frac{ Z \kappa^4}{\kappa^2 + k^2} - 2\fem \right) \avg{S_{ii}}\right]  .
\end{split}
\end{equation}

\subsection{Structure factors}

We can immediately substitute the above Green's function definitions, as well as the expression \eqref{eq_f_coulomb} that we derived for the force scaling factor, to get the structure factors from \eqref{eq_Sii_Gs}, \eqref{eq_Sie_Gs}, and \eqref{eq_See_Gs}. For each pair of species we have

\begin{equation}
\begin{split}
    \avg{S_{ii}}(\bs k) &= \frac{k^2 + \kappa^2}{k^2 + \kappa^2 + \chi^2} ,
    \\
    \avg{S_{ie}}(\bs k) &= \frac{\sqrt{Z}\kappa^2}{k^2 + \kappa^2 +\chi^2} ,
    \\
    \avg{S_{ee}}(\bs k) &= \frac{k^2}{k^2 + \kappa^2} + \frac{1}{\tau}\frac{\kappa^2\chi^2}{(k^2 + \kappa^2)(k^2 + \kappa^2 + \chi^2)} .
\end{split}
\end{equation}

These expressions recover the known results for structure factors (and equivalently, position-space correlation functions) in two-temperature plasma \citep{Ecker_Kroll_1964,Salpeter_1963}. The structure factors can be made more physically transparent when written in terms of the deviation from those of an ideal gas, where correlations between particles vanish. In a system of uncorrelated particles, the structure factor for species $a, b$ is given by $S_{\sone\stwo} = \delta_{\sone\stwo}$ where $\delta$ is the Kronecker delta. This comes from the fact that in the formula for the structure factor,
\begin{equation}
    \avg{S_{\sone\stwo }(\bs k)} = \langle \trho^*_\sone \trho_{\stwo} \rangle = \frac{1}{\sqrt{N_\sone N_{b}}}\sum_j \e^{i\bs k \cdot r_{\sone,j}}\sum_\ell \e^{-i\bs k \cdot r_{\stwo,\ell}},
\end{equation}
the double summation includes $j=\ell$ terms, in which the same particle appears in both summations, in cases where $\sone = \stwo$. Each of these terms yields 1 while the other terms, which each describe a correlation between two distinct particles, average to zero in an ideal gas. We can write the structure factors as

\begin{equation}
\label{eq_Sss_coulomb}
\begin{split}
    \avg{S_{ii}(\bs k)} &= 1 - \frac{\chi^2}{k^2 + \kappa^2 + \chi^2} ,
    \\
    \avg{S_{ie}(\bs k)} &= \frac{\sqrt{Z}\kappa^2}{k^2 + \kappa^2 +\chi^2} ,
    \\
    \avg{S_{ee}(\bs k)} &= 1 - \frac{\kappa^2}{k^2 + \kappa^2} + \frac{1}{\tau}\frac{\kappa^2\chi^2}{(k^2 + \kappa^2)(k^2 + \kappa^2 + \chi^2)} .
\end{split}
\end{equation}

For finite $k$, the deviations from ideal gas behavior for both electron-electron and electron-ion structure factors vanish as expected when $\kappa  \rightarrow 0$, corresponding to large electron temperature. Similarly, the deviation in ion-ion structure factor vanishes as $\chi \rightarrow 0$, corresponding to large ion temperature. We note, however, that the electron-ion structure factor remains nonzero when $\chi \rightarrow 0$. This means that, although the ions become uncorrelated with each other in the large ion temperature limit, the finite-temperature electrons can still react to each ion configuration, allowing some electron-ion correlation to persist.

\subsection{Energy}
\label{sec_energy}

The ion electrostatic energy at equilibrium, calculated from \eqref{eq_U_phii_Gs}, is given by

\begin{equation}
\begin{split}
    U_{\phi i} &= T_i \frac{V}{4\upi^2} \chi^2 \int_0^\infty dk \left[\frac{-\chi^2}{k^2 + \kappa^2 + \chi^2}\right] ,
\end{split}
\end{equation}
which is easily integrated to find
\begin{equation}
\label{eq_Uphii_finite}
    U_{\phi i} = -T_i \frac{V\chi^4}{8\upi k_D} .
\end{equation}

For the electrons, the potential energy as a functional of $\ci$ is given from \eqref{eq_U_phie_Gs} by
\begin{equation}
\label{eq_U_phie_ci}
\begin{split}
    U_{\phi e}(T_e; S_{ii}) =& T_e \frac{V}{2}\int \frac{d^3k}{(2\upi)^3} \frac{\kappa^2}{k^2} \left[ \frac{-\kappa^2}{k^2 + \kappa^2} + \left(\frac{Z \kappa^4}{(k^2 + \kappa^2)^2}  - 2\frac{Z\kappa^4}{k^2(k^2 + \kappa^2)}\right)S_{ii}\right] .
\end{split}
\end{equation}

For equilibrium ions, i.e. an ion structure factor given by \eqref{eq_Sss_coulomb}, the electron potential energy becomes
\begin{equation}
\begin{split}
    U_{\phi e} &= T_e \frac{V}{4\upi^2} \kappa^2 \int_0^\infty dk \left[\frac{-\kappa^2}{k^2 + \kappa^2} + Z\frac{\kappa^4}{(k^2 + \kappa^2)(k^2 + \kappa^2 + \chi^2)} - 2Z\frac{\kappa^2}{k^2 + \kappa^2 +\chi^2}\right] 
\end{split}
\end{equation}
which reduces to
\begin{equation}
\label{eq_Uphie_finite}
\begin{split}
    U_{\phi e} &= \frac{V}{8\upi} T_e\kappa^3\left[-1 + Z \frac{\kappa^2 k_D - \kappa k_D^2 - \kappa \chi^2}{\chi^2 k_D}\right].
\end{split}
\end{equation}

Adding the energies of each species gives the total electrostatic potential energy as
\begin{equation}
\label{eq_Uphi_total}
    U_{\phi}(T_e, T_i) = \frac{V}{8\upi}\left[-T_e\kappa^3 - T_i(k_D^3 - \kappa^3)\right] .
\end{equation}

This result is identical to the one obtained by \cite{Foster_et}. It also matches calculations by \citet{Triola_2022}, discussed further below.

Although the partition of potential energy into the electron and ion subsystems given by \eqref{eq_Uphie_finite} and \eqref{eq_Uphii_finite} is not trivial, the combined energy has a straightforward physical interpretation. It is a well-known result \citep{Kelly_1963} that in weakly-coupled Coulomb-interacting plasma at thermal equilibrium, the potential energy is
\begin{equation}
    U_\phi^\mathrm{equilib}(T) = -\frac{V}{8\upi} T_e k_D^3 .
\end{equation}

\begin{sloppypar}
In \eqref{eq_Uphi_total}, we can identify the term ${-(V/8\upi)T_e\kappa^3}$ as the potential energy of a gas of electrons interacting directly through a Coulomb potential, and identify the term ${-(V/8\upi)T_i(k_D^3 - \kappa^3)}$ as the potential energy of a gas of ions interacting through a shielded potential.    
\end{sloppypar}

It is also useful to write the energy in terms of simple dimensionless parameters. As above, we write the temperature ratio as $\tau = T_e/T_i$. In terms of the plasma parameter $\Lambda = n\lambda_D^3$, where $N = (1 + Z)N_i$ and $n=N/V$, the total electrostatic energy is

\begin{equation}
\label{eq_Uphi_dimensionless}
\begin{split}
    U_\phi(T_e,T_i) &= -\frac{T_i N}{8\upi\Lambda}\left[1 + \frac{\tau - 1}{(1 + \tau Z)^{3/2}}\right] .
\end{split}
\end{equation}

\begin{sloppypar}
Here the energy includes a term with the same form as the equilibrium result ${U_\phi^\mathrm{equilib} = -TN/8\upi\Lambda}$, plus a term proportional to $(\tau - 1)$, which clearly vanishes in the equal-temperature case.
\end{sloppypar}

In the regime of interest, our \eqref{eq_Uphi_total} is equivalent to the potential energy derived in recent work \citep{Triola_2022} studying equations of state in two-temperature plasma. However, an advantage of the approach here is the fact that it affords a clear prescription for partitioning the potential energy into electron and ion components. Some methods exist, but they differ depending on one's chosen scheme for assigning the `cross-temperature' that applies to electron-ion interactions. Our approach, although more restricted in requiring the large mass ratio limit $m_e/m_i \ll 1$, has the advantage of providing a natural separation between electron and ion energies, given in \eqref{eq_Uphie_finite} and \eqref{eq_Uphii_finite}. This separation is necessary when we want to predict the evolution of electron and ion temperatures in response to heating or to volume changes. In the subsequent sections, we further develop the separate thermodynamic descriptions of the two species, and then apply these descriptions to calculate changes in electron and ion temperatures.

\subsection{Entropy}

In order to track heat flow into the two subsystems separately, it is necessary to compute the entropy separately for each species. If both species are individually in equilibrium, we should use the generalized Gibbs ensemble and apply \eqref{eq_S_species_general} to calculate both entropies. If $\ci$ is known instead, then we can calculate the electron entropy in the canonical ensemble using \eqref{eq_S_e_massieu}. The calculation in either case is similar and is outlined as follows.

In the generalized Gibbs ensemble, it is first necessary to calculate explicitly the Planck potential of each species from \eqref{eq_Psie_coulomb} and \eqref{eq_Psii_coulomb}. Because we will not need to take derivatives, we can substitute the force from \eqref{eq_f_coulomb} to obtain $\hat\Psi_i(\beta_e,\beta_i,V)$ and $\hat\Psi_e(\beta_e,\beta_i,V)$. For the ions first, we have

\begin{equation}
\begin{split}
    \hat\Psi_{\phi i} &= - \frac{V}{2}\int \frac{d^3k}{(2\upi)^3} \left[\ln(k^2 + \kappa^2 + \chi^2) - \ln(k^2 + \kappa^2) - \frac{\chi^2}{k^2}\right]
\end{split} 
\end{equation}
which evaluates to
\begin{equation}
    \hat\Psi_{\phi i} = \frac{V}{12\upi}(k_D^3 - \kappa^3) .
\end{equation}

The Planck potential for the electrons simplifies greatly when substituting the generalized force from \eqref{eq_f_coulomb}, giving
\begin{equation}
\label{eq_Psie_hat}
\begin{split}
    \hat\Psi_{\phi e} &= - \frac{V}{2}\int \frac{d^3k}{(2\upi)^3} \left[ \ln \left(\frac{\kappa^2}{k^2} + 1 \right) - \frac{\kappa^2}{k^2}\right] .
\end{split}
\end{equation}
and therefore
\begin{equation}
    \label{eq_Psii_hat}
\begin{split}
    \hat\Psi_{\phi e} &= \frac{V}{12\upi}\kappa^3 .
\end{split}
\end{equation}

\begin{comment}
    
Already with these explicit expressions for the Planck potentials, we can find the (entropic) pressures of each species. The electrostatic component of the entropic pressure $\varpi_{\phi \sone}$ of each species is
\begin{equation}
    \varpi_{\phi \sone} = \consfrac{\hat \Psi_{\phi \sone}}{V}{T_\sone T_\stwo}
\end{equation}
and the mechanical pressure is related as $P_\sone = T_\sone \varpi_\sone$.

The resulting pressures for each species are
\begin{equation}
\begin{split}
    P_e &= n_eT_e - \frac{1}{24\upi} \kappa^3
    \\
    P_i &= n_iT_i - \frac{1}{24\upi} (k_D^3 - \kappa^3)
\end{split}
\end{equation}
which correspond to the 
\end{comment}

For finding the entropy of each species, it is finally necessary to compute the term associated with the entropic force, which for species $s$ we will denote $\mc S_{fs}$. For the electrons, this term is defined as 
\begin{equation}
\begin{split}
    \mc S_{fe} &\doteq  \fem \cdot S_{ii} ,
\end{split}
\end{equation}
which integrates to
\begin{equation}
\begin{split}
    \mc S_{fe} &= \frac{V}{8\upi}Z\frac{\kappa^4}{k_D} .
\end{split}
\end{equation}

The entropic force contribution $\mc S_{fi} \doteq \fim \cdot \ci$ to the ion entropy results from a nearly identical calculation as 

\begin{equation}
    \mc S_{fi} = -\tau \frac{V}{8\upi}Z\frac{\kappa^4}{k_D} .
\end{equation}

Then in total, the electrostatic component $\mc S_{\phi e}$ of the electron entropy simplifies to

\begin{equation}
\label{eq_S_phi_e}
\begin{split}
    \mc S_{\phi e} &= -\frac{V}{24\upi}\kappa^3 - \frac{V}{ 8\upi} Z \frac{\kappa^4}{k_D + \kappa}
\end{split}
\end{equation}
and the electrostatic component $\mc S_{\phi i}$ of the ion entropy is
\begin{equation}
\label{eq_S_phi_i}
\begin{split}
    \mc S_{\phi i} &= \frac{V}{12\upi}(k_D^3 - \kappa^3) - \frac{V}{8\upi } \chi^2 k_D .
\end{split}
\end{equation}

We can also find the electron entropy in the canonical ensemble, corresponding to a known $\ci$. When the ions are at equilibrium this $\ci$ is the equilibrium structure factor $\savg$, and so we would have $\Phi_e(\beta_e, V, \ci) = \hat\Psi_e(\beta_e, V, \fev)$. For arbitrary ion correlations, we instead use \eqref{eq_S_e_massieu} for the electron entropy, which along with \eqref{eq_Phie_Gs} for the Massieu potential and \eqref{eq_U_phie_ci} for the electron energy gives
\begin{equation}
\begin{split}
    \mc S_{\phi e} =& -\frac{V}{2} \int \frac{d^3k}{(2\upi)^3} \Bigg[ \ln\left(\frac{\kappa^2}{k^2} + 1\right) - \frac{\kappa^2}{k^2} + \frac{\kappa^2}{k^2 + \kappa^2}  
    \\& - \left(\frac{Z^2\kappa^4}{k^2(k^2 + \kappa^2)}  + 2\frac{Z\kappa^4}{k^2(k^2 + \kappa^2)} - \frac{Z\kappa^4}{(k^2 + \kappa^2)^2}\right) \ci \Bigg] .
\end{split}
\end{equation}

We can now use these entropy expressions to study adiabatic processes, in which the entropy of one or both species remains constant.

%\subsection{Equation of State}

\section{Results}
\label{sec_results}

\subsection{Overview}
\label{sec_sub_overview}

In this section, we apply the equations of state for two-temperature electron-ion Coulomb plasma to describe what happens if the plasma is compressed or heated. Many of the central analytical results of this work are presented here. Namely, \eqref{eq_dTdV_e} and \eqref{eq_dTdV_i} describe the temperature change for each species as the system undergoes adiabatic compression. In \eqref{eq_correlation_DeltaTi} and \eqref{eq_DTi_suddenheating}, we present the general and specific cases, respectively, of a new, simple formula for the change in ion temperature due to correlation heating after heat is deposited suddenly into the electrons.

\subsection{Compression}
\label{sec_compression}

In experimental conditions that give rise to moderately-coupled two-temperature plasma, it is common that the system also undergoes expansion or compression \citep{Koenig_Benuzzi-Mounaix_Ravasio_Vinci_Ozaki_Lepape_Batani_Huser_Hall_Hicks_et,Beule_Ebeling_Forster_1997,Loisel_Bailey_Liedahl_Fontes_Kallman_Nagayama_Hansen_Rochau_Mancini_Lee_2017, Eliezer_Henis_Nissim_Pinhasi_Val_2015, Kodama_Norreys_Mima_Dangor_Evans_Fujita_Kitagawa_Krushelnick_Miyakoshi_Miyanaga_et,Fortov_Ilkaev_Arinin_Burtzev_Golubev_Iosilevskiy_Khrustalev_Mikhailov_Mochalov_Ternovoi_et}. It is therefore of interest to predict how the electron and ion temperatures will evolve during volume changes. We consider here a reversible process in which no heat is transferred to or from the system. Because heat transfer is additionally forbidden between the electron and ion subsystems, the entropy of each system remains constant throughout the process. This provides two constraints on the temperatures, which are sufficient to solve for the change in electron and ion temperature in response to a change in volume. 

For brevity, we define the partial derivatives of entropy to be

\begin{equation}
\begin{split}
    \upi_\sone &\doteq \left( \frac{\partial\mc S_\sone}{\partial V} \right) ,
    \\
    \sigma_{\sone\stwo} &\doteq \left( \frac{\partial\mc S_\sone}{\partial T_\stwo} \right) .
\end{split}
\end{equation}

We then calculate temperature changes by enforcing constant entropy to each order in $\Lambdainv$. The calculation is outlined in detail in Appendix~\ref{sec_compression_temperature_changes}. Applying the results, we have to leading order for both species simply
\begin{equation}
\label{eq_dTdV_0th}
    \left( \frac{dT_\sone}{dV}\right)^{(0)} = -\frac{2}{3}\frac{T_\sone}{V} ,
\end{equation}
which is the ideal gas result for adiabatic compression. To next order, we have the temperature change for the electrons as

\begin{equation}
\label{eq_dTdV_e}
    \left( \frac{dT_e}{dV}\right)^{(1)} = -\frac{2}{3}\frac{T_e}{N_e} \frac{\kappa^3}{48\upi}\left[-1 -3Z\frac{\kappa}{k_D + \kappa}\right] ,
\end{equation}
while for the ions the temperature change is
\begin{equation}
\label{eq_dTdV_i}
    \left( \frac{dT_i}{dV}\right)^{(1)} = -\frac{2}{3}\frac{T_i}{N_i} \frac{\kappa^3}{48\upi}\left[-1 - \frac{k_D^3 + \kappa^3}{\kappa^3} + 3\frac{k_D}{\kappa}\right] .
\end{equation}

For both species, the first-order correction to the temperature change during expansion is positive for all electron and ion temperatures. Since the zeroth-order contribution \eqref{eq_dTdV_0th} is negative, the result is that in a weakly-coupled plasma, the magnitude of the temperature decrease during expansion is smaller than that of an ideal gas. Similarly, compression of a weakly-coupled plasma will increase the temperature of both species, but slightly less than it would for an ideal gas. It is evident that the first-order corrections to the two species' temperatures are different. We further explore the implications of this finding in \S\ref{sec_discussion}.

\subsection{Reversible correlation heating}
\label{sec_rev_heating}

Our results can also be used to determine how changes in electron temperature and ion temperature couple to each other. There exist many processes in two-temperature plasma that transfer energy unequally to the two species. These include X-rays, which heat electrons through reverse Bremsstrahlung; radiative cooling, which primarily cools electrons; cyclotron resonance heating, which can heat either species depending on the frequency; shocks, which primarily heat ions; and viscous dissipation, which generally heats ions \citep{Haines_LePell_Coverdale_Jones_Deeney_Apruzese_2006}. Eventually, collisions will transfer energy between species in order to equalize the electron and ion temperatures. However, if the energy input happens much faster than electron-ion equilibration, we may ask whether energy is nevertheless transferred between species on a fast timescale.

Working at constant volume, we apply temperature change $dT_\sone^{(0)}$ to species $\sone$ slowly enough that the entropy of species $\stwo$ is constant, following a similar procedure to that used above to study compression. Heuristically, where $\omega$ is the rate of change of $T_\sone$ and $\nu_{\stwo\stwo}$ is the collision frequency for particles of species $\stwo$ with each other, we say that $\omega \ll \nu_{\stwo\stwo}$ is a sufficient condition for the process to be treated as adiabatic for the dynamics of species $\stwo$. The first-order conservation of entropy condition gives
\begin{equation}
    0 = \sigma^{(1)}_{\stwo\sone}dT_\sone^{(0)} + \sigma^{(0)}_{\stwo\stwo}dT^{(1)}_\stwo .
\end{equation}

Then using the expressions for derivatives of entropy that we have in \eqref{eq_pis_sigmas}, it is straightforward to find that
\begin{equation}
\label{eq_crossspecies_differential}
\begin{split}
    \left(\frac{dT_i}{dT_e}\right)_V^{(1)} &= \frac{T_i}{T_e}\frac{1}{N_i}\frac{V}{24\upi} \kappa^2k_D\left(1 - \frac{\kappa}{k_D}\right)^2 ,
    \\
    \left(\frac{dT_e}{dT_i}\right)_V^{(1)} &= \frac{T_e}{T_i}\frac{1}{N_e}\frac{V}{24\upi} \kappa^2k_D\left(1 - \frac{\kappa}{k_D}\right)^2 .
\end{split}
\end{equation}

Therefore, as the temperature of one species is increased slowly, the temperature of the other species increases. This comes about because $\sigma_{ei}, \sigma_{ie} < 0$ and so the potential component of the entropy of species $\stwo$ decreases as $T_\sone$ increases; the kinetic component of $\mc S_\stwo$ then has to increase in order to keep the entropy constant, which requires $T_\stwo$ to increase. In statistical terms, species $\stwo$ becomes more spatially correlated as species $\sone$ is heated; in order to preserve the entropy $\mc S_\stwo$, the disorder must increase in some other degrees of freedom of $\stwo$. In this case, those other degrees of freedom are the velocities, and increasing their entropy requires the temperature to increase.

This `correlation heating' is collisionless in that it occurs due to spatial rearrangement of particles of each species, without reference to the collision rate between species. We have put no constraints on the inter-species energy exchange rate $\nuie$, and so collisional energy exchange can be made much slower than the rate $\omega_\stwo$ at which we expect the spatial distribution of species $\stwo$ to rearrange itself. The physical mechanisms of correlation heating become more transparent when we study the effects of suddenly depositing a large amount of heat into one species, rather than the slow and reversible heating that we have just analyzed.

\subsection{Sudden heating}
\label{sec_irrev_heating}

The formalism in this work can be applied in some cases where the plasma is not even in a two-temperature quasi-equilibrium state. For example, we can make well-defined thermodynamic statements about the electron subsystem before the ion subsystem has had time to equilibrate. We can use this to answer the question of how energy is partitioned when we suddenly transfer heat only to the electrons. For example, a sudden burst of x-rays may transfer energy to the electrons while barely affecting the ions; we may then ask how energy flows into the ion distribution over a longer timescale. We analyze a process consisting of the following steps:

\begin{enumerate}
    \item Instantaneously, energy $Q_e$ is deposited into the electron velocity distribution.
    \item \label{enum_elecequilib} On a timescale $\sim \nuee^{-1}$, the electrons come to equilibrium at their new temperature. The ion distribution has not had time to evolve and so the ion-ion structure factor remains at the original value $S_{ii,0}$ .
    \item \label{enum_ionequilib} On a longer timescale $\sim \omega_{pi}^{-1}$, the ion distribution relaxes to its new equilibrium structure factor $S_{ii,2}$ in response to the new electron screening.
    \item On a much longer timescale $\sim \nuie ^{-1}$, collisions transfer energy from the electrons to the ions and bring the two species' temperatures into equilibrium.
\end{enumerate}

We study the dynamics during Step~\ref{enum_ionequilib}, before collisions have had enough time to transfer any heat between species. Energy can, however, be transferred between species by an `effective work' defined as  
\begin{equation}
\label{eq_Ws_defn}
    \mc W_s = \mc F_s\cdot d\ci .
\end{equation}Physically, this corresponds to the work\footnote{As defined, $\mc W_s$ is an entropic quantity, but can be converted to an energetic quantity by multiplying by $T_s$.} required to shift ion correlations while working against (or with) the effective potential generated by the electrons. We show in this section that the relaxation dynamics can be captured by treating the ions as a Yukawa one-component plasma and requiring that their energy be conserved; this known result, however, is specific to the Coulomb interactions and first-order expansion to which we specialized in \S\ref{sec_electron_ion}. The approach outlined here is more general and could be applied to compute correlation heating for other effective interaction models.

\begin{figure}
    \centering
    \includegraphics[width=0.9\textwidth]{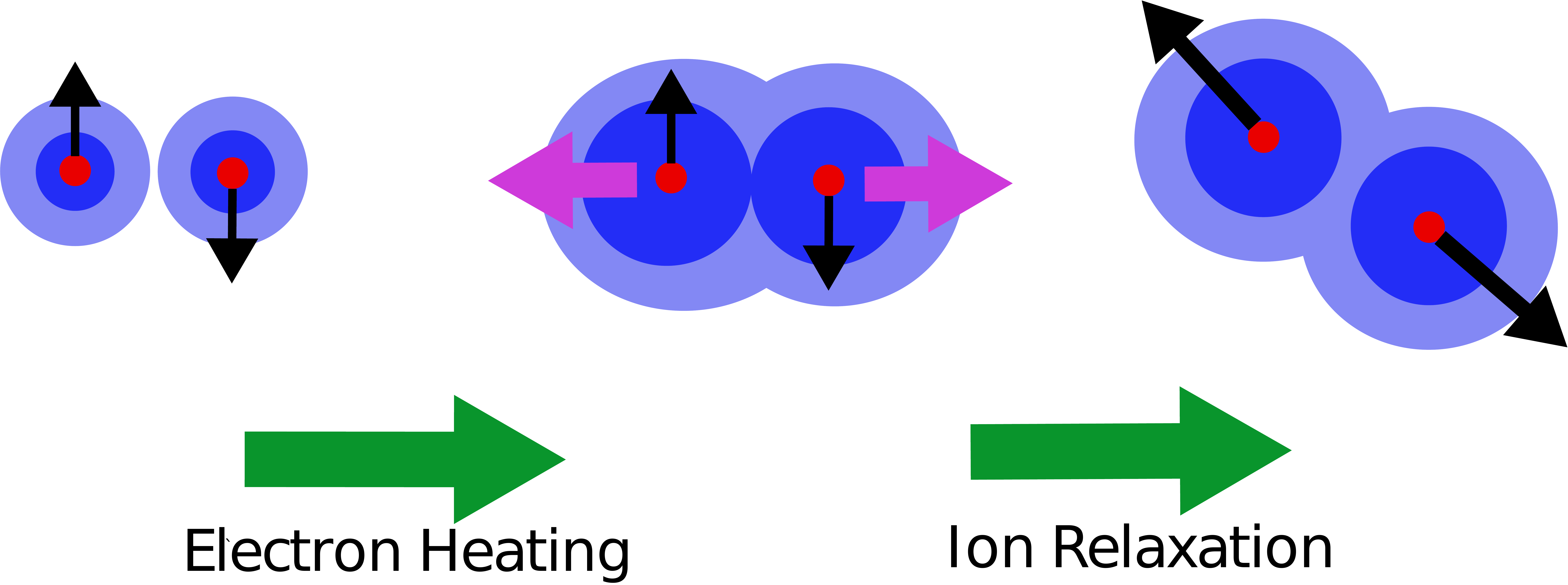}
    \caption{Steps of the correlation heating process. Two ions, closely shielded by electron clouds, pass near each other by chance. At that moment, some fast process heats the electrons without directly transferring energy to the ions; the electron screening clouds therefore expand. Under their new reduced shielding, the ions are suddenly subjected to a force pushing them apart, causing them to gain kinetic energy.}
    \label{fig_heating_clouds}
\end{figure}

The steps of this process are shown schematically in Fig.~\ref{fig_heating_clouds}. Ions are shown in red, and the characteristic size of electron screening clouds is represented by concentric circles. In a weakly-coupled plasma, each screening cloud should include many ions, but a simplified picture is given here for illustration. In Step~\ref{enum_elecequilib}, the electron clouds expand so that each ion suddenly feels the electric field due to nearby ions more significantly than before. While well-shielded ions were previously allowed to be relatively close to each other, the less-shielded electric field now pushes ions apart, imparting kinetic energy to the ion distribution on average. This process has been predicted theoretically, and observed in ultracold neutral plasma experiments \citep{Killian_Pattard_Pohl_Rost_2007, Kuzmin_ONeil_2002, Morawetz_Bonitz_Morozov_Ropke_Kremp_2001}. The phenomenon has been given several names, including `disorder-induced heating' and `correlation heating.' However, little work has been done on first-principles analytical models to predict the magnitude of this heating effect. The effect is largest in a more strongly-coupled regime, which is generally intractable for the kind of analytical work presented here, but we can gain insight into the process by studying the leading-order effects in $\Lambdainv$.

The calculation of the change in ion temperature is worked out in detail Appendix~\ref{sec_sub_sudden_heating}. In brief, the calculation relies on conservation of total energy during Step~\ref{enum_elecequilib} and Step~\ref{enum_ionequilib}. Additionally, because this constraint is not sufficient to specify the final ion temperature, the calculation treats the ion relaxation in Step~\ref{enum_ionequilib} as adiabatic for the electron subsystem. It is straightforward to determine the change in structure factor $\Delta S_{ii} = S_{ii,2} - S_{ii,0}$ during the process. As a function of $\Delta S_{ii}$, we find that the change in ion temperature $\Delta T_i = T_{i2} - T_{i0}$ is
\begin{equation}
\label{eq_correlation_DeltaTi}
    \Delta T_i = -\frac{2}{3} T_{i0} V\int\frac{d^3k}{(2\upi)^3} \left(\Gii - \tau \mc F_e\right) \Delta S_{ii} .
\end{equation}

The right side of this expression is just the change in an effective ion potential energy $E_i^\mathrm{eff} = T_i\Gii^\prime \cdot \ci$ corresponding to the effective potential that we found previously. The increase in ion kinetic energy is equal to the decrease in effective potential energy. Since the effective Green's function $\Gii^\prime$ corresponds to a Yukawa potential, this result indicates that the ion subsystem can be treated as a Yukawa one component plasma (YOCP), with screening length given by the electron inverse Debye length $\kappa$, and whose energy is conserved during equilibration. This is in line with results obtained elsewhere demonstrating that the ions can be treated as a YOCP \citep{Gericke_Murillo_2003}. Parallel work shows that, under the same assumptions, the ion collision operator is the same as that determined from a collection of screened ions \citethomas. Therefore, in addition to the equilibrium state found here, the dynamical evolution of the ion distribution is described by a YOCP model.

\begin{sloppypar}
To write this temperature change explicitly, we will denote the electron screening lengths before and after the initial heating step as ${\kappa_0^2 = 4\upi e^2 n_e/T_{e0}}$ and ${\kappa_1^2 = 4\upi e^2 n_e/T_{e1}}$. To the relevant order, changes to the ion screening length are irrelevant, so ${\chi^2 = 4\upi^2 Z^2 e^2 n_i/T_{i0}}$ can be used at all stages. Finally, the total screening lengths will be ${k_{D0}^2 = \kappa_0^2 + \chi^2}$ and ${k_{D1}^2 = \kappa_1^2 + \chi^2}$. Using \eqref{eq_f_coulomb} and \eqref{eq_Sss_coulomb} the temperature change becomes
\end{sloppypar}
\begin{equation}
    \Delta T_i = \frac{2}{3} \frac{T_{i0}}{N_i} \frac{V}{2} \chi^2 \int\frac{d^3k}{(2\upi)^3} \frac{1}{k^2}\frac{k^2}{k^2 + \kappa_1^2} \left(\frac{\chi^2}{k^2 + k_{D1}^2} - \frac{\chi^2}{k^2 + k_{D0}^2}\right) .
\end{equation}

After integrating, we obtain the following formula for the change in ion temperature:

\begin{equation}
\label{eq_DTi_suddenheating}
    \Delta T_i = \frac{2}{3}\frac{T_{i0}}{N_i}  \frac{V\chi^4}{8\upi} \left(\frac{1}{k_{D1} + \kappa_1} - \frac{1}{k_{D0} + \kappa_1}\right) .
\end{equation}

This formula holds for arbitrarily large changes in $\kappa_0 \rightarrow \kappa_1$, within the constraint of weak coupling. We can also find its differential form by imposing that the initial heating is small and then expanding to first order in $dT_e$. The expression for $dT_i/dT_e$ obtained in this way matches exactly that in \eqref{eq_crossspecies_differential}, found previously through the conservation of ion entropy. However, when evaluated for a finite $\Delta T_e$, \eqref{eq_DTi_suddenheating} is clearly unequal to the differential expression in \eqref{eq_crossspecies_differential} integrated over the same temperature interval. This is unsurprising because these equations describe different physical processes. The differential case is reversible by construction; the ion entropy is forced to be constant, and although the electron entropy increases due to the external heating, this may be done reversibly. By contrast, the relaxation process in the sudden heating case is irreversible; the ion subsystem departs from its thermal equilibrium state in Step~1, and the final ion entropy is greater than that at the beginning.

\section{Discussion}

\label{sec_discussion}

\subsection{Energy partition}

\label{sec_sub_disc_energy_partition}

Many interesting features of the physics of two-temperature plasma can be seen by analyzing the potential energy formula \eqref{eq_Uphi_total} in the appropriate limits. In all plots in this section, energies are normalized to the characteristic energy scale 
\begin{equation}
    \energynorm = \frac{e^2}{a_e}
\end{equation}
representing the characteristic potential energy of interaction between electrons, where $a_e = (4\upi n_e/ 3)^{-1/3}$ is the Wigner-Seitz radius, a characteristic interparticle distance. 
 
\begin{figure}
    \begin{subfigure}{0.5\columnwidth}
        \centering
        \includegraphics[width=\columnwidth]{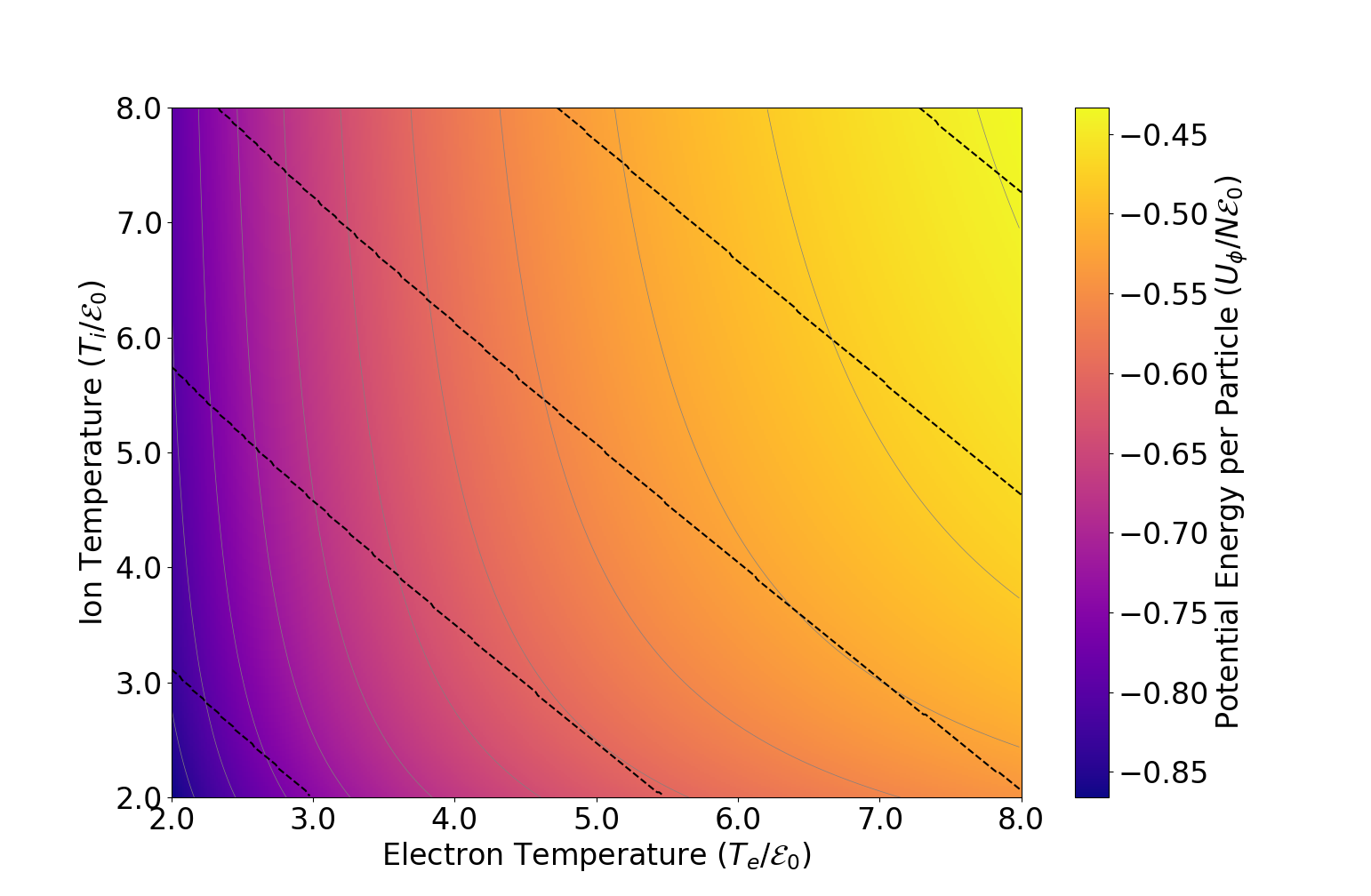}
        \caption{Ion charge $Z = 1$ }
        \label{fig_temps_heatmapZ1}
    \end{subfigure}
    \begin{subfigure}{0.5\columnwidth}
        \centering
        \includegraphics[width=\columnwidth]{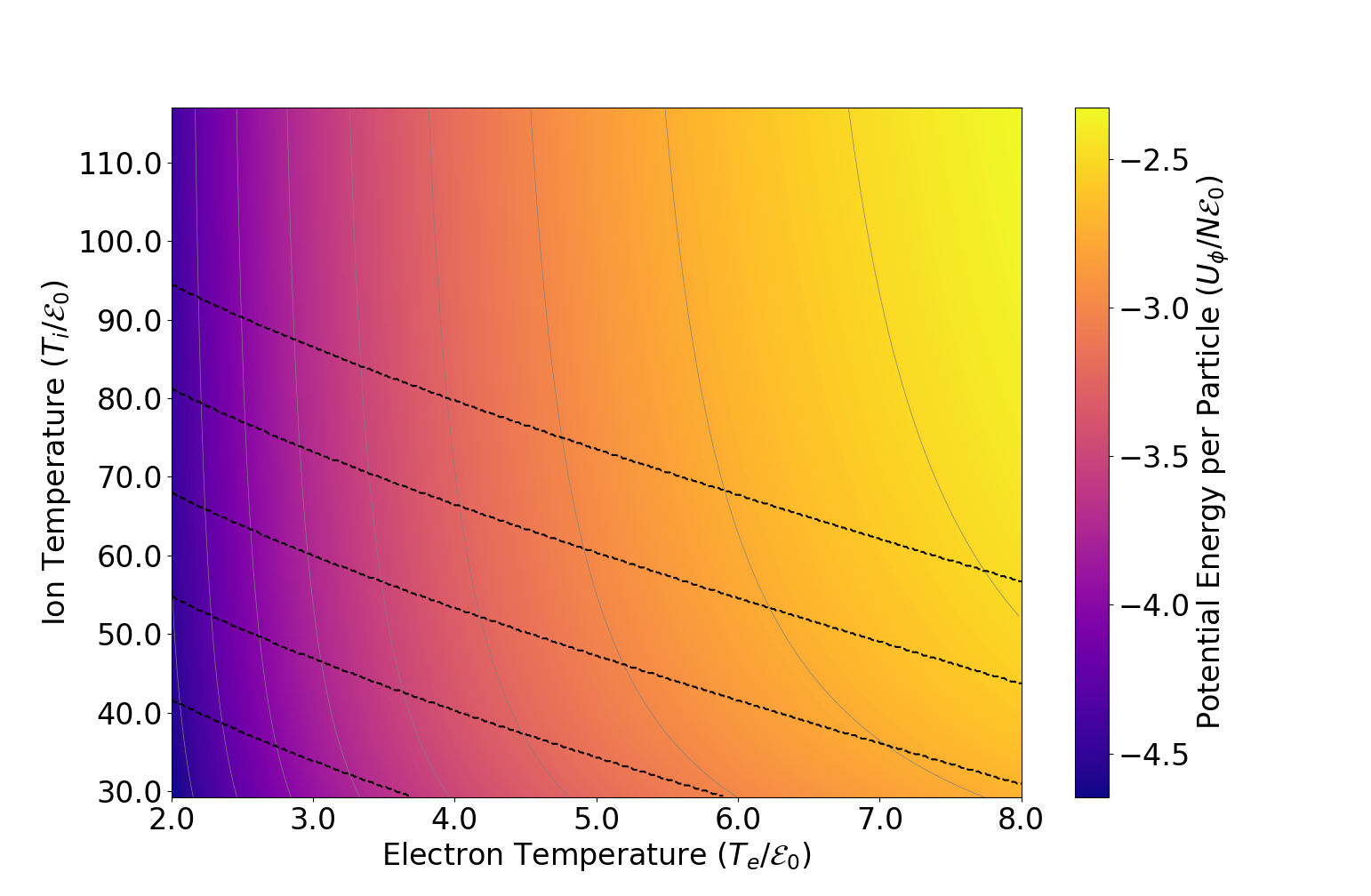}
        \caption{Ion charge $Z = 5$}
        \label{fig_temps_heatmapZ5}
    \end{subfigure}
    \caption{Electrostatic potential energy per particle in weakly-coupled two-temperature plasma, with equipotential contours shown in grey. All energies are listed in units of $\energynorm = e^2 / a_e$. Dashed black lines are curves of constant total energy.}
    \label{fig_temps_heatmap}
\end{figure}

In Fig.~\ref{fig_temps_heatmapZ1} we demonstrate the partition of energy into kinetic and potential degrees of freedom in a $Z=1$ plasma at constant temperature. The horizontal and vertical axes represent the electron temperature and ion temperature, respectively. The heatmap shows the electrostatic potential energy, which is negative in all cases. The axis scale is different in the $Z=1$ and $Z=5$ cases because $\energynorm$ is independent of $Z$, and higher ion charge leads to more potential energy and stronger coupling.

Even if a system is constrained by constant energy, it may still move around the energy plot by exchanging energy between species. The dashed black curves show trajectories of constant $U = U_K + U_\phi$, where $U_K$ is the total kinetic energy in the plasma. The total energy can be written as 
\begin{equation}
    U = \frac{3}{2}N_eT_e + \frac{3}{2}N_iT_i - T_i \frac{Vk_D^3}{8\upi}\left[1 + \frac{\tau - 1}{(1 + Z\tau)^{3/2}}\right] ,
\end{equation}
and we can determine symbolically what happens when one temperature is made much larger than the other while holding the volume and particle numbers fixed.

First taking the hot electron limit, $\tau \rightarrow \infty$ at fixed $T_i$, the potential energy reduces to 
\begin{equation}
    U_\phi \rightarrow -T_i\frac{V\chi^3}{8\upi} ,
\end{equation} which, as we argued in \S \ref{sec_energy}, is just the potential energy of an ion OCP at equilibrium. The effects of electron screening vanish and the electrons become a uniform neutralizing background.

Next taking the hot ion limit, $\tau \rightarrow 0$ at fixed $T_e$, the potential energy reduces to
\begin{equation}
    U_\phi \rightarrow -T_e \frac{V\kappa^3}{8\upi}\left[1 + \frac{3}{2}Z\right] ,
\end{equation}
which is almost the potential energy of an electron OCP at equilibrium. However, the ion contribution supplies an additional $3Z/2$ term because the ions do not simply act as a uniform neutralizing background. Although ion positions are uncorrelated with each other in the large $T_i$ limit, any realization of the ion distribution still consists of $N_i$ discrete charges. Because the electrons equilibrate much faster than the ion distribution rearranges itself, the electrons have time to cluster around the ions, allowing the system to reach a lower energy state than would be found in a one-component electron plasma. This asymmetry is visible in Fig.~\ref{fig_temps_heatmapZ1}.

We can additionally study the effect of varying the ion charge, where for simplicity we continue to assume that all ions in the plasma have the same charge. In Fig.~\ref{fig_temps_heatmapZ5}, we plot the partition of energy in the same way in a $Z=5$ plasma. Following this trend into the $Z \rightarrow \infty$ limit, our potential energy expression reduces to
\begin{equation}
    U_\phi \rightarrow -T_i \frac{V\chi^3}{8\upi} ,
\end{equation}
which is just the energy of a coulomb-interacting ion gas. In this limit, the ion-ion interaction is strong enough that the energy in the system is dominated by direct interactions between ions, so electron screening provides only a negligible correction, and the effect of varying the electron temperature becomes unimportant. In the opposite limit, where $Z \rightarrow 0$, we find
\begin{equation}
    U_\phi \rightarrow -T_e \frac{V\kappa^3}{8\upi} .
\end{equation}
Here, the ions have become a uniform neutralizing background, and so their influence on the potential energy vanishes. This limit is, of course, not physically realizable because it would entail fractional ion charge. However, we can still evaluate it mathematically and expect the theory to be well-behaved in this limit. The $Z \to 0$ limit simply means that there are many particles of the heavy species and fewer particles of the light species, which (by quasineutrality) have greater charge.

\subsection{Compressional temperature separation}

\label{sec_sub_disc_compression}

\begin{figure}
    \begin{subfigure}{0.5\columnwidth}
        \centering
        \includegraphics[width=\columnwidth]{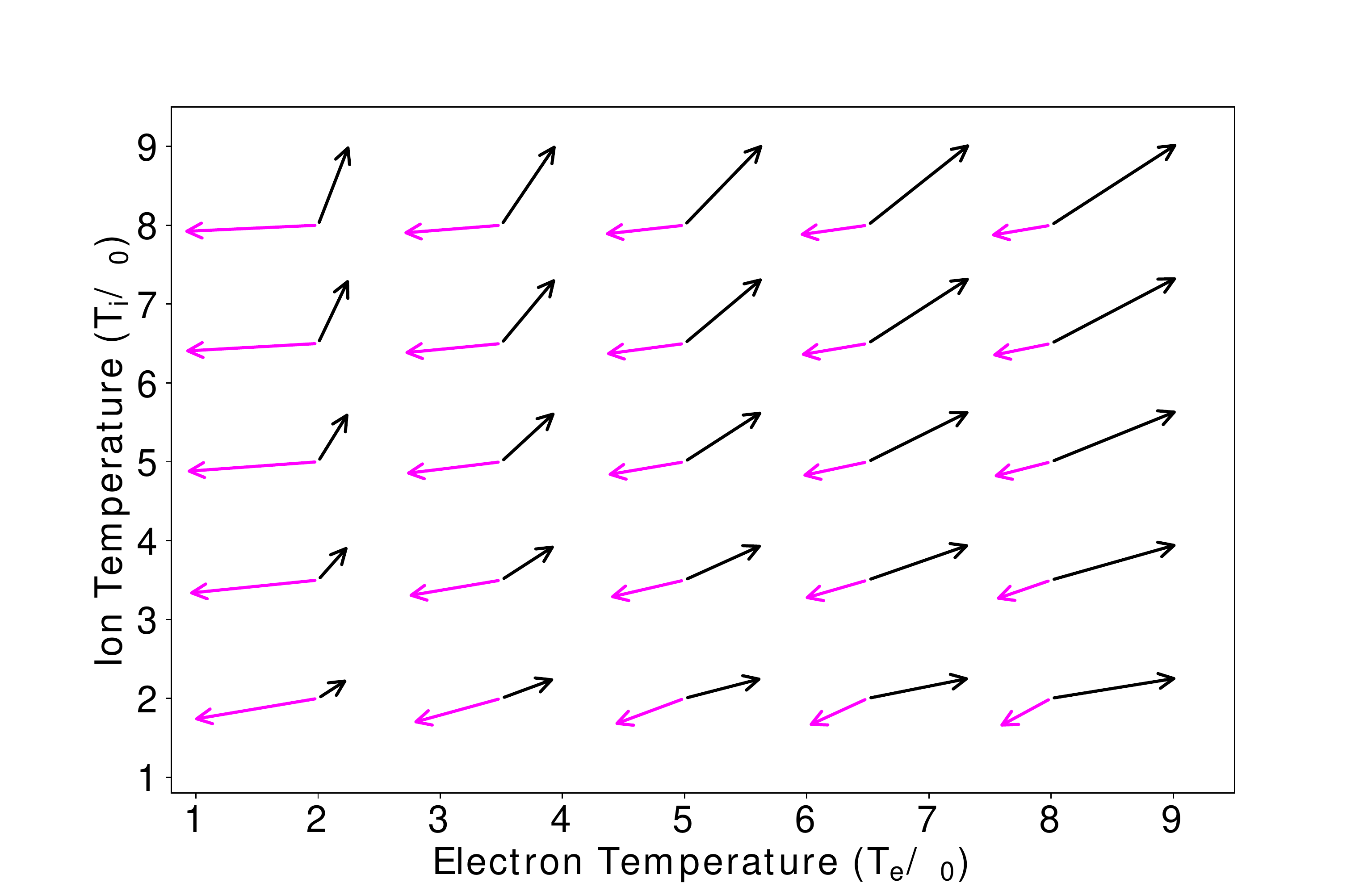}
        \caption{Ion charge $Z = 1$ }
        \label{fig_compressionZ1}
    \end{subfigure}
    \begin{subfigure}{0.5\columnwidth}
        \centering
        \includegraphics[width=\columnwidth]{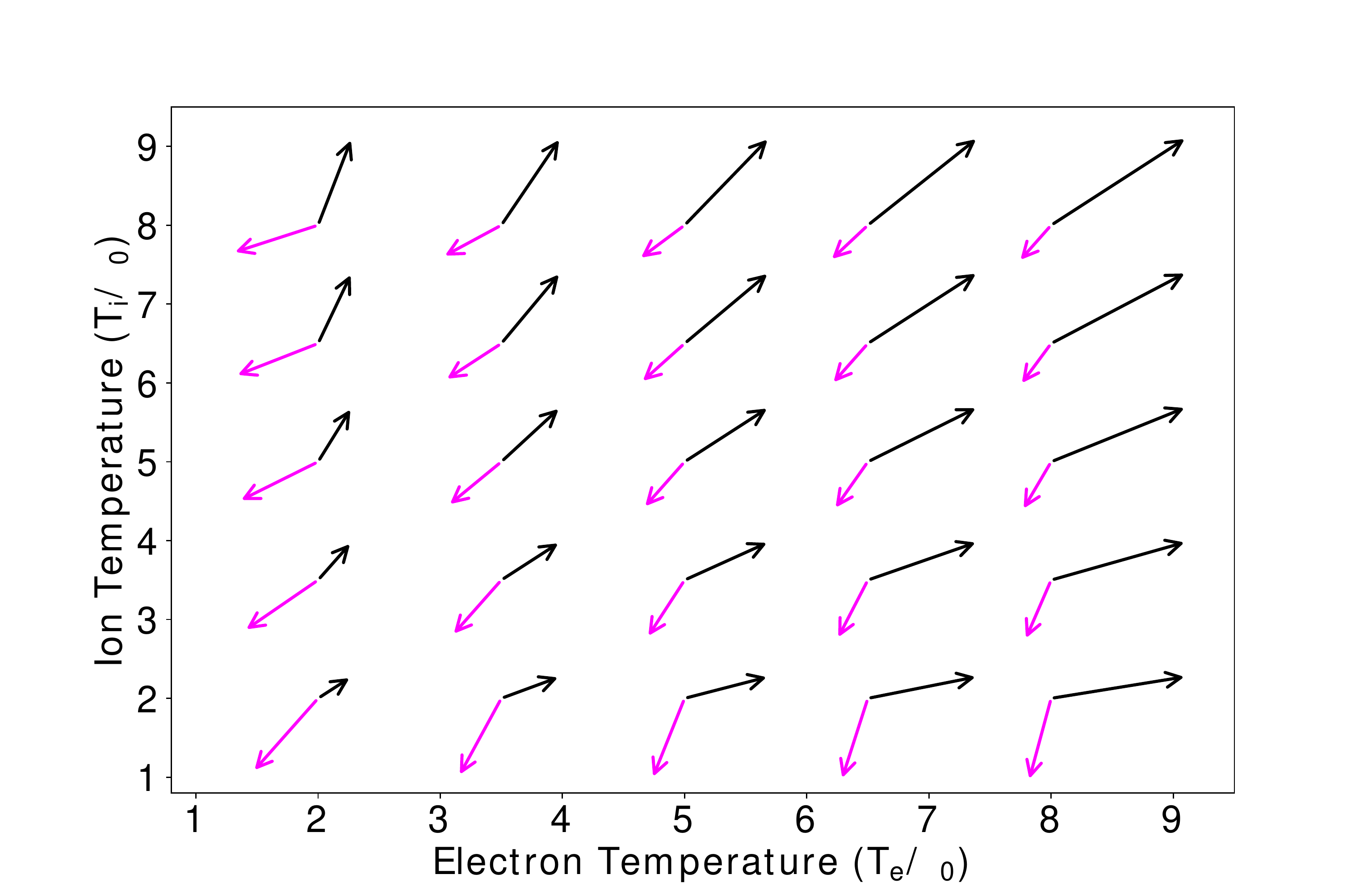}
        \caption{Ion charge $Z = 5$}
        \label{fig_compressionZ5}
    \end{subfigure}
    \caption{Temperature change during compression shown in $T_e, T_i$ space. Black arrows show the direction of the temperature change $(dT_e, dT_i)$ of an ideal gas associated with some small compressional volume change $(dV < 0)$. The magenta arrows show the first-order (in $\Lambdainv$) correction to the temperature change in a moderately coupled plasma with ion charge $Z=1$ (left) and $Z=5$ (right). Both temperature axes are normalized to the characteristic energy scale $\energynorm$. Arrow lengths are normalized to arbitrary units; the ideal gas gas and plasma correction arrows are normalized separately.}
    \label{fig_compression_2T}
\end{figure}

The plots in Fig.~\ref{fig_temps_heatmap} apply to a plasma held at constant volume. As we found in \S\ref{sec_compression}, the temperatures of both species change by a different amount than would be expected of an ideal gas during compression and expansion. Fig.~\ref{fig_compression_2T} displays the differential temperature change vector $(dT_e, dT_i)$ associated with a differential step of adiabatic compression for plasmas with ion charge states $Z=1$ and $Z=5$. 
%It is worth noting that the axis scales, in units of temperature normalized to $\energynorm = e^2/a_e$, are different in the two plots. This is necessary to ensure that both species remain moderately coupled throughout the domain of the plot.

The standard ideal gas result (i.e. infinitely weakly coupled) is shown in black, with the corrections in moderately-coupled plasma in magenta. The black and magenta vectors are normalized separately. For many conditions, e.g. high $T_e$ in the $Z=1$ case, the plasma correction is small, but reduces the magnitude of the temperature change. In this case, the species' temperatures change by (approximately) the same ratio as in an ideal gas, but the plasma coupling causes both species to heat less during compression than ideal gases would. For other conditions, namely low electron temperature and comparatively high ion temperature, the two species' temperatures are affected differently by compression. In the $Z=1$ case, starting at $T_i > T_e$ causes the ion temperature to increase more than the electron temperature, driving an even greater temperature separation than would be generated by ideal-gas dynamics. This is of interest in fusion applications, where operating in a hot-ion mode is often desirable. The hot ion mode can be particularly beneficial in magnetic fusion devices \citep{Clarke_1980, Fisch_Herrmann_1994}, which operate in a weakly-coupled regime. In that regime, since the fusion alpha particles slow down on electrons, the method to achieve hot ion mode relies upon a wave-induced alpha channeling effect \citep{Fisch_Rax_1992}. In contrast, we show here that hot ion mode might in fact be accomplished through a compression effect.

Conversely, in the $Z=5$ case, the most dramatic temperature separation can be seen in the $T_e > T_i$ region, in which the electron temperature is driven even higher. This final effect can be easily explained as follows. We can see from \eqref{eq_dTdV_e} and \eqref{eq_dTdV_i} that the temperature change for each species includes a prefactor that is inversely proportional to the number of particles of that species. This is primarily a result of the fact that the amount of energy transferred to each species $\sone$ depends only weakly on $N_\sone$, while the heat capacity is proportional to $N_\sone$. Therefore, for greater $Z$, there is a smaller number of ions, and the first-order moderate-coupling correction to $(dT_i/dV)$ is more significant. Because compressing plasmas are a promising option for intense x-ray sources \citep{Chin_Ruby_Nilson_Bishel_Coppari_Ping_Coleman_Craxton_Rygg_Collins_2022}, the hot electron mode achieved under these conditions may be of practical interest.

\begin{figure}
    \begin{subfigure}{0.5\columnwidth}
        \centering
        \includegraphics[width=\columnwidth]{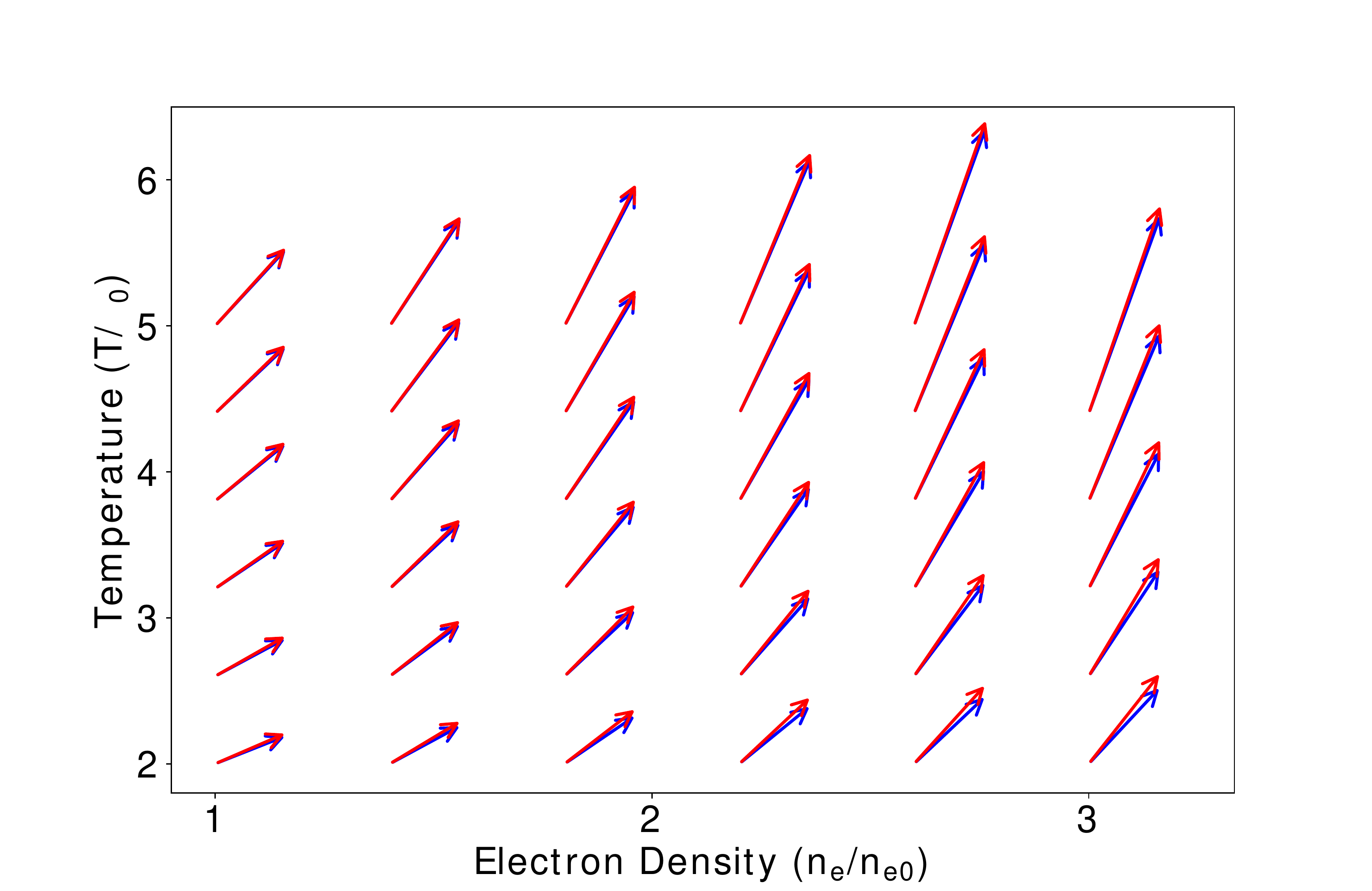}
        \caption{Ion charge $Z = 1$ }
        \label{fig_compression_nt_Z1}
    \end{subfigure}
    \begin{subfigure}{0.5\columnwidth}
        \centering
        \includegraphics[width=\columnwidth]{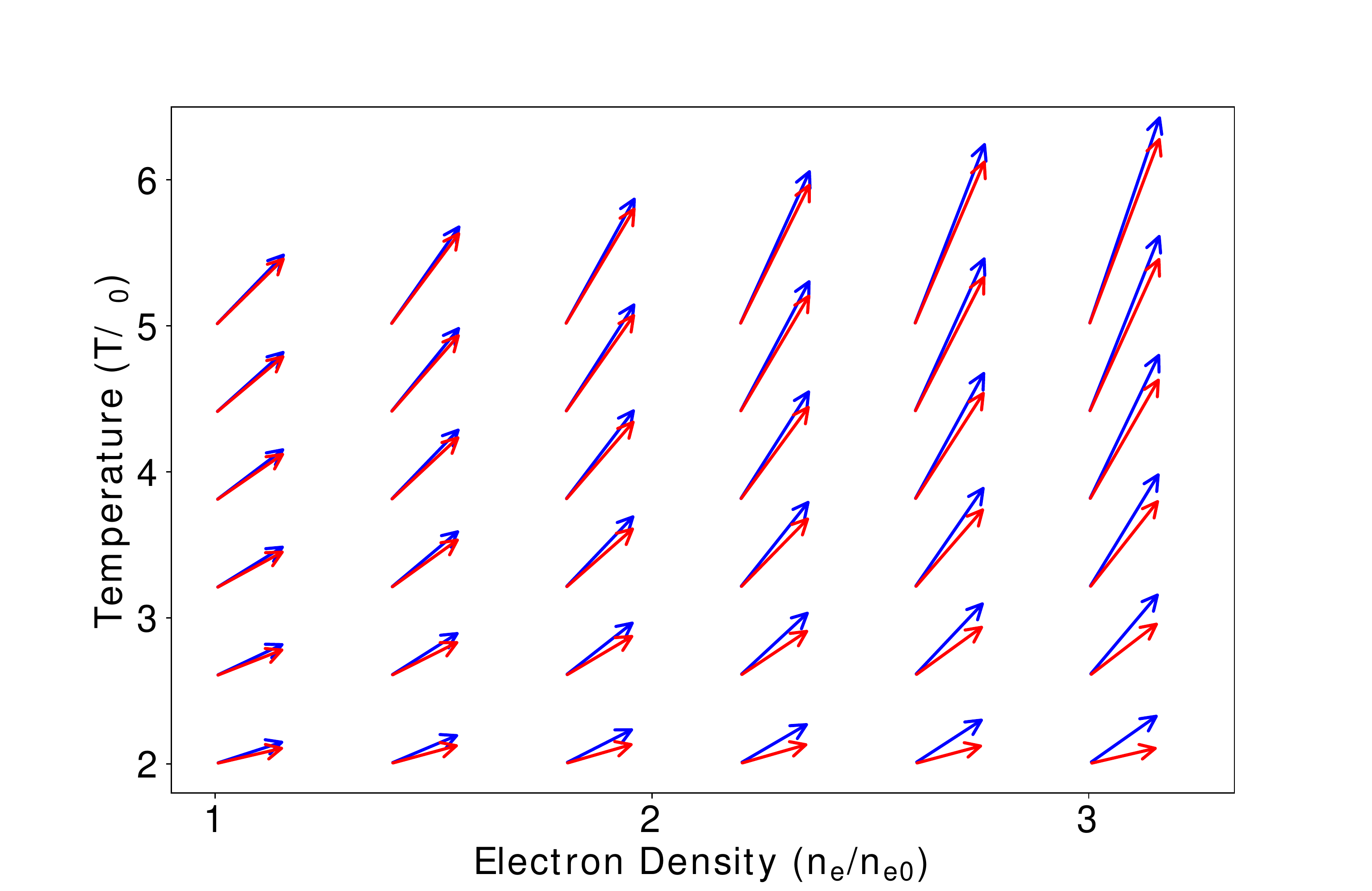}
        \caption{Ion charge $Z = 5$}
        \label{fig_compression_nt_Z5}
    \end{subfigure}
    \caption{Changes in electron and ion temperature as a function of density and temperature, starting in an equal temperature state $\tau = 1$. For given density change $dn$, blue arrows show $dT_e$ and red arrows show $dT_i$. The temperature axis is normalized to the characteristic energy scale $\energynorm$.}
    \label{fig_compression_nT}
\end{figure}

Interestingly, even when the ion and electron temperatures start equal, compression of a moderately-coupled plasma can induce a spontaneous separation of the temperatures. Fig.~\ref{fig_compression_nT} shows the evolution of electron (blue) and ion (red) temperatures for a variety of starting densities $n_e$ and starting temperatures $T = T_e = T_i$. Results are shown for ion charge states $Z=1$ and $Z=5$. In the $Z=1$ case, adiabatic compression generally drives the ion temperatures higher than the electron temperature. The effect is small, but is most pronounced at high density and low temperature as expected. In the $Z=5$ case, adiabatic compression from equal temperatures drives the electron temperature higher than the ion temperature for all conditions shown.

These results may be seen as unexpected because, for equal pressures and densities, the electron pressure is higher than the ion pressure in moderately-coupled plasma and the electrons could therefore be expected to heat more. To identify the mechanisms at work, it is useful to examine the entropy derivatives $\upi_\sone, \sigma_{\sone\stwo}$. In the case of ionization $Z=1$ and equal species temperatures, these are evaluated explicitly in Appendix~\ref{sec_evaluation_coefficients}.

The term in the entropy change expressions with largest magnitude is the one associated with $\sigma_{ee}$, causing the first-order changes in correlation entropy to be more negative for electrons than for ions when $dV > 0$ (and $\tau=1,Z=1$). Thus, if the process is to be adiabatic for both species, the first-order correction to electron temperature has to be greater (more positive) than the first-order correction to ion temperature during expansion. Correspondingly, during compression, where $dV < 0$, the electron temperature increases less than the ion temperature.

\subsection{Correlation heating}

\label{sec_sub_disc_correlation_heating}

\label{sec_sub_disc_compression}

\begin{figure}
    \begin{subfigure}{0.5\columnwidth}
        \centering
        \includegraphics[width=\columnwidth]{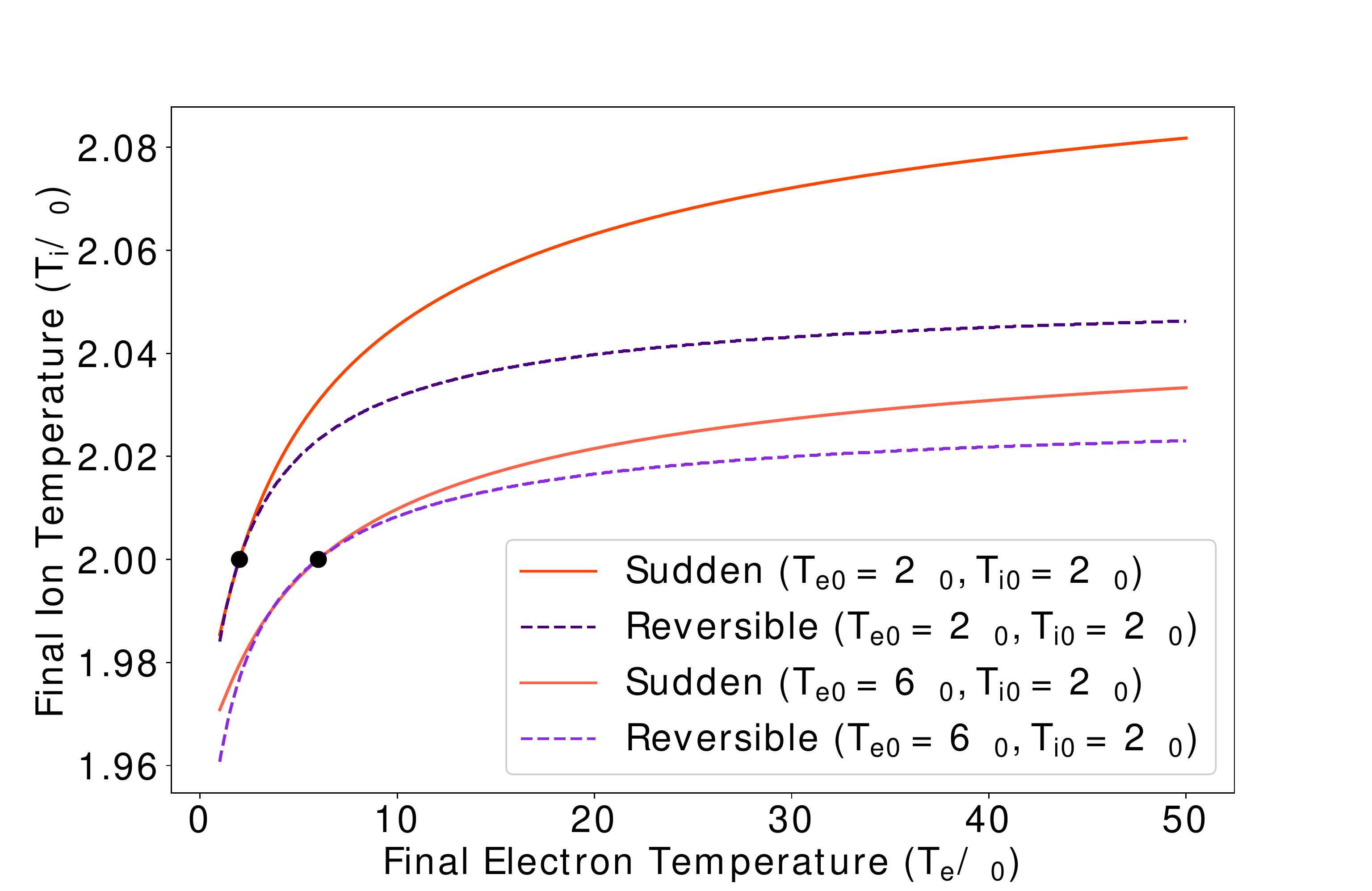}
        \caption{Heating}
    \end{subfigure}
    \begin{subfigure}{0.5\columnwidth}
        \centering
        \includegraphics[width=\columnwidth]{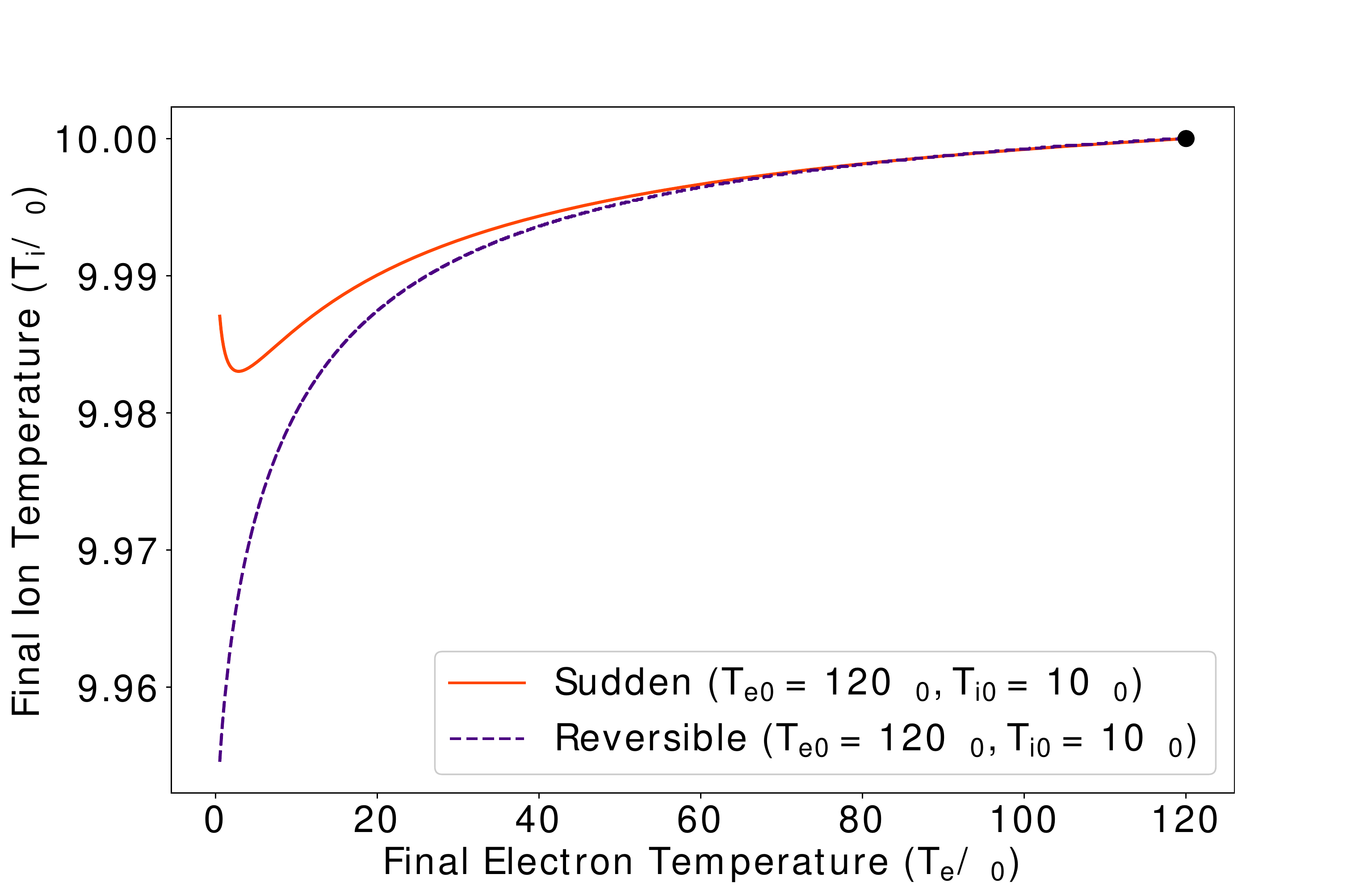}
        \caption{Cooling}
    \end{subfigure}
    \caption{Final $T_i$ as a function of $T_e$ after heating or cooling. The black circles show the initial conditions for each process. The purple curves (dashed lines) show reversible heating, where the electron heating is adiabatic for the ions. The orange curves (solid lines) show irreversible heating, where the electron heating is rapid compared to ion timescales. In panel (a), electrons and ions start cold and are heated; in (b), ions start cold but electrons start hot and are cooled.}
    \label{fig_cross_heating}
\end{figure}

In cases where heat is deposited into the electrons only, we derived formulas \eqref{eq_crossspecies_differential} and \eqref{eq_DTi_suddenheating} for when the electron heating was slow (meaning reversible ion relaxation) and sudden (meaning irreversible ion relaxation) respectively. The results of both heating processes are shown in Fig.~\ref{fig_cross_heating}, with temperatures again normalized to the energy scale $\energynorm$. For both heating processes in panel (a), we consider a starting ion temperature as low as possible within the bounds considered in Appendix~\ref{sec_sub_coupling_strength}, $T_{i0} = 2\energynorm$. The final ion temperature after heating (and subsequent relaxation, in the case of sudden heating) is plotted as a function of the final electron temperature for a $Z=1$ plasma. 

For small amounts of heating and cooling, the reversible and irreversible formulas yield similar results. As the electron heating increases, the final ion temperature in the irreversible process becomes larger than in the reversible process. This is sensible; the irreversible ion relaxation should generate entropy in the ion subsystem, much of which must go into the ion kinetic degrees of freedom, forcing $T_i$ to increase more than in the reversible case. 

When the electrons are instead cooled as in panel (b) of Fig.~\ref{fig_cross_heating}, the two expressions eventually diverge from each other. This can be seen directly from the equations for the two processes. In \eqref{eq_crossspecies_differential}, $dT_i/dT_e$ is positive-definite so the ion temperature must decrease monotonically as the electron temperature decreases. Meanwhile, in \eqref{eq_DTi_suddenheating} $\Delta T_i$ clearly asymptotes to $0$ as $\kappa \rightarrow \infty$, corresponding to a low electron temperature. Physically, the result can be justified as follows. If the electron temperature is decreased slowly, then the screening cloud around each ion slowly shrinks toward the ion, decreasing the effective potential that each ion feels from other ions. This opens more volume for the ion distribution to explore. The process is then effectively adiabatic expansion, which we expect to cool a gas; if the change in the available volume is done at constant entropy, the kinetic energy of the ions must decrease in order to balance the increasing volume contribution to the total entropy.

By contrast, if the electrons are suddenly cooled, the screening clouds will immediately collapse toward the ions. This situation would be difficult to achieve physically, but we briefly discuss its implications because the model captures interesting physical behavior in this regime. In the extreme case, we can take an initial condition where the plasma starts in a very hot state, such as the $T_e \rightarrow \infty$ limit discussed above, in which electron screening is negligible. If we suddenly and drastically cool the electrons to the point that they shield the ions from a distance shorter than the inter-ion spacing, the final state is effectively a neutral gas (albeit a classical analogue, as we have neglected quantum effects). Every ion can then continue unperturbed on the trajectory that it was following at the moment of the cooling and will experience only occasional interactions with other ions in this gas limit. The kinetic energy of the ion distribution is therefore unchanged, and so the ion temperature should not change. The potential contribution to the ion entropy in the initial state was negative, and this contribution vanishes in the nearly-ideal gas final state. Thus the total ion entropy increases due to the sudden electron heating, which is what we expect of an irreversible process.

\section{Conclusions}

In this work, we have derived analytical equations of state for the electrons and ions in a moderately-coupled two-temperature plasma. The physical quantities that we predict, such as structure factors and internal energy, are consistent with those derived in other works. We have used our coupled quasi-equations of state \eqref{eq_Psie_coulomb} and \eqref{eq_Psii_coulomb} to predict the evolution of both species' temperatures in response to external forcing. During compression and expansion, we have found that for both species the magnitude of the temperature change in moderately-coupled plasma \eqref{eq_dTdV_e} and \eqref{eq_dTdV_i} is smaller than that in ideal gas. In the moderate coupling regime, we have described a novel effect that contributes to interspecies temperature separation under adiabatic compression and expansion. We have also derived a correlation heating effect, whereby increasing the temperature of one species causes the temperature of the other species to increase, even without collisional energy exchange between species, per \eqref{eq_crossspecies_differential} and \eqref{eq_DTi_suddenheating}.

All of the analysis in this work was done for non-magnetized plasma. It has been suggested based on simulations that applying a strong magnetic field reduces the magnitude of disorder-induced heating, at least in UCP \citep{Tiwari_Baalrud_2018}. In addition, we have treated only a single fixed charge state. On timescales where inelastic ionization and recombination processes become relevant, the ion charge states will introduce an important new variable. Ionization equilibrium in two-temperature plasma has been a subject of ongoing debate, and the dynamics of ionization and recombination at multiple temperatures have been suggested to introduce a variety of novel phenomena \citep{Beule_Ebeling_Forster_1997,Andre_1995,Chen_Han_1999,Gleizes_Chervy_Gonzalez_1999,Crowley_2014}.

With these simplifications aside, the derivation of the EOS of both species is done with a high level of generality before specializing to the case of interest. Our general expression \eqref{eq_Psie_Gs} for the electron Massieu potential relies on weak electron coupling and a timescale separation between species, and \eqref{eq_Psii_Gs} for the ion Planck potential additionally relies on weak ion coupling. However, both expressions allow general interaction potentials. Although we use the bare Coulomb potential in this work, this could easily be replaced with other effective interactions. For example, in a dusty plasma, the ions and electrons in this work could be replaced with the dust and the background plasma ions respectively. In that case, it would be valuable to account for the contribution of the electrons without treating them explicitly. That could be done by replacing the Coulomb potential by a Yukawa potential to capture electron screening \citep{Mukherjee_Jaiswal_Shukla_Hakim_Thomas_2020, Shukla_Avinash_Mukherjee_Ganesh_2017}. Moderately-coupled plasma often appears near the warm dense matter (WDM) regime, in which degeneracy becomes important \citep{Koenig_Benuzzi-Mounaix_Ravasio_Vinci_Ozaki_Lepape_Batani_Huser_Hall_Hicks_et,Bonitz_Dornheim_Moldabekov_Zhang_Hamann_Kählert_Filinov_Ramakrishna_Vorberger_2020}. Although the treatment in this work is inherently classical, the generality of the results means that it may be extensible some distance into the WDM regime using the variety of semi-classical effective potentials that exist for approximating the leading-order effects of electron degeneracy. The Coulomb potentials in this work can easily be replaced with, for example, the Deutsch potential in parameter regimes where degeneracy is, like coupling strength, small but non-negligible \citep{Triola_2022}.

One implication for inertial confinement fusion is the ability to heat ions more than electrons under compression of a fuel in certain moderately-coupled regimes. While the effect is by necessity small in the moderate coupling regime that allowed the perturbative calculations of this work, a small increase in ion temperature near the point of ignition can have a highly nonlinear effect on the overall fusion yield. In an expanding plasma, such that found during the propagating burn phase in an ICF implosion, there are regimes in which electrons cool faster than ions, which could be exploited to reduce radiation, increase confinement time, and even enhance fusion rates by Salpeter screening \citep{Salpeter_van_Horn_1969,Ichimaru_1993}. In fast ignition scenarios, a high-power ignitor is used to heat electrons rapdily, which then transfer their energy to ions as a `spark' for fusion \citep{Kodama_Norreys_Mima_Dangor_Evans_Fujita_Kitagawa_Krushelnick_Miyakoshi_Miyanaga_et, Tabak_Hinkel_Atzeni_Campbell_Tanaka_2006}. Correlation heating could serve as an additional channel for rapidly heating ions in this scenario.
The design of a fusion scheme to profit from any of these effects would require more detailed study, but the analytical results in this work could inform the choice of regimes.

This work was supported by DOE Grant No. DE-SC0016072 and DOE-NNSA Grant No. 83228–10966 [Prime No. DOE (NNSA) DE- NA0003764].

Competing interests: the authors declare none.

\appendix

\section{Regimes of Applicability}
\label{sec_regimes}

\subsection{Timescales}

In all cases, the collisions that exchange energy within each species must be much faster than the rate of inter-species energy exchange in order to treat the electrons and ions as equilibrium systems at different temperatures, so $\nuie \ll \nuee, \nuii$.
Our calculation additionally requires that the timescale of rearrangement of the electron subsystem is much faster than the timescale of rearrangement of the ion subsystem so that every microstate of the ion subsystem sees an average over many electron configurations. We require therefore that $\omega_{pi}, \nuii \ll \omega_{pe}, \nuee$.

\subsection{Coupling strength}
\label{sec_sub_coupling_strength}

In this work, we follow one common convention in defining the coupling between species $\sone, \stwo$ to be 
\begin{equation}
\label{eq_Gamma_def}
    \Gamma_{\sone\stwo} = \left |\frac{q_\sone q_\stwo}{a_\sone T_\sone}\right|
\end{equation} 
where $q_s,n_s,T_s$ are respectively the charge, density, and temperature of species $s$ and $a_\sone = (4\upi n_\sone / 3)^{-1/3}$ is the characteristic interparticle spacing. Our expression for the electron Planck potential \eqref{eq_Psie_Gs} relies on weak electron coupling, meaning heuristically that $\Gamma_{ee} \ll 1$. The expression for the ion Planck potential \eqref{eq_Psii_Gs} additionally relies on weak ion-ion coupling, roughly $\Gamma_{ii} \ll 1$.

The results of our calculations, however, do not appear as integer powers of $\Gamma$ but rather as a power series in $\Lambdainv = 1/4\upi\Lambda$. The `plasma parameter' is defined as the number of particles in a Debye sphere\footnote{The common definition, quoted here, might as well be said to give the number of particles in a `Debye cube,' but the geometrical factor $4\upi/3\approx 4$ is often irrelevant. Here, we keep track of the factors to provide clearer bounds on our parameters.} $\Lambda = n\lambda_D^3$. Although it would be natural to expand simply in $1/\Lambda$, it is of interest to capture the largest effects that are consistent with the parameter ordering in this theory. It is clear from, for example, \eqref{eq_Uphi_dimensionless} that the leading-order effects are suppressed by a numerical factor. For the potential energy, this factor is around $0.04$, which is important when determining the expected error. Our choice of $\Lambdainv = 1/4\upi\Lambda$ is in keeping with \citet{O’Neil_Rostoker_1965}, in which three-body correlations are calculated, providing expressions for the plasma potential energy to next order\footnote{It is interesting to note that these corrections include terms of order $\Lambdainv^2\ln\Lambdainv$ as well as $\Lambdainv^2$, but their presence doesn't affect the discussion here.} for single-temperature plasma. Remarkably, \citet{Abe_1959_03}, who uses the same expansion parameter and agrees with \citet{O’Neil_Rostoker_1965}, shows that the higher-order corrections cause a reasonably small deviation from the Debye-H\"uckel law (the first-order result as recovered in our work) for values as high as $\Lambdainv = 1/2$. Although these works treat only single-temperature plasmas, it is clear that analagous orderings should hold when the species' temperatures are different but comparable.

In the two-temperature case, we can define a plasma parameter for each species, where as a reminder $\kappa^2 = 4\upi e^2 n_e/T_e, \; \chi^2 = 4\upi Z^2e^2 n_i/T_i$ are the inverse squared screening lengths, as
\begin{equation}
\begin{split}
    \Lambda_e =& n_e\kappa^{-3} = \frac{3}{4\upi} \left(3\Gamma_{ee}\right)^{-3/2}
    \\
    \Lambda_i =& n_i\chi^{-3} = \frac{3}{4\upi} \left(3\Gamma_{ii}\right)^{-3/2} .
\end{split}
\end{equation}
We now have two small parameters, $\Lambdainv_e = 1/4\upi\Lambda_e$ and $\Lambdainv_i = 1/4\upi\Lambda_i$, but for $T_e \sim T_i$ and $Z \sim 1$, we have $\Lambdainv \sim \Lambdainv_e \sim \Lambdainv_i$ and so the distinction between these parameters should not affect our orderings.

As a concrete bound\footnote{While this bound on the expansion parameter is more generous than is typical of asymptotic theories, its validity is corroborated by the releatively small error demonstrated in Fig.~12 of \citet{Abe_1959_08}.} 
for the purposes of most figures in the text, we require that $1/4\upi\Lambda_e < 1/2$ and $1/4\upi\Lambda_i < 1/2$. It is then reasonable to ask what the maximum correlation heating is within this regime, consistent with the assumptions of our work. From \eqref{eq_DTi_suddenheating}, starting with some temperature ratio $\tau_0 = T_{e0}/T_{i0}$ and heating the electrons to arbitrarily high temperature, we find that
\begin{equation}
    \frac{\Delta T_i}{T_i} = \frac{\Lambdainv}{3}\left(1 - \frac{1}{\sqrt{Z^{-1}\tau_0^{-1} + 1}}\right).
\end{equation}

For a $Z=1$ plasma, requring that the electrons also start moderately coupled, this means that, for $1/4\upi\Lambda_i = 1/2$, the maximum correlation heating is $\Delta T_i/T_i \approx 4.9\%$. Allowing $Z$ to vary but continuing to impose the moderate-coupling condition, it can be shown that the maximum correlation heating is achieved for very large $Z$ and is $\Delta T_i/T_i \approx 16.7\%$. While high-$Z$ atomic states would be hard to achieve in a relatively cold and dense plasma, this bound could be relevant in dusty plasmas.

\section{Computing Partition Functions}

In this appendix, we calculate the partition function of the electron subsystem. Because of the general functional form in which we are working, the result is readily applied to give the ion partition function as well. 

\subsection{Components of the potential energy}
\label{sec_sub_energy_defns}

It is first helpful to break the physically-relevant potential energy (each particle interacting with every particle except itself) into a term that includes self-interactions and a constant self-energy term. We write the potential potential energy due to same-species electrostatic interactions as $E_{\phi s} = E^\mathrm{full}_{\phi s} - E_{\phi s}^\mathrm{self}$. Here we call $E_{\phi s}$ the `physical electrostatic energy,' i.e. the one that will be relevant for all thermodynamic calculations. We call $E^\mathrm{full}_{\phi s}$ the `full electrostatic energy' because it includes self-interactions, and we denote the difference between these by the $E_{\phi s}^\mathrm{self}$, the self-energy. The full electrostatic energy is given by
\begin{equation}
    E^\mathrm{full}_{\phi s} = \sum_j^{N_s}q_s\phi_s(\bs r_{s,j}),
\end{equation}
and we subtract the self-energy, given by
\begin{equation}
    E_{\phi s}^\mathrm{self} = \sum_j^{N_s} q_s\varphi(0) ,
\end{equation}
to ensure that the physical energy $E_s$ is finite even if this self-energy (which may diverge with, for example, the Coulomb potential) is infinite. For the electrons the full electrostatic energy for the electrons can be written using the above definitions as
\begin{equation}
    E^\mathrm{full}_{\phi e}  =\frac{1}{T_e}\sum_j -e\phi(\bs r_j) .
\end{equation}

Fourier transforming the potential gives
\begin{equation}
     E^\mathrm{full}_{\phi e}  = -\frac{e}{T_e}\sum_j \int \frac{d^3k}{(2\upi)^3} \widetilde\phi(\bs k) \e^{\im \bs k\cdot \bs r_j} ,
\end{equation}
and then using the definition of the electron distribution in Fourier space, and imposing the reality condition $\trho_e(-\bs k) = \trho_e^*(\bs k)$, we have
\begin{equation}
    E^\mathrm{full}_{\phi e}  =-\frac{e}{T_e}\sqrt{N_e}\int \frac{d^3k}{(2\upi)^3} \widetilde\phi(\bs k) \trho_e^*(\bs k) .
\end{equation}

Finally, using the potential in terms of Green's functions as given in \eqref{eq_phi_greensfunctions}, the energy is
\begin{equation}
\label{eq_E_phi_e}
    E^\mathrm{full}_{\phi e} = T_e\Gee\trho_e\cdot\trho_e + T_e\Gei\trho_i\cdot\trho_e ,
\end{equation}
where the product here is the shorthand notation for integration over $\bs k$ defined in \eqref{eq_dot_defn}.

It bears repeating that these expressions include self-energy due to particles interacting with their own potentials. The self-energy term is independent of the electron configuration and so can be factored out of the integrals in the partition function. The electron partition function is the product of the kinetic term, the subtracted self-energy term, and the full electrostatic term including self-interactions:
\begin{equation}
    \mathcal{Z}_e = \frac{1}{N_e!}\left(\frac{m_e}{2\upi\hbar}\right)^{3N_e}\mathcal{Z}_{Ke}\mc Z_{\phi e}^\mathrm{self} \mathcal{Z}_{\phi e}^\mathrm{full} ,
\end{equation}
where the additional factor comes from the phase space integration measure. As usual, the kinetic term is
\begin{equation}
\begin{split}
    \mathcal{Z}_{Ke} &= \int d^{3N_e}v\exp\left\{ -\beta_e \half m_ev^2\right\} ,
\end{split}
\end{equation}
while the subtracted self-energy term is
\begin{equation}
\label{eq_Z_phie_self}
    \mc Z_{\phi e}^\mathrm{self} = \exp \left\{\beta_e E_{\phi e}^\mathrm{self}\right\}
\end{equation}
and lastly, the full electrostatic part of the electron partition function is 
\begin{equation}
\label{eq_Z_phi_e}
\begin{split}
    \mathcal{Z}^\mathrm{full}_{\phi e}(\beta_i, V, \civ)  & =\int_V d^{3N_e} r \exp \left\{ - \Gee\trho_e\cdot\trho_e - \Gei\trho_e\cdot\trho_i\right\} .
\end{split}
\end{equation}

For convenience we will often write only the electrostatic parts of the thermodynamic potentials, defined as $\Phi_{\phi e} = \ln (\mc Z^\mathrm{self}_{\phi e}\mc Z^\mathrm{full}_{\phi e}), \Psi_{\phi e} = \ln (\mc Z^\mathrm{self}_{\phi e} \zeta^\mathrm{full}_{\phi e}), \Psi_{\phi i} = \ln (\mc Z^\mathrm{self}_{\phi i} \zeta^\mathrm{full}_{\phi i})$.

\subsection{Configuration integral}
\label{sec_sub_partition_calculation}

With these expressions for the potential energy, we are prepared to calculate the partition function of the electron subsystem. It is useful to convert the integral over discrete particle positions $d^{3N_e}r$ into a smooth integral over a field $\mc D \trho_e$. Using a procedure based on that in \cite{Ariel_Diamant_2020}, we start by inserting a delta function, so that

\begin{equation}
\begin{split}
    \mathcal{Z}^\mathrm{full}_{\phi e}   = & \int d^{3N_e} r \int \mathcal{D}\trho_e \prod_{\bs k^\prime} \delta\left(\trho_e(\bs k^\prime) - \frac{1}{\sqrt{N_e}}\sum_j \e^{-\im \bs k^\prime\cdot \bs r_j}\right) 
    \\ &\times \exp \left\{ V\int \frac{d^3k}{(2\upi)^3} \left[ -\Gee|\trho_e(\bs k)|^2 - \Gei\trho_e(-\bs k)\trho_i(\bs k)    \right]  \right\} .
\end{split}
\end{equation}

We then rewrite the delta function in terms of an integral over a dummy field $\widetilde \psi$. For intermediate steps in this calculation, it will be useful to define the number of allowed Fourier modes in the system as 
\begin{equation}
    \Omega = V \int_{\widetilde V} \frac{d^3k}{(2\upi)^3}
\end{equation} 
where the subscript $\widetilde V$ on the integral indicates that the domain of integration is the entire dual space\footnote{The lower limits of this integral would be set by the system size, and the upper limits could be set by some arbitrary cutoff, but we will find the bounds to be irrelevant as $\Omega$ does not appear in the final expressions.}. The resulting partition function is

\begin{equation}
\begin{split}
    \mathcal{Z}^\mathrm{full}_{\phi e}  =& \int d^{3N_e} r \int \mathcal{D}\trho_e \int \frac{\mathcal{D}\widetilde\psi}{(2\upi)^\Omega} \exp\left\{\im V \int\frac{d^3k}{(2\upi)^3} \widetilde\psi(\bs k)\left(\trho_e(\bs k) - \frac{1}{\sqrt{N_e}}\sum_j\ e^{-\im \bs k\cdot \bs r_j}\right) \right\}
    \\
    &\times \exp \left\{ V\int \frac{d^3k}{(2\upi)^3} \left[ -\Gee|\trho_e(\bs k)|^2 - \Gei\trho_e(-\bs k) \trho_i(\bs k)   \right]  \right\} .
\end{split}
\end{equation}

Now we group the parts of the integral that depend on the individual particle positions into a $\widetilde\psi$-dependent factor, which we simplify as 
\begin{equation}
\begin{split}
    \int d^{3N_e}r \e^{-\im V\int \frac{d^3k}{(2\upi)^3} \widetilde\psi\frac{1}{\sqrt{N_e}}\sum_j \e^{-\im kr_j}} =& \left[V + \int d^3r (-1 + \e^{-\im V\int \frac{d^3k}{(2\upi)^3} \widetilde\psi \frac{1}{\sqrt{N_e}} \e^{-\im \bs k \cdot \bs r}})\right]^{N_e}
    \\
    \rightarrow & V^{N_e}\mathcal{I}[\widetilde\psi]
\end{split}
\end{equation}
where $\mathcal{I}[\widetilde\psi]$ is defined as
\begin{equation}
\begin{split}
    \mathcal{I}[\widetilde\psi] =& \exp\left\{{n_{e}\int d^3r(\e^{-\im V\int \frac{d^3k}{(2\upi)^3} \widetilde\psi \frac{1}{\sqrt{N_e}}\e^{-\im \bs k \cdot \bs r}} - 1)}\right\} .
\end{split}
\end{equation}

The last step holds exactly only in the thermodynamic limit $N_e \rightarrow \infty$, but makes no direct assumptions about the density or coupling strength. However, we note that using this limit form of the exponential is only valid when the condition
\begin{equation}
    n_{e}\int d^3r\left(\e^{-\im V\int \frac{d^3k}{(2\upi)^3} \widetilde\psi \frac{1}{\sqrt{N_e}} \e^{-\im \bs k \cdot \bs r}} - 1\right) \ll 1
\end{equation}
is satisfied. It can be shown that this is consistent with $\widetilde \psi \cdot \widetilde \psi \ll 1$. In other words, this approximation involves expanding in weak particle correlations. To evaluate the partition function, we'll want to approximate $\mathcal{I}[\widetilde\psi]$ so that we can integrate it. To second order in $\widetilde\psi$ we have 
\begin{equation}
\label{eq_I_psi_approx}
\begin{split}
    \mathcal{I}[\widetilde\psi] \approx& \exp{\left\{-\im \sqrt{N_e}\widetilde\psi(0) -\frac{1}{2}V\int \frac{d^3k}{(2\upi)^3} |\widetilde\psi(\bs k)|^2\right\}} .
\end{split}
\end{equation}

The $\widetilde \psi(0)$ term can be dropped because it is dual to $\trho_e(0) = n_{e}/\sqrt{N_e}$, which is a fixed parameter of the system and not a variable in the configuration integral. For the potential component of the electron canonical partition function we are now left with
\begin{equation}
\label{eq_Ze_pre-gaussian-integration}
\begin{split}
    \mathcal{Z}^\mathrm{full}_{\phi e} = & \frac{V^{N_e}}{(2\upi)^\Omega} \int \mathcal{D}\trho_e \int \mathcal{D}\widetilde\psi \mathcal{I}[\widetilde\psi]
    \\ & \times \exp \left\{ V\int \frac{d^3k}{(2\upi)^3} \left[\im \widetilde \psi(\bs k)\trho_e(\bs k) -\Gee(\bs k)|\trho_e(\bs k)|^2 - \Gei(\bs k)\trho_i(\bs k)\trho_e(-\bs k) \right]  \right\} .
\end{split}
\end{equation}

We will first integrate over $\widetilde\psi$, which, using the approximation in \eqref{eq_I_psi_approx}, only requires evaluating a Gaussian integral. The result has the form 

\begin{equation}
\begin{split}
    \mathcal{Z}^\mathrm{full}_{\phi e} = & \frac{V^{N_e}}{(2\upi)^\Omega} \int \mathcal{D}\trho_e(\bs k) \exp\left\{V \int \frac{d^3k}{(2\upi)^3}\left[ -\Gee(\bs k)|\trho_e(\bs k)|^2 - \Gei(\bs k)\trho_i(\bs k)\trho_e(-\bs k) \right]\right\}
    \\&\times
    \left(\sqrt{2\upi}\right)^\Omega
     \exp \left\{ -\frac{V}{2}\int \frac{d^3k}{(2\upi)^3} |\trho_e(\bs k)|^2  \right\} .
\end{split}
\end{equation}

Now integrating over the electron spatial distribution $\trho_e$ by the same process as for $\widetilde \psi$, we have

\begin{equation}
\begin{split}
    \mathcal{Z}^\mathrm{full}_{\phi e}  & =  \frac{V^{N_e}}{(2\upi)^{\Omega/2}}
          \left( \prod_{\bs k^\prime} \sqrt{\frac{\upi}{\Gee(\bs k^\prime) + \frac{1}{2}}}\right)
          \exp\left\{\int \frac{d^3k}{(2\upi)^3} \frac{ \frac{1}{4}\Gei(\bs k)^2}{\Gee(\bs k)+\frac{1}{2}} |\trho_i(\bs k)|^2\right\} .
\end{split}
\end{equation}

We note that $\mc Z_{\phi e}^\mathrm{full}$ is now a functional of $\ci$ and not of the individual $\trho_i$, justifying the choice to use $\ci$ as the thermodynamic variable. Finally, some simplification yields

\begin{equation}
\begin{split}
    \mathcal{Z}^\mathrm{full}_{\phi e}  & =  V^{N_e}
          \exp\left\{V\int \frac{d^3k}{(2\upi)^3} \left[ -\frac{1}{2}\ln\left(2\Gee(\bs k) + 1 \right) + \frac{ \frac{1}{2}\Gei(\bs k)^2}{2\Gee(\bs k)+1}\ci(\bs k) \right]\right\} .
\end{split}
\end{equation}

The self-energy correction \eqref{eq_Z_phie_self} gives a contribution of 
\begin{equation}
    \mc Z_{\phi e}^\mathrm{self} = \exp \left\{ V\int \frac{d^3k}{(2\upi)^3} \Gee(\bs k) \right\}
\end{equation}

Then in total, the electron partition function in the canonical ensemble is

\begin{equation}
\label{eq_Ze_solved}
\begin{split}
    \mathcal{Z}_e  &=  \frac{1}{N_e!}\left[\left(\frac{m_eT_e}{2\upi\hbar^2}\right)^{3/2}V\right]^{N_e}
    \exp\left\{V\int \frac{d^3k}{(2\upi)^3} \left[\Gee -\frac{1}{2}\ln\left(2\Gee + 1 \right) + \frac{ \frac{1}{2} \Gei^2}{2\Gee+1}\ci \right]\right\}
\end{split}
\end{equation}
where we have dropped the $\bs k$ arguments for compactness of notation when there is no ambiguity, but the functions $\Gonetwo$ and $\ci$ continue to depend on $\bs k$.

In expanding to second order in $\widetilde\psi$, we have implicitly made the weak coupling assumption $1/4\upi\Lambda_e \ll 1$. This approximation would be more transparent (but more algebraically convoluted) if we had integrated with respect to $\trho_e$ first. We would then have had an exponential of the form $-(\frac{1}{\Gee} + 1)|\widetilde\psi|^2 - iO(\widetilde\psi^3)$. The higher-order term can be dropped if the coefficient on the quadratic term is much greater than unity. Otherwise, we would have to retain higher-order terms in the exponential, making the integral much more complicated. Since $\Gee \propto \Gamma_{ee}$, the quadratic term is largest at weak coupling. Equivalently, as the dual field to $\trho_e$, $\widetilde \psi$ represents the deviation from the uncorrelated ideal-gas-like system. Expanding in small $|\widetilde \psi|^2$ therefore means looking at the limit of weak interparticle (in this case electron-electron) correlations, which is equivalent to weak coupling in classical plasma.

So far, we have assumed weak coupling only for the electrons and made no assumptions about ion dynamics, except that the relevant dynamical timescales for ions are slower than those for electrons. Therefore, \eqref{eq_Ze_solved} could be used to find thermodynamic properties of the electron gas interacting in some fixed ion distribution, which may be prescribed based on some other theory or on observations, and needs not be in thermal equilibrium. In the remainder of the derivation, we will apply a similar treatment to the ions at equilibrium, in which we will assume that $\Gamma_{ii} \ll 1$. This treatment is presented, and the results summarized for both electrons and ions, in \S\ref{sec_twotemp_partition}.

\section{Compression and Heating}

\subsection{Temperature changes under compression}
\label{sec_compression_temperature_changes}

We here describe the calculation of the temperature change for both species during compression using the entropy expressions derived in this work. As a reminder, we define the partial derivatives of the entropies to be
\begin{equation}
\begin{split}
    \upi_\sone &\doteq \left( \frac{\partial\mc S_\sone}{\partial V} \right) ,
    \\
    \sigma_{\sone\stwo} &\doteq \left( \frac{\partial\mc S_\sone}{\partial T_\stwo} \right) .
\end{split}
\end{equation}

For each species $\sone$ (with the other species denoted $\stwo$), the constraint of constant entropy means that

\begin{equation}
    0 = \upi_\sone dV + \sigma_{\sone \sone} dT_\sone + \sigma_{\sone \stwo} dT_\stwo
\end{equation}
where $dV, dT_e, dT_i$ are differential changes in volume, electron temperature, and ion temperature respectively. We break down each quantity by orders in $\Lambdainv$ and solve for the temperature change order-by-order. We note that the interspecies interaction does not appear at lowest order, so $\sigma^{(0)}_{\sone \stwo} = 0$ when $\sone \neq \stwo$. This simplifies the calculation and to zeroth order, the temperature change of each species is
\begin{equation}
    \left( \frac{dT_\sone}{dV}\right)^{(0)} = -\frac{\upi_\sone^{(0)}}{\sigma_{\sone \sone}^{(0)}}
\end{equation}
while to next order, the temperature change of each species is

\begin{equation}
    \left( \frac{dT_\sone}{dV}\right)^{(1)} = -\frac{\upi_\sone^{(1)} + \sigma_{\sone \sone}^{(1)}\left(\frac{dT_\sone}{dV}\right)^{(0)} + \sigma^{(1)}_{\sone \stwo} \left( \frac{dT_\stwo}{dV}\right)^{(0)}}{\sigma_{\sone \sone}^{(0)}} .
\end{equation}

Using\eqref{eq_S_phi_e} and \eqref{eq_S_phi_i} we obtain the following expressions for $\upi_\sone, \sigma_{\sone \stwo}$:

\begin{equation}
\label{eq_pis_sigmas}
\begin{split}
    \upi_e &= \frac{N_e}{V} + \frac{1}{48\upi}\kappa^3 \left[ 1 + 3Z\frac{\kappa}{\kappa + k_D} \right] ,
    \\
    \upi_i &= \frac{N_i}{V} - \frac{1}{24\upi} (k_D^3 - \kappa^3) + \frac{1}{16\upi}\chi^2k_D ,
    \\
    \sigma_{ee} &= \frac{3N_e}{2T_e} + \frac{3}{2T_e}\frac{V}{24\upi}\kappa^3\left[ 1 + 3Z\frac{\kappa}{k_D + \kappa} + Z\frac{\kappa}{k_D\chi^2}\left(k_D - \kappa \right)^2\right] ,
    \\
    \sigma_{ei} &= -\frac{3}{2T_i}\frac{V}{24\upi}Z\frac{\kappa^4 }{k_D\chi^2}\left(k_D - \kappa\right)^2 ,
    \\
    \sigma_{ii} &= \frac{3N_i}{2T_i} + \frac{3}{2T_i}\frac{V}{24\upi}\frac{\chi^4}{k_D} ,
    \\
    \sigma_{ie} &= -\frac{3}{2T_e}\frac{V}{24\upi} \frac{\kappa^2}{k_D}\left(k_D - \kappa\right)^2 .
\end{split}
\end{equation}

Notably, we have from these expressions that $\sigma^{(1)}_{ei} = \sigma^{(1)}_{ie}$. In other words, the ion entropy decreases when the electron temperature is raised (and the ion temperature is held constant), and the electron entropy decreases by an equal amount when the ion temperature is raised (and the electron temperature is held constant). The remaining quantities ($\upi_e^{(1)}, \upi_i^{(1)}, \sigma_{ee}^{(1)}, \sigma_{ii}^{(1)}$) are all positive-definite, meaning that the electron and ion entropies increase when the system volume, or the temperature of the respective subsystem, increases.

\subsection{Evaluation of entropy derivatives}
\label{sec_evaluation_coefficients}

In the case of $Z=1, \tau = 1$, the expressions in \eqref{eq_pis_sigmas} for the first-order components of each term simplify to
\allowdisplaybreaks
\begin{subequations}
\begin{equation}
\begin{split}
    \upi_e^{(1)} &\rightarrow \frac{\kappa^3}{48\upi}\left[3\sqrt{2} - 2\right] ,
    \\
    \upi_i^{(1)} &\rightarrow \frac{\kappa^3}{48\upi}\left[2 - \sqrt{2}\right] ,
    \\
\end{split}
\end{equation}

\begin{equation}
\begin{split}
    \sigma_{ee}^{(1)}\left(\frac{dT_e}{dV}\right)^{(0)} &\rightarrow \frac{\kappa^3}{48\upi}\left[-9\sqrt{2} + 8\right] ,
    \\
    \sigma_{ei}^{(1)}\left(\frac{dT_i}{dV}\right)^{(0)} &\rightarrow \frac{\kappa^3}{48\upi}\left[3\sqrt{2} - 4\right] ,
    \\
    \sigma_{ii}^{(1)}\left(\frac{dT_i}{dV}\right)^{(0)} &\rightarrow \frac{\kappa^3}{48\upi}\left[-\sqrt{2}\right] ,
    \\
    \sigma_{ie}^{(1)}\left(\frac{dT_e}{dV}\right)^{(0)} &\rightarrow \frac{\kappa^3}{48\upi}\left[3\sqrt{2} - 4\right] .
\end{split}
\end{equation}
\end{subequations}

The physical importance of these quantities is discussed in \S\ref{sec_sub_disc_compression}. In brief, it is evident in this form that most of the partial derivatives are positive. The exceptions are the first-order same-species temperature derivatives $\sigma_{ee}^{(1)}, \sigma_{ii}^{(1)}$.

\subsection{Sudden heating calculation}

\label{sec_sub_sudden_heating}

In this section, we work through the details of the correlation heating process outlined in \S\ref{sec_irrev_heating}. To calculate the heating, we start by finding the new electron temperature $T_{e1}$ at the end of Step~\ref{enum_elecequilib} to first order in $\Lambdainv$. We expand each temperature in powers of $\Lambdainv$ such that
\begin{equation}
\begin{split}
    T &= T^{(0)} + T^{(1)}
\end{split}
\end{equation}
where $T^{(1)}\sim \Lambdainv T^{(0)}$. It is also worth noting that the Coulomb interaction effects enter only at sub-leading order because $U_{\phi} \sim \Lambdainv U_K$.

We define the system to start in a state with temperatures $T_{e0}, T_{i0}$. We then deposit energy $Q_e$ into the electron subsystem and use energy conservation order-by-order to determine the new temperature $T_{e1} = T_{e0} + \Delta T_e$. We leave $S_{ii}$ fixed because the ion distribution has no time to evolve on the timescale of the initial heating.

To leading order, the temperature change is that of an ideal gas:

\begin{equation}
    \frac{3}{2}N_e T_{e1}^{(0)} = \frac{3}{2}N_eT_{e0}^{(0)} + Q_e .
\end{equation}

To next order, the heating does not appear explicitly, but the change in kinetic energy $\Delta U_K ^{(1)} \sim \Delta T_e^{(1)}$ has to balance the change in potential energy, which gives

\begin{equation}
    \frac{3}{2}N_eT_{e1}^{(1)} = -U_\phi (T_{e1}^{(0)}; S_{ii,0}) + U_\phi (T_{e0}^{(0)}; S_{ii,0}) .
\end{equation}

Now, from Step~\ref{enum_elecequilib} to Step~\ref{enum_ionequilib}, the ion distribution evolves irreversibly while electron entropy is conserved. The electron temperature evolves from $T_{e1}$ to $T_{e2}$, while the ion temperature evolves from $T_{i0}$ to $T_{i2}$. Because all temperature changes during this step are sub-leading order in $\Lambdainv$, the final ion structure factor $S_{ii,2}$ can be written as $S_{ii}(T_{e1}, T_{i0})$, which is the equilibrium structure factor at the immediate post-heating electron temperature and the original ion temperature.

Conservation of electron entropy requires that

\begin{equation}
    0 = \Phi_e(T_{e2}; S_{ii,2}) - \Phi_e(T_{e1}; S_{ii,0}) + \Delta(\beta_e U_e).
\end{equation}
 
 In the electron Massieu potential, there are two non-constant terms that survive to the order of interest. One is the kinetic entropy term, which we expand as

\begin{equation}
    N_e \ln (V T_e^{3/2}) \approx N_e \ln (V {T_e^{(0)}}^{3/2}) + \frac{3}{2}N_e \frac{T_e^{(1)}}{T_e^{(0)}} .
\end{equation}

The other term comes from the part of the Massieu potential proportional to $\ci$, and so the change can be associated with the same `generalized work' defined in \eqref{eq_Ws_defn}. In total, the change in the electron temperature during the ion relaxation is
\begin{equation}
    \frac{3}{2}N_e (T_{e2}^{(1)} - T_{e1}^{(1)}) = -T_e^{(0)}\Delta(\beta_e U_e) - T_{e0}^{(0)} V \int\frac{d^3k}{(2\upi)^3} \mc F_e(T_{e1}, T_{i0}) \Delta S_{ii} .
\end{equation}

Here we have treated $\mc F_e$ as constant because, again, corrections due to $T_e^{(1)}$ enter only at higher order. 

Conservation of energy during the relaxation requires that
\begin{equation}
    0 = U_e(T_{e2}; S_{ii,2}) - U_e(T_{e1}; S_{ii,0}) + U_i(T_{e2}, T_{i2};S_{ii,2}) - U_i(T_{e1},T_{i0};S_{ii,0}) .
\end{equation}

Here we denote by $U_i(T_e, T_i; S_{ii})$ the energy of the nonequilibrium ion subsystem, where the ions have a Maxwellian velocity distribution characterized by $T_i$, but spatial correlations given by $S_{ii}$; the potential energy is then given by integrating over this structure factor.

To leading order, the total energy conservation condition can be written as

\begin{equation}
    0 = \frac{3}{2}N_e (T_{e2}^{(1)} - T_{e1}^{(1)}) + T_e^{(0)}\Delta(\beta_e U_e) + \frac{3}{2}N_iT_{i2}^{(1)} + V\int\frac{d^3k}{(2\upi)^3} T_i \Gii \Delta S_{ii}
\end{equation}
where we have used the fact that there is no change in the ion temperature to zeroth order. 

All together, the change in ion temperature $\Delta T_i = T_{i2} - T_{i0}$ is
\begin{equation}
    \Delta T_i = -\frac{2}{3} T_{i0} V\int\frac{d^3k}{(2\upi)^3} \left(\Gii - \tau \mc F_e\right) \Delta S_{ii} .
\end{equation}

\section{Entropic Potentials}

%\subsection{General Ensembles}
\label{sec_statistical_ensembles}

In this appendix, we briefly review ensemble theory in statistical mechanics. We work in the formalism of entropic potentials outlined in \citet{Planes_Vives_2002}, but present the derivations differently. For completeness, we begin on a very elementary level. All of the thermodynamic calculations presented in this work could be done with the standard thermodynamic potentials (Helmholtz free energy, Gibbs free energy, etc). Although less widely adopted, the entropic equivalents are useful in our case because they make fewer references to temperature. In the two-temperature systems that interest us, the entropic potentials of two subsystems can then, for example, simply be added together.

We work with a set of macroscopic variables labeled $U, V, \sety$. For the sake of convention, $U$ is the internal energy and $V$ is the volume. The remaining $\sety = \{Y_1, Y_2, ...\}$ could be any extensive properties of the system. These variables are also referred to as generalized displacements; a common example of an additional extensive variable is the particle number $N$, but here we work in generality. We call all of these variables macroscopic because, in general, they are large-scale properties, which hide the microscopic details. Since each microstate is a full description of every relevant degree of freedom (e.g. for an ideal gas, the 6-d phase space position of every particle), there are many large sets of microstates that correspond to a single macrostate and cannot be distinguished through macroscopic observations (measurements that reveal $U,V,\sety$).
%We choose our macroscopic variables such that they are extensive

We work first in the microcanonical ensemble, in which our extensive variables $U, V, \sety$ are held constant.   
%We start with the entropy functional $\mc S(U, V, \sety)$, which is maximized at equilibrium. 
The microcanonical `partition function' is just the multiplicity $\Omega$, given by

\begin{equation}
    \Omega = \int_{U,V,\sety} dX .
\end{equation}

Here, $dX$ represents the integration over all degrees of freedom in the microscopic phase space, and the subscript on the integral indicates that the domain of integration is restricted to microstates with the imposed values of energy and volume, and of the general extensive variables $\sety$. Implicit in this integration is the ergodic hypothesis, the postulate that all accessible microstates are equally likely, i.e. that all phase space volumes $dX$ (within the domain of integration) should have equal measure. In a discrete context, the multiplicity is the `number of accessible microstates.' Setting Boltzmann's constant $k_B = 1$, Boltzmann's formula gives the entropy as $\mc S = \ln \Omega$.

Broadly speaking, the role of each subsequent step is to relax constraints on the system, solving for its average properties when we don't know the exact values of $U, V, \sety$ but instead know something about how the system interacts with the environment.

It is often most physically relevant to work in the canonical ensemble, in which the system can exchange heat with the environment but the temperature is held constant. We work with inverse temperature $\beta = 1/T$. 

The environment could be some external heat bath, but we can alternatively break our full system into smaller subsystems and focus on one subsystem characterized by extensive variables $\{U_s, V_s, \sety_s\}$. The subsystem is placed in thermal contact with the rest of the system, which is much larger and so functions as a heat bath; therefore, $U_s$ is no longer held externally fixed. We write the energy of the subsystem $E_s = E_s(X_s)$ and the energy of the bath $E_b = E_b(X_s)$ as functions of the subsystem microstate $X_s$. This is to distinguish microstate-specific energies $E_s, E_b$ from the thermodynamic, averaged energies $U_s, U_b$, which we will proceed to calculate.

Regardless of what the bath is, we assign it an inverse temperature $\beta_b$, which specifies how the entropy $\mc S_b$ of the bath is related to the energy $E_b$ of the bath. We define inverse temperature by the relation
\begin{equation}
\label{eq_beta_defn}
    \beta_b = \consfrac{S_b}{E_b}{V, \sety_b}
\end{equation}
and take $\beta_b$ to be constant because the bath is so much larger than the subsystem that the exchange of energy has negligible effect on the intensive properties of the bath. We write the full system inverse temperature $\beta \doteq \beta_b$, which we will find to apply to both subsystems.

We denote by $\Omega_b(X_s)$ the multiplicity of the bath when the subsystem is in microstate $X_s$. Then the total multiplicity of the combined subsystem and bath is given by an integral over subsystem microstates $dX_s$ of the form 
\begin{equation}
\label{eq_Stot_Omegab}
    \Omega = \int_{V_s, \sety_s} dX_s \Omega_b(X_s) ,
\end{equation}
where the bound on the integral indicates that the domain of integration has been expanded to include microstates of arbitrary subsystem energy, while still constraining volume and the remaining intensive variables $\sety$. Using Boltzmann's entropy formula again, this integral can also be written as 
\begin{equation}
    \Omega = \int_{V_s, \sety_s} dX_s \e^{S_b(X_s)} .
\end{equation}

From \eqref{eq_beta_defn}, we have (for constant $V_s, \sety_s$ and thus constant $V_b, \sety_b$) that $d\mc S_b = \beta dU_b$. We impose conservation of total energy by requiring
\begin{equation}
    E_{tot} = E_b + E_s = \textit{const} .
\end{equation} 

Energy conservation implies that $d\mc S_b = -\beta dE_s$ and Euler's theorem  gives $\mc S_b = -\beta E_s$ \citep{Planes_Vives_2002}.  Finally, rather than dealing with the multiplicity of the combined system, we would like to work with potentials that only make reference to the internal properties of the subsystem. We relabel the left side of \eqref{eq_Stot_Omegab} as $\mc Z$, the partition function of the canonical ensemble, which gives 

\begin{equation}
    \mc Z = \int_{V_s, \sety_s} dX_s \e^{-\beta E_s(X_s)} .
\end{equation}

We see that subsystem microstates $X_s$ are now not all equally likely; they are weighted by different probabilities because they correspond to different numbers $\Omega_b$ of bath microstates. As in the main text, we identify $p_s(X_s) = \mc Z^{-1} \exp\left\{ -\beta E_s(X_s)\right\}$ as the probability density of microstate $X_s$. The average value of any quantity $q$ can then be defined as
\begin{equation}
    \langle q \rangle = \int_{V_s, \sety_s} dX_s q(X_s) p_s(X_s) .
\end{equation}
The thermodynamic energy $U_s$ is the average over the energies of microstates, i.e. ${U_s = \langle E_s \rangle}$.

The entropic potential corresponding to the canonical partition function is the Massieu potential $\Phi$, fulfilling the same role as the entropy in the microcanonical ensemble and defined by

\begin{equation}
    \Phi = \ln \mc Z .
\end{equation} 

The Massieu potential is related to the Helmholtz free energy $A$ as $\Phi = -TA$, and this relation holds between the other entropic potentials below and the free energies corresponding to their ensembles (e.g. Planck potential and Gibbs free energy).

In the canonical ensemble, the entropy of the subsystem $\mc S_s$ is no longer necessarily maximized at equilibrium. However, since the combined system has fixed energy and so is effectively in the microcanonical ensemble, we know that when the subsystem and bath are in equilibrium, the total entropy $\mc S = \mc S_s + \mc S_b$ will be maximized (with respect to changes in $U_s$). Equivalently, $\Phi$ is maximized at equilibrium. Knowing that entropy is maximized and that $dE_s = - dE_b$, we have 
\begin{equation}
    \consfrac{\mc S_s}{E_s}{V_s,\sety_s} = \consfrac{\mc S_b}{E_b}{V_s,\sety_s}
\end{equation}
which shows that the temperatures are equal, i.e. using the definition $\beta_s = (\partial \mc S_s/\partial E_s)_{V_s,\sety_s}$, we have $\beta_s = \beta_b \doteq \beta$.

We could also have transformed from the microcanonical to canonical ensemble on thermodynamic grounds, without reference to the microphysics. The natural variables of the canonical ensemble are $\beta, V, \sety$, and so the Massieu potential $\Phi(\beta, V, \sety)$ is related to the entropy $\mc S(U, V, \sety)$ by a Legendre transform

\begin{equation}
    \Phi = \mc S - \beta U .
\end{equation}

In standard thermodynamics with energetic potentials, the Legendre transform is guaranteed to yield a single-valued function because $U(\mc S, V, \sety)$ is convex for stable systems. By contrast, $\mc S(U, V, \sety)$ is concave; the Legendre transform remains single-valued, with the associated entropic potentials now being concave in their extensive variables and convex in their intensive variables \citep{Balian_1991}. In this case, we have exchanged an extensive variable ($U$) for its conjugate intensive variable ($\beta$). We can also choose to exchange the volume $V$ for the conjugate entropic pressure $\varpi$, defined by

\begin{equation}
    \varpi = \consfrac{\Phi}{V}{\beta, \sety} .
\end{equation}

The resulting isothermal-isobaric Gibbs ensemble describes a system at fixed temperature and pressure. The partition function in this ensemble is

\begin{equation}
    \mc Z_G = \int_{\sety} dX \e^{-\beta E(X) - \varpi V(X)} .
\end{equation}

The corresponding entropic potential, the isothermal-isobaric Planck potential $\Xi(\beta, \varpi, \sety)$, is defined in statistical mechanics and thermodynamics, respectively, as

\begin{equation}
\begin{split}
    \Xi &= \ln \mc Z_G ,
    \\
    \Xi &= \mc S - \beta U - \varpi V .
\end{split}
\end{equation}

We can do the same procedure for any other pair of conjugate variables, transforming to a new ensemble in which the intensive variable is held constant. For any of the extensive variables (generalized displacements) $Y_j$, we define the conjugate intensive variable (entropic force) $\mc F_j$ in the canonical ensemble by

\begin{equation}
\label{eq_Fj_general}
    \mc F_j = \consfrac{\Phi}{Y_j}{\beta, V, \{Y_{i\neq j}\}} .
\end{equation}

Now we can choose to work in an ensemble where $\mc F_j$ is held fixed, which we refer to as a generalized Gibbs ensemble. We choose here to fix $\beta, V$, but the generalization is straightforward. Following an identical procedure to the one above, the generalized Gibbs partition function $\zeta$ is

\begin{equation}
\label{eq_zeta_Gibbs_general}
    \zeta = \int_{V, \{Y_{i\neq j}\}} dX \e^{-\beta E(X) - \mc F_j Y_j(X)} .
\end{equation}

Finally, the corresponding entropic potential is the generalized Planck potential $\Psi (\beta, V, \mc F_j, \{Y_{i\neq j}\})$, which is expressed as
\begin{equation}
\label{eq_Psi_Gibbs_general}
\begin{split}
    \Psi &= \ln \zeta ,
    \\
    \Psi &= \mc S - \beta U - \mc F_j Y_j .
\end{split}
\end{equation}

\begin{figure}
    \begin{subfigure}[t]{0.48\columnwidth}
        \centering
        \begin{tikzpicture}
            \draw[thick,->] (0,0) -- (4.5,0) node[anchor=north west] {$Y$};
            \draw[thick,->] (0,0) -- (0,4.5) node[anchor=south east] {$\Phi(Y)$};
            \draw[red, thick] (0.5,0.5) .. controls (1,3.2) and (3,3.7) .. (4.5,4);
            
            \draw (2.5,-0.1) -- (2.5,0.3);
            \node at (2.52,-0.3) {$Y_0$};
        	\draw[blue, thick] [-{stealth}](2.1,3.32) -- (2.9,3.32);
            \node[blue] at (2.5,3.1) {$\mc F$};
        \end{tikzpicture}
        \caption{}
        \label{fig_forces_gradient}
    \end{subfigure}
    \begin{subfigure}[t]{0.48\columnwidth}
        \centering
        \begin{tikzpicture}
            \draw[thick,->] (0,0) -- (4.5,0) node[anchor=north west] {$Y$};
            \draw[thick,->] (0,0) -- (0,4.5) node[anchor=south east] {$\Phi(Y)$};
            \draw[red, thick] (0.5,0.5) .. controls (1,3.2) and (3,3.7) .. (4.5,4);
            
            \draw (3.5,-0.1) -- (3.5,0.3);
            \node at (3.52,-0.3) {$Y_f$};
        	\draw[blue, thick] [-{stealth}](3.25,3.74) -- (3.75,3.74);
            \node[blue] at (3.5,3.52) {$\mc F$};
        	\draw [-{stealth}](4.4,3.74) -- (3.9,3.74);
            \node at (4.3,3.52) {$\mc F_{ext}$};
        \end{tikzpicture}
        \caption{}
        \label{fig_forces_transform}
    \end{subfigure}
    \caption{The Massieu potential $\Phi$ depends on a thermodynamic variable $Y$, and for fixed $Y_0$ the entropic force exerted by the system is given by $\partial \Phi(Y_0)/\partial Y$. If an external force $\mc F_{ext}$ is applied, then the system will shift to some $Y = Y_f$ where the system's entropic force $\mc F$ balances the external force.}
\end{figure}
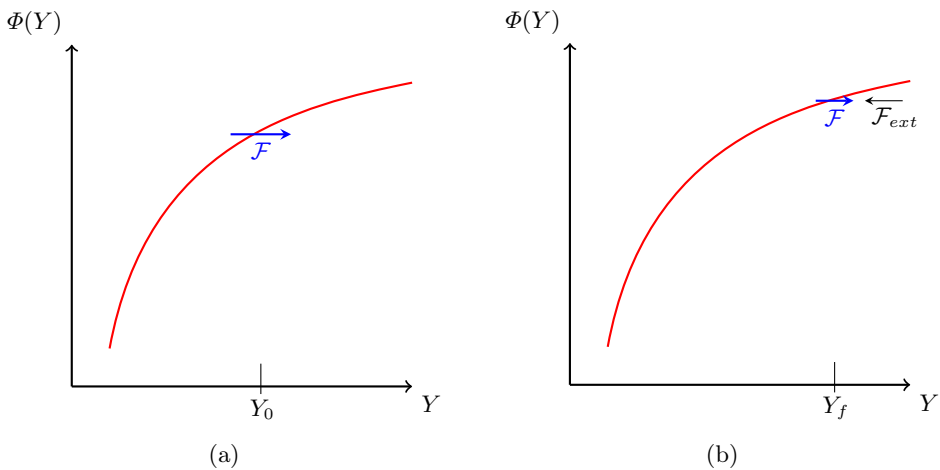

This transformation can be visualized as follows. Suppose that, as in Fig.~\ref{fig_forces_gradient}, we have a system characterized by an extensive variable $Y$. We place the system at temperature $T$ and ignore (treat as fixed) any other extensive variables that the system might have. We start the system at some fixed position $Y = Y_0$. The system `wants' to maximize $\Phi$, which is to say that if $Y$ were allowed to vary, the system would evolve toward an equilibrium state that maximizes $\Phi(Y)$. This tendency is captured by the entropic force $\mc F$, which gives the gradient of $\Phi$ along which the system would move if it were allowed; the entropic force is defined as
\begin{equation}
    \mc F \doteq \consfrac{\Phi}{Y}{} ,
\end{equation}
meaning that $\mc F$ is the `conjugate variable' to $Y$\footnote{In the formalism adopted here, the product of conjugate pairs is unitless (with $k_B$ = 1), as opposed to many thermodynamic formalisms in which the product has units of energy.}.

Entropic forces are related to forces by $F = T\mc F$, which we can easily see as follows:
\begin{equation}
    F = -\consfrac{A}{Y}{} = -\consfrac{(-T\Phi)}{Y}{} = T\mc F .
\end{equation}

If we hold $Y = Y_0$ fixed, then the surroundings must exert whatever force is necessary to hold the system at $Y_0$. For example, a rigid box is often employed in thermodynamic treatments and applies whatever force is necessary to prevent the gas from expanding under its own pressure. If instead the external force is some fixed $F_{ext}$ and $Y$ is allowed to vary, then the system will evolve through different values of $Y$ in order to increase $\Phi$, until it reaches a value $Y_f$ where the force $F(Y_f)$ balances $F_{ext}$ as in Fig.~\ref{fig_forces_transform}.

There is a small notational subtlety that we include here to highlight the analogy to the calculation in the body of this paper. In many systems, the exchange of an extensive variable with the environment is governed by $\Delta Y_{subsystem} + \Delta Y_{environment} = 0$, e.g. a gas in a box where the total volume of the gas and environment is conserved. Here, we choose to use the same variable $Y$ to parametrize the system and the environment, so we have $\Delta Y_{subsystem} = \Delta Y_{environment}$. This is not a physically obscure scenario; for instance, one could be interested in the position of a piston being pushed by two different gases, both acting in the same direction. Then the mechanical equilibrium condition is naturally $F + F_{ext} = 0$.

The external force can be associated with some external potential $\Phi_{ext}$. In many cases, we are interested in external systems that act like a bath, meaning that their extensive properties ($\mc F_{Ext}$) remain constant as they exchange some extensive quantity with the subsystem. This can happen when the external system is so much larger than the subsystem that changes in $Y$ are negligible, e.g. the heat bath considered above for the canonical ensemble, or the Earth's atmosphere exerting pressure on some experimental system of interest, where volume changes due to the experiment are negligibly small compared to the volume of the atmosphere. A constant force can also be achieved by choosing variables such that $(\partial^2 \Phi_{ext}/\partial Y^2) = 0$.

Rather than large external baths, we are ultimately interested in systems where the environment depends nontrivially on the subsystem's state. One such system is the `adiabatic piston,' consisting of two chambers of gas (perhaps initially at different temperatures or pressures) separated by a piston that is allowed to move but not to exchange heat; the thermodynamics of this system have been the subject of controversy \citep{Gruber_1999,Kestemont_Broeck_Mansour_2000,Crosignani_DiPorto_Segev_1996,Cencini_Palatella_Pigolotti_Vulpiani_2007,Gislason_2010}. In this work, we use the separation of timescales between our subsystems to determine the conditions of mechanical equilibrium. This is best illustrated through a toy model, which we discuss in the following appendix.

\section{Toy Model}
\label{sec_strings_model}
\subsection{Strings model}

In this appendix, we analyze a model system whose thermodynamic description bears many similarities to the two-temperature plasma.

\begin{figure}
    \begin{subfigure}[t]{0.32\columnwidth}
        \centering
        \includegraphics[width=0.95\columnwidth]{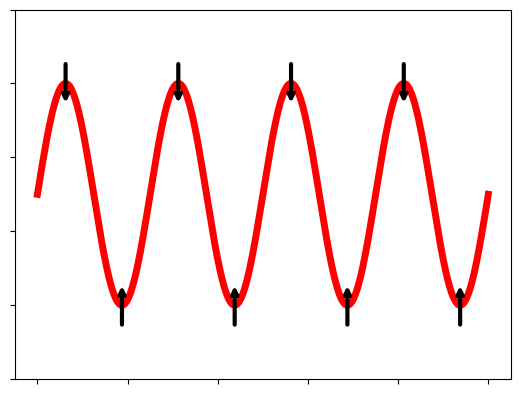}
        \caption{}
        \label{fig_string_ions}
    \end{subfigure}
    \begin{subfigure}[t]{0.32\columnwidth}
        \centering
        \includegraphics[width=0.95\columnwidth]{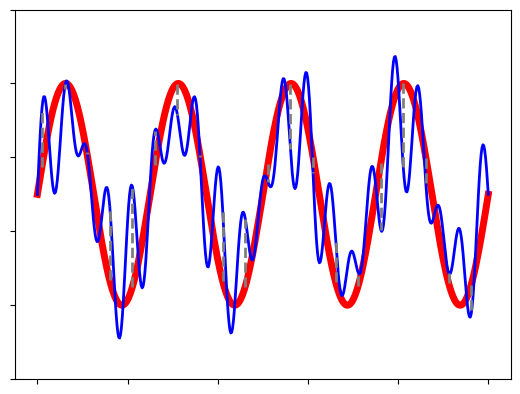}
        \caption{}
        \label{fig_string_coupled}
    \end{subfigure}
    \begin{subfigure}[t]{0.32\columnwidth}
        \centering
        \includegraphics[width=0.95\columnwidth]{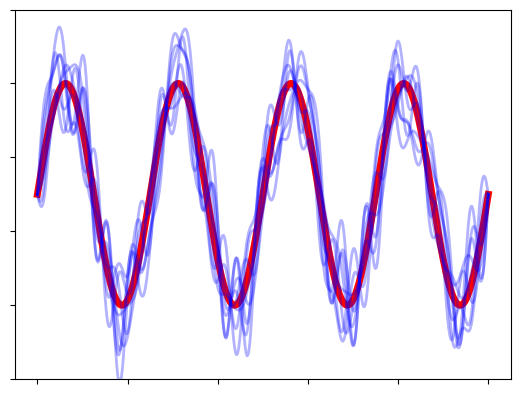}
        \caption{}
        \label{fig_string_realizations}
    \end{subfigure}
    \caption{Coupled strings evolving on separate timescales. (a) the `ion string' alone, with the restoring forces shown. (b) the `electron string' follows the position of the `ion string' but with additional fluctuations superimposed. (c) on ion timescales, the system passes through many electron configurations and the effective force on the ion string comes from the average over these configurations.}
    \label{fig_string_model}
\end{figure}

We consider a pair of strings of length $L$ under tension with fixed endpoints. We label them the `ion string' and the `electron string' for the sake of analogy. We consider linear oscillations of both strings. We decompose each string's oscillations into modes of wavenumber $k$ and denote by $n(k)$ the amplitude of the $k$ mode, where $n(k)$ is the fourier transform of the string's vertical displacement $y(x)$. We locate the origin such that, for consistency with the boundary conditions, each $n(k)$ may be negative but must be real. The reality condition for $y(x)$ requires that $n(k) = n(-k)$. Then the potential energy in each string $s$ is
\begin{equation}
    E_s = \frac{L}{2\upi}\int dk \half\lambda_s k^2 n_s^2(k)
\end{equation}
where $\lambda_s$ is a constant. Fig.~\ref{fig_string_ions} shows the ion string alone, with a single vibrational mode excited.

If the electron and ion string are uncoupled, we can analyze them independently. For each string, we expect oscillations to be more pronounced at the low-$k$ modes, and to fall off as something like $1/k^2$ at higher $k$. We can see this, for example, from the equipartition theorem.

We introduce a coupling term between the strings, which we model as a spring connecting the electron and ion strings at every horizontal position, with energy proportional to $\int dx (n_e(x) - n_i(x))^2$. In terms of a coupling constant $\gamma$, this interaction energy $E_x$ is
\begin{equation}
    E_x = \frac{L}{2\upi}\int dk \half\gamma \left[n_i^2(k) - 2 n_i(k)n_e(k) + n_e^2(k)\right] .
\end{equation}
In Fig.~\ref{fig_string_coupled}, we show the electron string coupled to an ion string where a single mode is excited; we can expect that the electron string will primarily match the modes of the ion string in order to minimize the coupling energy, but thermal fluctuations will sustain additional electron modes.

We impose that the mass of the electron string is much lower (or the tension much greater) than that of the ion string, so that the electron string oscillates on a much faster timescale. On the timescales on which the ion string evolves, the electron subsystem passes through many microstates in its ensemble, and so the electron string can be taken to apply an effective force to the ion string given by a thermal average over electron configurations.

\begin{figure}
    \centering
    \includegraphics[width = 0.5\textwidth]{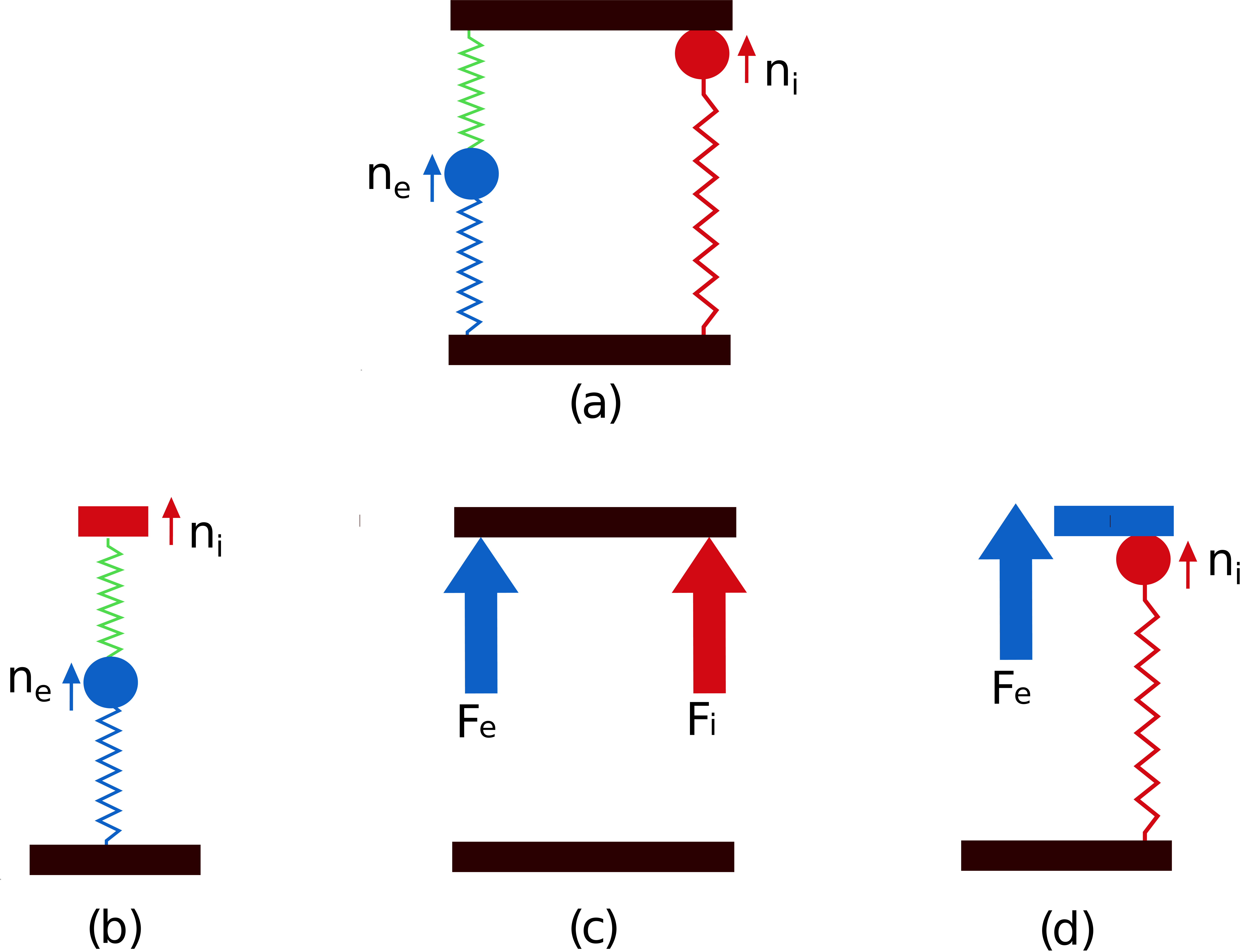}
    \caption{Stages of the statistical approach to the two-temperature spring problem. On electron timescales, the electron subsystem comes to equilibrium for some fixed ion position. Averaging over electron configurations yields an effective force, which must balance the force due to the ions. On ion timescales, the ion subsystem comes to equilibrium, with the electrons applying an effective force.}
    \label{fig_model_springs}
\end{figure}

\subsection{Simplified springs model}

Because the modes are independent, we can apply an even simpler model. At each wavenumber $k$, the system can be represented as shown in Fig.~\ref{fig_model_springs}a. The electron and ion modes oscillate separately as if on springs with spring constants that we call $\kappa_e = \lambda_e k^2$ and $\kappa_i = \lambda_i k^2$ respectively; an additional spring of constant $\gamma$ couples the oscillations. We impose that $m_e \ll m_i$ so that the springs cannot exchange energy with each other and the system is allowed to reach a true two-temperature steady state.

We can now proceed to a statistical description of this system by the same process as for the electron-ion plasma. We place the electron spring in contact with a reservoir of inverse temperature $\beta_e$ and the ion spring in contact with a reservoir of $\beta_i$. On the timescale of electron dynamics, the ion spring can be considered fixed in place and so the electron system can be considered as a mass attached by springs to two fixed points ($0$ and $n_i$). Then, treating the ion spring location as an external parameter as depicted in Fig.~\ref{fig_model_springs}b, the potential part of the electron  partition function is
\begin{equation}
    \mc Z_{\phi e} = \int_{-\infty}^\infty d n_e e^{-\beta_e\half\kappa_e n_e^2 - \beta_e\half\gamma (n_i - n_e)^2} .
\end{equation}

The partition function can readily be evaluated, yielding
\begin{equation}
    \mc Z_{\phi e} = \sqrt{\frac{2\upi}{\beta_e\kappa_e + \beta_e\gamma}} \exp\left\{-\beta_e\half\frac{\kappa_e\gamma}{\kappa_e + \gamma} n_i^2\right\} .
\end{equation}

We can find the corresponding Massieu potential $\Phi(\beta_e, n_i^2)$. Here we have chosen to use $n_i^2$ as the generalized displacement. Although the linear displacement $n_i$ is a somewhat more physically intuitive choice, it is useful to choose an extensive quantity so that the standard thermodynamic relations can readily be generalized. We have so far used the canonical ensemble, in which $n_i^2$ is held fixed. The entropic force $\mc F_e$ captures the tendency of the system to move to higher $\Phi$, which it is prevented from doing by some external force holding $n_i^2$ constant. Applying the same procedure as we did for the plasma case, we can find $\mc F_e$ for the electron spring to be
\begin{equation}
    \mc F_e = -\beta_e \half\frac{\kappa_e\gamma}{\kappa_e + \gamma}
\end{equation}
and the corresponding force is $F_e = T_e \mc F_e$.

Because the electrons are in equilibrium, the force $F_e$ generated by the electron subsystem (which we just found from the electron partition function) should exactly balance the externally-imposed force, as in Fig.~\ref{fig_model_springs}c. In this case, the external force is the force $F_i$ due to the ions, and so our force balance condition is $F_e = -F_i$\footnote{Note that, as in Appendix~\ref{sec_statistical_ensembles}, the sign convention here differs from the one we would use for e.g. a gas pushing against a piston. We define both forces to act in the same direction.}. The entropic force from the ion subsystem is then
\begin{equation}
\label{eq_spring_Fi}
    \mc F_i = \beta_i \half \frac{\kappa_e\gamma}{\kappa_e + \gamma} .
\end{equation}

Now we are ready to evaluate the ion partition function. We don't know the value of $n_i^2$ \textit{a priori}, so we can't start in the canonical ensemble. However, we know the value of $\mc F_i$ that the ion subsystem needs to exert based on the calculation above. This suggests we should do our calculations in a generalized Gibbs ensemble to find a thermodynamic potential (Planck potential) $\Psi(\beta_i, \mc F_i)$ in terms of the variables we know.

The ion partition function $\zeta_{\phi i}$ in this ensemble is represented in Fig.~\ref{fig_model_springs}d. Its potential part is
\begin{equation}
\label{eq_sping_zetai}
\begin{split}
    \zeta_{\phi i} &= \int_{-\infty}^\infty dn_i \exp\left\{ -\left(\beta_i\half\kappa_i +  \mc F_i\right) n_i^2 \right\} .
\end{split}
\end{equation}

Including the kinetic component, and grouping the irrelevant constants into a separate term $\Psi_{i0}$, the ion Planck potential is
\begin{equation}
    \label{eq_spring_Psi_i}
    \Psi_i= \Psi_{i0} - \half\ln\beta_i - \half\ln(\beta_i\kappa_i + 2\mc F_i)
\end{equation}
while in the same generalized Gibbs ensemble, the electron Planck potential is
\begin{equation}
    \label{eq_spring_Psi_e}
    \Psi_{e} = \Psi_{e0} - \half\ln\beta_e - \half\ln(\beta_e\kappa_e + \beta_e\gamma) + \left(-\beta_e\half\frac{\kappa_e\gamma }{\kappa_e + \gamma} - \mc F_e\right) n_i^2
\end{equation}
where we have again grouped irrelevant constants into $\Psi_{e0}$.

Having now computed the partition functions for electrons and ions at any wavenumber, we could now use the fact that the wavenumbers are all independent to write the partition functions for the electron and ion string models above as
\begin{equation}
    \mc Z^\textit{strings} = \prod_{\bs k} \mc Z^\textit{springs}(\bs k) .
\end{equation}

In the form of its answer, the string model is then a close analogue to the electron-ion plasma. The difference in the two expressions, other than a different $k$ scaling because of the Coulomb potential, comes primarily from the temperature-independent configurational terms in the plasma partition function, which arise from converting the integral over discrete particle positions into an integral over continuous fields. However, we can still find the key physical insight about the two-temperature statistical procedure in the reduced models; in the remainder of this appendix, we discuss the simple two-spring model.

We can derive various thermal averages from \eqref{eq_spring_Psi_e} and \eqref{eq_spring_Psi_i}. The average potential energy of the electron subsystem is
\begin{equation}
    U_{\phi e} = \half T_e + \half \frac{\kappa_e \gamma}{\kappa_e + \gamma}\langle n_i^2 \rangle .
\end{equation}
The first term is the energy of a spring in equipartition, as expected. The second term  represents the energy required to stretch the combined electron spring ($\kappa_e$) and coupling spring ($\gamma$) from their rest length of zero to the new length $n_i$ required to match the ion displacement. The coefficient $\kappa_e \gamma / (\kappa_e + \gamma)$ is just the spring constant of the two springs in series.

The average squared displacement $\langle n_i^2\rangle$ is given by differentiating the ion Planck potential with respect to $\mc F_i$ and then substituting the known expression for the entropic force from \eqref{eq_spring_Fi}. The result is
\begin{equation}
    \langle n_i^2 \rangle =  \frac{T_i}{\kappa_i + \kappa_e \gamma / (\kappa_e + \gamma)}
\end{equation}
which is the displacement of a system at temperature $T_i$ consisting of two springs in series: the ion spring and the combined electron/coupling spring above. 

The potential energy of the ion subsystem is
\begin{equation}
    U_{\phi i} =  \half T_i\frac{\kappa_i}{\kappa_i + \kappa_e \gamma /(\kappa_e + \gamma)}
\end{equation}
which is the usual $\half T_i$ from equipartition, but attenuated by a factor that is less than unity. Of the energy contained in the full system, with the ion spring added in parallel to the series-combined electron/coupling spring, this factor corresponds to the fraction of the energy in the ion spring alone. Adding together the electron energy and ion energy, the terms simplify nicely; if we now include the kinetic energy component, we have that the total energy is
\begin{equation}
    U = T_e + T_i
\end{equation}
which is, again, the result that we expect by equipartition.

It is straightfoward, if tedious, to solve the dynamics of the two-spring system exactly by standard techniques. In such a calculation, we must take $m_e/m_i \rightarrow 0$ and keep only leading-order terms in this mass ratio in order to prevent energy exchange between the two subsystems. If we apply stochastic forcing (as well as infinitesimal damping) to the electron and ion masses, then impose that the statistics of each forcing function are those that we would obtain from heat baths of temperatures $T_e$ and $T_i$, the result is thermal averages exactly matching the quantities above.

\begin{comment}
\section{Entropy Integrals}
\label{sec_integrals}

\begin{equation}
\begin{split}
    \hat\Psi_{\phi i} &= - \frac{V}{2}\int \frac{d^3k}{(2\upi)^3} \left[\ln(k^2 + \kappa^2 + \chi^2) - \ln(k^2 + \kappa^2) - \frac{\chi^2}{k^2}\right]
\end{split} 
\end{equation}

As with the potential energy, the Planck potential has infinite terms associated with the infinite ion self-energy, but they are constant and have no effect on the thermodynamics of the system. The finite part integrates to

\begin{equation}
    \hat\Psi_{\phi i}^\mathrm{full} = \frac{V}{12\upi}(k_D^3 - \kappa^3) .
\end{equation}

The Planck potential for the electrons simplifies greatly when substituting in the generalized force from \eqref{eq_f_coulomb}, giving
\begin{equation}
\label{eq_Psie_hat}
\begin{split}
    \hat\Psi_{\phi e} &= - \frac{V}{2}\int \frac{d^3k}{(2\upi)^3} \left[ \ln \left(\frac{\kappa^2}{k^2} + 1 \right) - \frac{\kappa^2}{k^2}\right] .
\end{split}
\end{equation}

\end{comment}

\bibliography{tempseparation}

\begin{thebibliography}{80}
\expandafter\ifx\csname natexlab\endcsname\relax\def\natexlab#1{#1}\fi
\def\au#1{#1} \def\ed#1{#1} \def\yr#1{#1}\def\at#1{#1}\def\jt#1{\textit{#1}}
  \def\bt#1{#1}\def\bvol#1{\textbf{#1}} \def\vol#1{#1} \def\pg#1{#1}
  \def\publ#1{#1}\def\arxiv#1{#1}\def\org#1{#1}\def\st#1{\textit{#1}}

\bibitem[Abe(1959{\natexlab{{\em a\/}}})]{Abe_1959_03}
{\sc \au{Abe, R.}} \yr{1959{\natexlab{{\em a\/}}}}  \at{Equation of state of
  classical electron gas}.  \jt{Progress of Theoretical Physics}
  \bvol{21}~(3),  \pg{475–476}.

\bibitem[Abe(1959{\natexlab{{\em b\/}}})]{Abe_1959_08}
{\sc \au{Abe, R.}} \yr{1959{\natexlab{{\em b\/}}}}  \at{Giant cluster expansion
  theory and its application to high temperature plasma*}.  \jt{Progress of
  Theoretical Physics}  \bvol{22}~(2),  \pg{213–226}.

\bibitem[Andre(1995)]{Andre_1995}
{\sc \au{Andre, P.}} \yr{1995}  \at{Partition functions and concentrations in
  plasmas out of thermal equilibrium}.  \jt{IEEE Transactions on Plasma
  Science}  \bvol{23}~(3),  \pg{453–458}.

\bibitem[Ariel \& Diamant(2020)]{Ariel_Diamant_2020}
{\sc \au{Ariel, G.} \& \au{Diamant, H.}} \yr{2020}  \at{Inferring entropy from
  structure}.  \jt{Physical Review E}  \bvol{102}~(2),  \pg{022110}.

\bibitem[Balian(1991)]{Balian_1991}
{\sc \au{Balian, R.}} \yr{1991} {\em From Microphysics to Macrophysics\/}.
  \publ{Berlin, Heidelberg: Springer Berlin Heidelberg}.

\bibitem[Bergeson {\em et~al.\/}(2019)Bergeson, Baalrud, Ellison, Grant,
  Graziani, Killian, Murillo, Roberts \&
  Stanton]{Bergeson_Baalrud_Ellison_Grant_Graziani_Killian_Murillo_Roberts_Stanton_2019}
{\sc \au{Bergeson, S.~D.}, \au{Baalrud, S.~D.}, \au{Ellison, C.~L.}, \au{Grant,
  E.}, \au{Graziani, F.~R.}, \au{Killian, T.~C.}, \au{Murillo, M.~S.},
  \au{Roberts, J.~L.} \& \au{Stanton, L.~G.}} \yr{2019}  \at{Exploring the
  crossover between high-energy-density plasma and ultracold neutral plasma
  physics}.  \jt{Physics of Plasmas}  \bvol{26}~(10),  \pg{100501}.

\bibitem[Beule {\em et~al.\/}(1997)Beule, Ebeling \&
  Förster]{Beule_Ebeling_Forster_1997}
{\sc \au{Beule, D.}, \au{Ebeling, W.} \& \au{Förster, A.}} \yr{1997}
  \at{Adiabatic equation of state and ionization equilibrium of dense plasma}.
  \jt{Physica A: Statistical Mechanics and its Applications}  \bvol{241}~(3),
  \pg{719–728}.

\bibitem[Boercker \& More(1986)]{Boercker_More_1986}
{\sc \au{Boercker, D.~B.} \& \au{More, R.~M.}} \yr{1986}  \at{Statistical
  mechanics of a two-temperature, classical plasma}.  \jt{Physical Review A}
  \bvol{33}~(3),  \pg{1859–1869}.

\bibitem[Bonitz {\em et~al.\/}(2020)Bonitz, Dornheim, Moldabekov, Zhang,
  Hamann, Kählert, Filinov, Ramakrishna \&
  Vorberger]{Bonitz_Dornheim_Moldabekov_Zhang_Hamann_Kählert_Filinov_Ramakrishna_Vorberger_2020}
{\sc \au{Bonitz, M.}, \au{Dornheim, T.}, \au{Moldabekov, Z.~A.}, \au{Zhang,
  S.}, \au{Hamann, P.}, \au{Kählert, H.}, \au{Filinov, A.}, \au{Ramakrishna,
  K.} \& \au{Vorberger, J.}} \yr{2020}  \at{Ab initio simulation of warm dense
  matter}.  \jt{Physics of Plasmas}  \bvol{27}~(4),  \pg{042710}.

\bibitem[Braginskii(1958)]{Braginskii_1958}
{\sc \au{Braginskii, S.~I.}} \yr{1958}  \at{Transport phenomena in a completely
  ionized two-temperature plasma}.  \jt{Soviet Phys. JETP}  \bvol{Vol: 6}.

\bibitem[Cencini {\em et~al.\/}(2007)Cencini, Palatella, Pigolotti \&
  Vulpiani]{Cencini_Palatella_Pigolotti_Vulpiani_2007}
{\sc \au{Cencini, M.}, \au{Palatella, L.}, \au{Pigolotti, S.} \& \au{Vulpiani,
  A.}} \yr{2007}  \at{Macroscopic equations for the adiabatic piston}.
  \jt{Physical Review E}  \bvol{76}~(5),  \pg{051103}.

\bibitem[Chen \& Han(1999)]{Chen_Han_1999}
{\sc \au{Chen, X.} \& \au{Han, P.}} \yr{1999}  \at{On the thermodynamic
  derivation of the saha equation modified to a two-temperature plasma}.
  \jt{Journal of Physics D: Applied Physics}  \bvol{32}~(14),  \pg{1711}.

\bibitem[Chen {\em et~al.\/}(2004)Chen, Simien, Laha, Gupta, Martinez,
  Mickelson, Nagel \&
  Killian]{Chen_Simien_Laha_Gupta_Martinez_Mickelson_Nagel_Killian_2004}
{\sc \au{Chen, Y.~C.}, \au{Simien, C.~E.}, \au{Laha, S.}, \au{Gupta, P.},
  \au{Martinez, Y.~N.}, \au{Mickelson, P.~G.}, \au{Nagel, S.~B.} \&
  \au{Killian, T.~C.}} \yr{2004}  \at{Electron screening and kinetic-energy
  oscillations in a strongly coupled plasma}.  \jt{Physical Review Letters}
  \bvol{93}~(26),  \pg{265003}.

\bibitem[Chin {\em et~al.\/}(2022)Chin, Ruby, Nilson, Bishel, Coppari, Ping,
  Coleman, Craxton, Rygg \&
  Collins]{Chin_Ruby_Nilson_Bishel_Coppari_Ping_Coleman_Craxton_Rygg_Collins_2022}
{\sc \au{Chin, D.~A.}, \au{Ruby, J.~J.}, \au{Nilson, P.~M.}, \au{Bishel,
  D.~T.}, \au{Coppari, F.}, \au{Ping, Y.}, \au{Coleman, A.~L.}, \au{Craxton,
  R.~S.}, \au{Rygg, J.~R.} \& \au{Collins, G.~W.}} \yr{2022}  \at{Emission
  phases of implosion sources for x-ray absorption fine structure
  spectroscopy}.  \jt{Physics of Plasmas}  \bvol{29}~(5),  \pg{052702}.

\bibitem[Christensen-Dalsgaard \&
  Däppen(1992)]{ChristensenDalsgaard_Dappen_1992}
{\sc \au{Christensen-Dalsgaard, J.} \& \au{Däppen, W.}} \yr{1992}  \at{Solar
  oscillations and the equation of state}.  \jt{The Astronomy and Astrophysics
  Review}  \bvol{4}~(3),  \pg{267–361}.

\bibitem[Clarke(1980)]{Clarke_1980}
{\sc \au{Clarke, J.~F.}} \yr{1980}  \at{Hot-ion-mode ignition in a tokamak
  reactor}.  \jt{Nuclear Fusion}  \bvol{20}~(5),  \pg{563}.

\bibitem[Craxton {\em et~al.\/}(2015)Craxton, Anderson, Boehly, Goncharov,
  Harding, Knauer, McCrory, McKenty, Meyerhofer, Myatt, Schmitt, Sethian,
  Short, Skupsky, Theobald, Kruer, Tanaka, Betti, Collins, Delettrez, Hu,
  Marozas, Maximov, Michel, Radha, Regan, Sangster, Seka, Solodov, Soures,
  Stoeckl \&
  Zuegel]{Craxton_Anderson_Boehly_Goncharov_Harding_Knauer_McCrory_McKenty_Meyerhofer_Myatt_et}
{\sc \au{Craxton, R.~S.}, \au{Anderson, K.~S.}, \au{Boehly, T.~R.},
  \au{Goncharov, V.~N.}, \au{Harding, D.~R.}, \au{Knauer, J.~P.}, \au{McCrory,
  R.~L.}, \au{McKenty, P.~W.}, \au{Meyerhofer, D.~D.}, \au{Myatt, J.~F.},
  \au{Schmitt, A.~J.}, \au{Sethian, J.~D.}, \au{Short, R.~W.}, \au{Skupsky,
  S.}, \au{Theobald, W.}, \au{Kruer, W.~L.}, \au{Tanaka, K.}, \au{Betti, R.},
  \au{Collins, T. J.~B.}, \au{Delettrez, J.~A.}, \au{Hu, S.~X.}, \au{Marozas,
  J.~A.}, \au{Maximov, A.~V.}, \au{Michel, D.~T.}, \au{Radha, P.~B.},
  \au{Regan, S.~P.}, \au{Sangster, T.~C.}, \au{Seka, W.}, \au{Solodov, A.~A.},
  \au{Soures, J.~M.}, \au{Stoeckl, C.} \& \au{Zuegel, J.~D.}} \yr{2015}
  \at{Direct-drive inertial confinement fusion: A review}.  \jt{Physics of
  Plasmas}  \bvol{22}~(11),  \pg{110501}.

\bibitem[Crosignani {\em et~al.\/}(1996)Crosignani, Di~Porto \&
  Segev]{Crosignani_DiPorto_Segev_1996}
{\sc \au{Crosignani, B.}, \au{Di~Porto, P.} \& \au{Segev, M.}} \yr{1996}
  \at{Approach to thermal equilibrium in a system with adiabatic constraints}.
  \jt{American Journal of Physics}  \bvol{64}~(5),  \pg{610–613}.

\bibitem[Crowley(2014)]{Crowley_2014}
{\sc \au{Crowley, B. J.~B.}} \yr{2014}  \at{Continuum lowering – a new
  perspective}.  \jt{High Energy Density Physics}  \bvol{13},  \pg{84–102}.

\bibitem[Davidovits \& Fisch(2019)]{Davidovits_Fisch_2019}
{\sc \au{Davidovits, S.} \& \au{Fisch, N.~J.}} \yr{2019}  \at{Understanding
  turbulence in compressing plasma as a quasi-{EOS}}.  \jt{Physics of Plasmas}
  \bvol{26}~(6),  \pg{062709}.

\bibitem[Dharma-Wardana \& Perrot(1998)]{Dharma-wardana_Perrot_1998}
{\sc \au{Dharma-Wardana, M. W.~C.} \& \au{Perrot, F.}} \yr{1998}  \at{Energy
  relaxation and the quasiequation of state of a dense two-temperature
  nonequilibrium plasma}.  \jt{Physical Review E}  \bvol{58}~(3),
  \pg{3705–3718}.

\bibitem[Ecker \& Kröll(1964)]{Ecker_Kroll_1964}
{\sc \au{Ecker, G.} \& \au{Kröll, W.}} \yr{1964}  \at{Correlation funktions of
  a system with different temperatures of the particle components}.
  \jt{Zeitschrift für Naturforschung A}  \bvol{19}~(13),  \pg{1447–1451}.

\bibitem[Edwards(2013)]{Edwards_Patel_Lindl_Atherton_Glenzer_Haan_Kilkenny_Landen_Moses_Nikroo_et}
{\sc \au{Edwards, M. J.~{\etal}.}} \yr{2013}  \at{Progress towards ignition on
  the national ignition facility}.  \jt{Physics of Plasmas}  \bvol{20}~(7),
  \pg{070501}.

\bibitem[Eliezer {\em et~al.\/}(2015)Eliezer, Henis, Nissim, Pinhasi \&
  Val]{Eliezer_Henis_Nissim_Pinhasi_Val_2015}
{\sc \au{Eliezer, S.}, \au{Henis, Z.}, \au{Nissim, N.}, \au{Pinhasi, S.~V.} \&
  \au{Val, J. M.~M.}} \yr{2015}  \at{Introducing a two temperature plasma
  ignition in inertial confined targets under the effect of relativistic shock
  waves: The case of {DT} and {pB11}}.  \jt{Laser and Particle Beams}
  \bvol{33}~(3),  \pg{577–589}.

\bibitem[Essén(1977)]{Essen_1977}
{\sc \au{Essén, H.}} \yr{1977}  \at{The physics of the born–oppenheimer
  approximation}.  \jt{International Journal of Quantum Chemistry}
  \bvol{12}~(4),  \pg{721–735}.

\bibitem[Fan {\em et~al.\/}(2016)Fan, Liu, Liu, Yu \&
  He]{Fan_Liu_Liu_Yu_He_2016}
{\sc \au{Fan, Z.}, \au{Liu, J.}, \au{Liu, B.}, \au{Yu, C.} \& \au{He, X.~T.}}
  \yr{2016}  \at{Ignition conditions relaxation for central hot-spot ignition
  with an ion-electron non-equilibrium model}.  \jt{Physics of Plasmas}
  \bvol{23}~(1),  \pg{010703}.

\bibitem[Fisch \& Herrmann(1994)]{Fisch_Herrmann_1994}
{\sc \au{Fisch, N.~J.} \& \au{Herrmann, M.~C.}} \yr{1994}  \at{Utility of
  extracting alpha particle energy by waves}.  \jt{Nuclear Fusion}
  \bvol{34}~(12),  \pg{1541}.

\bibitem[Fisch \& Rax(1992)]{Fisch_Rax_1992}
{\sc \au{Fisch, N.~J.} \& \au{Rax, J.-M.}} \yr{1992}  \at{Interaction of
  energetic alpha particles with intense lower hybrid waves}.  \jt{Physical
  Review Letters}  \bvol{69}~(4),  \pg{612–615}.

\bibitem[Fortov {\em et~al.\/}(2007)Fortov, Ilkaev, Arinin, Burtzev, Golubev,
  Iosilevskiy, Khrustalev, Mikhailov, Mochalov, Ternovoi \&
  Zhernokletov]{Fortov_Ilkaev_Arinin_Burtzev_Golubev_Iosilevskiy_Khrustalev_Mikhailov_Mochalov_Ternovoi_et}
{\sc \au{Fortov, V.~E.}, \au{Ilkaev, R.~I.}, \au{Arinin, V.~A.}, \au{Burtzev,
  V.~V.}, \au{Golubev, V.~A.}, \au{Iosilevskiy, I.~L.}, \au{Khrustalev, V.~V.},
  \au{Mikhailov, A.~L.}, \au{Mochalov, M.~A.}, \au{Ternovoi, V.~Y.} \&
  \au{Zhernokletov, M.~V.}} \yr{2007}  \at{Phase transition in a strongly
  nonideal deuterium plasma generated by quasi-isentropical compression at
  megabar pressures}.  \jt{Physical Review Letters}  \bvol{99}~(18),
  \pg{185001}.

\bibitem[Foster {\em et~al.\/}(2023)Foster, Fetsch \& Fisch]{Foster_et}
{\sc \au{Foster, T.~E.}, \au{Fetsch, H.} \& \au{Fisch, N.~J.}} \yr{2023}
  \at{Fast correlation heating in moderately coupled electron-ion plasmas}.
  \jt{(Under review in Journal of Plasma Physics)} .

\bibitem[Gericke \& Murillo(2003)]{Gericke_Murillo_2003}
{\sc \au{Gericke, D.} \& \au{Murillo, M.}} \yr{2003}  \at{Disorder-induced
  heating of ultracold plasmas}.  \jt{Contributions to Plasma Physics}
  \bvol{43}~(5–6),  \pg{298–301}.

\bibitem[Geyko \& Fisch(2017)]{Geyko_Fisch_2017}
{\sc \au{Geyko, V.~I.} \& \au{Fisch, N.~J.}} \yr{2017}  \at{Compressibility and
  heat capacity of rotating plasma}.  \jt{Physics of Plasmas}  \bvol{24}~(2),
  \pg{022113}.

\bibitem[Gislason(2010)]{Gislason_2010}
{\sc \au{Gislason, E.~A.}} \yr{2010}  \at{A close examination of the motion of
  an adiabatic piston}.  \jt{American Journal of Physics}  \bvol{78}~(10),
  \pg{995–1001}.

\bibitem[Gleizes {\em et~al.\/}(1999)Gleizes, Chervy \&
  Gonzalez]{Gleizes_Chervy_Gonzalez_1999}
{\sc \au{Gleizes, A.}, \au{Chervy, B.} \& \au{Gonzalez, J.~J.}} \yr{1999}
  \at{Calculation of a two-temperature plasma composition: bases and
  application to {SF6}}.  \jt{Journal of Physics D: Applied Physics}
  \bvol{32}~(16),  \pg{2060}.

\bibitem[Gregori {\em et~al.\/}(2007)Gregori, Ravasio, Höll, Glenzer \&
  Rose]{Gregori_Ravasio_Holl_Glenzer_Rose_2007}
{\sc \au{Gregori, G.}, \au{Ravasio, A.}, \au{Höll, A.}, \au{Glenzer, S.~H.} \&
  \au{Rose, S.~J.}} \yr{2007}  \at{Derivation of the static structure factor in
  strongly coupled non-equilibrium plasmas for x-ray scattering studies}.
  \jt{High Energy Density Physics}  \bvol{3}~(1),  \pg{99–108}.

\bibitem[Gruber(1999)]{Gruber_1999}
{\sc \au{Gruber, C.}} \yr{1999}  \at{Thermodynamics of systems with internal
  adiabatic constraints: time evolution of the adiabatic piston}.  \jt{European
  Journal of Physics}  \bvol{20}~(4),  \pg{259}.

\bibitem[Haines {\em et~al.\/}(2006)Haines, LePell, Coverdale, Jones, Deeney \&
  Apruzese]{Haines_LePell_Coverdale_Jones_Deeney_Apruzese_2006}
{\sc \au{Haines, M.~G.}, \au{LePell, P.~D.}, \au{Coverdale, C.~A.}, \au{Jones,
  B.}, \au{Deeney, C.} \& \au{Apruzese, J.~P.}} \yr{2006}  \at{Ion viscous
  heating in a magnetohydrodynamically unstable $z$ pinch at over
  $2\ifmmode\times\else\texttimes\fi{}{10}^{9}$ kelvin}.  \jt{Physical Review
  Letters}  \bvol{96}~(7),  \pg{075003}.

\bibitem[Han {\em et~al.\/}(2022)Han, Park, Sung, Kang, Lee, Chung, Hahm, Kim,
  Park, Bak, Cha, Choi, Choi, Gwak, Hahn, Jang, Lee, Kim, Kim, Kim, Ko, Ko,
  Lee, Lee, Lee, Lee, Lee, Lee, Park, Seo, Yang, Yoon \&
  Na]{Han_Park_Sung_Kang_Lee_Chung_Hahm_Kim_Park_Bak_et}
{\sc \au{Han, H.}, \au{Park, S.~J.}, \au{Sung, C.}, \au{Kang, J.}, \au{Lee,
  Y.~H.}, \au{Chung, J.}, \au{Hahm, T.~S.}, \au{Kim, B.}, \au{Park, J.-K.},
  \au{Bak, J.~G.}, \au{Cha, M.~S.}, \au{Choi, G.~J.}, \au{Choi, M.~J.},
  \au{Gwak, J.}, \au{Hahn, S.~H.}, \au{Jang, J.}, \au{Lee, K.~C.}, \au{Kim,
  J.~H.}, \au{Kim, S.~K.}, \au{Kim, W.~C.}, \au{Ko, J.}, \au{Ko, W.~H.},
  \au{Lee, C.~Y.}, \au{Lee, J.~H.}, \au{Lee, J.~K.}, \au{Lee, J.~P.}, \au{Lee,
  K.~D.}, \au{Lee, K.~D.}, \au{Park, Y.~S.}, \au{Seo, J.}, \au{Yang, S.~M.},
  \au{Yoon, S.~W.} \& \au{Na, Y.-S.}} \yr{2022}  \at{A sustained
  high-temperature fusion plasma regime facilitated by fast ions}.  \jt{Nature}
   \bvol{609}~(79267926),  \pg{269–275}.

\bibitem[Harbour {\em et~al.\/}(2018)Harbour, Förster, Dharma-Wardana \&
  Lewis]{Harbour_Forster_Dharma-wardana_Lewis_2018}
{\sc \au{Harbour, L.}, \au{Förster, G.~D.}, \au{Dharma-Wardana, M. W.~C.} \&
  \au{Lewis, L.~J.}} \yr{2018}  \at{Ion-ion dynamic structure factor, acoustic
  modes, and equation of state of two-temperature warm dense aluminum}.
  \jt{Physical Review E}  \bvol{97}~(4),  \pg{043210}.

\bibitem[Heckler(1994)]{Heckler_1994}
{\sc \au{Heckler, A.~F.}} \yr{1994}  \at{Astrophysical applications of quantum
  corrections to the equation of state of a plasma}.  \jt{Physical Review D}
  \bvol{49}~(2),  \pg{611–617}.

\bibitem[Hu {\em et~al.\/}(2011)Hu, Militzer, Goncharov \&
  Skupsky]{Hu_Militzer_Goncharov_Skupsky_2011}
{\sc \au{Hu, S.~X.}, \au{Militzer, B.}, \au{Goncharov, V.~N.} \& \au{Skupsky,
  S.}} \yr{2011}  \at{First-principles equation-of-state table of deuterium for
  inertial confinement fusion applications}.  \jt{Physical Review B}
  \bvol{84}~(22),  \pg{224109}.

\bibitem[Hummer \& Mihalas(1988)]{Hummer_Mihalas_1988}
{\sc \au{Hummer, D.~G.} \& \au{Mihalas, D.}} \yr{1988}  \at{The equation of
  state for stellar envelopes. {I} - an occupation probability formalism for
  the truncation of internal partition functions}.  \jt{The Astrophysical
  Journal}  \bvol{331},  \pg{794}.

\bibitem[Ichimaru(1993)]{Ichimaru_1993}
{\sc \au{Ichimaru, S.}} \yr{1993}  \at{Nuclear fusion in dense plasmas}.
  \jt{Reviews of Modern Physics}  \bvol{65}~(2),  \pg{255–299}.

\bibitem[Kelly(1963)]{Kelly_1963}
{\sc \au{Kelly, D.~C.}} \yr{1963}  \at{Plasma equation of state}.  \jt{American
  Journal of Physics}  \bvol{31}~(11),  \pg{827–828}.

\bibitem[Kestemont {\em et~al.\/}(2000)Kestemont, Broeck \&
  Mansour]{Kestemont_Broeck_Mansour_2000}
{\sc \au{Kestemont, E.}, \au{Broeck, C. V.~d.} \& \au{Mansour, M.~M.}}
  \yr{2000}  \at{The “adiabatic” piston: And yet it moves}.
  \jt{Europhysics Letters}  \bvol{49}~(2),  \pg{143}.

\bibitem[Killian {\em et~al.\/}(2007)Killian, Pattard, Pohl \&
  Rost]{Killian_Pattard_Pohl_Rost_2007}
{\sc \au{Killian, T.~C.}, \au{Pattard, T.}, \au{Pohl, T.} \& \au{Rost, J.~M.}}
  \yr{2007}  \at{Ultracold neutral plasmas}.  \jt{Physics Reports}
  \bvol{449}~(4),  \pg{77–130}.

\bibitem[Kodama {\em et~al.\/}(2001)Kodama, Norreys, Mima, Dangor, Evans,
  Fujita, Kitagawa, Krushelnick, Miyakoshi, Miyanaga, Norimatsu, Rose, Shozaki,
  Shigemori, Sunahara, Tampo, Tanaka, Toyama, Yamanaka \&
  Zepf]{Kodama_Norreys_Mima_Dangor_Evans_Fujita_Kitagawa_Krushelnick_Miyakoshi_Miyanaga_et}
{\sc \au{Kodama, R.}, \au{Norreys, P.~A.}, \au{Mima, K.}, \au{Dangor, A.~E.},
  \au{Evans, R.~G.}, \au{Fujita, H.}, \au{Kitagawa, Y.}, \au{Krushelnick, K.},
  \au{Miyakoshi, T.}, \au{Miyanaga, N.}, \au{Norimatsu, T.}, \au{Rose, S.~J.},
  \au{Shozaki, T.}, \au{Shigemori, K.}, \au{Sunahara, A.}, \au{Tampo, M.},
  \au{Tanaka, K.~A.}, \au{Toyama, Y.}, \au{Yamanaka, T.} \& \au{Zepf, M.}}
  \yr{2001}  \at{Fast heating of ultrahigh-density plasma as a step towards
  laser fusion ignition}.  \jt{Nature}  \bvol{412}~(68496849),  \pg{798–802}.

\bibitem[Koenig {\em et~al.\/}(2005)Koenig, Benuzzi-Mounaix, Ravasio, Vinci,
  Ozaki, Lepape, Batani, Huser, Hall, Hicks, MacKinnon, Patel, Park, Boehly,
  Borghesi, Kar \&
  Romagnani]{Koenig_Benuzzi-Mounaix_Ravasio_Vinci_Ozaki_Lepape_Batani_Huser_Hall_Hicks_et}
{\sc \au{Koenig, M.}, \au{Benuzzi-Mounaix, A.}, \au{Ravasio, A.}, \au{Vinci,
  T.}, \au{Ozaki, N.}, \au{Lepape, S.}, \au{Batani, D.}, \au{Huser, G.},
  \au{Hall, T.}, \au{Hicks, D.}, \au{MacKinnon, A.}, \au{Patel, P.}, \au{Park,
  H.~S.}, \au{Boehly, T.}, \au{Borghesi, M.}, \au{Kar, S.} \& \au{Romagnani,
  L.}} \yr{2005}  \at{Progress in the study of warm dense matter}.  \jt{Plasma
  Physics and Controlled Fusion}  \bvol{47}~(12B),  \pg{B441}.

\bibitem[Kraeft {\em et~al.\/}(1998)Kraeft, Schlanges, Kremp, Riemann \&
  DeWitt]{Kraeft_Schlanges_Kremp_Riemann_DeWitt_1998}
{\sc \au{Kraeft, W.~D.}, \au{Schlanges, M.}, \au{Kremp, D.}, \au{Riemann, J.}
  \& \au{DeWitt, H.~E.}} \yr{1998}  \at{Equation of state of strongly coupled
  plasmas}.  \jt{Zeitschrift für Physikalische Chemie}  \bvol{204}~(1–2),
  \pg{199–212}.

\bibitem[Kuzmin \& O’Neil(2002)]{Kuzmin_ONeil_2002}
{\sc \au{Kuzmin, S.~G.} \& \au{O’Neil, T.~M.}} \yr{2002}  \at{Numerical
  simulation of ultracold plasmas: How rapid intrinsic heating limits the
  development of correlation}.  \jt{Physical Review Letters}  \bvol{88}~(6),
  \pg{065003}.

\bibitem[Lindl {\em et~al.\/}(2004)Lindl, Amendt, Berger, Glendinning, Glenzer,
  Haan, Kauffman, Landen \&
  Suter]{Lindl_Amendt_Berger_Glendinning_Glenzer_Haan_Kauffman_Landen_Suter_2004}
{\sc \au{Lindl, J.~D.}, \au{Amendt, P.}, \au{Berger, R.~L.}, \au{Glendinning,
  S.~G.}, \au{Glenzer, S.~H.}, \au{Haan, S.~W.}, \au{Kauffman, R.~L.},
  \au{Landen, O.~L.} \& \au{Suter, L.~J.}} \yr{2004}  \at{The physics basis for
  ignition using indirect-drive targets on the national ignition facility}.
  \jt{Physics of Plasmas}  \bvol{11}~(2),  \pg{339–491}.

\bibitem[Loisel {\em et~al.\/}(2017)Loisel, Baily, Liehdahl, Fontes, Kallman,
  Nagayama, Hansen, Rochau, Mancini \&
  Lee]{Loisel_Bailey_Liedahl_Fontes_Kallman_Nagayama_Hansen_Rochau_Mancini_Lee_2017}
{\sc \au{Loisel, G.}, \au{Baily, J.}, \au{Liehdahl, D.}, \au{Fontes, C.},
  \au{Kallman, T.}, \au{Nagayama, T.}, \au{Hansen, S.}, \au{Rochau, G.},
  \au{Mancini, R.} \& \au{Lee, R.}} \yr{2017}  \at{Benchmark experiment for
  photoionized plasma emission from accretion-powered x-ray sources}.
  \jt{Physical Review Letters}  \bvol{119}~(7),  \pg{075001}.

\bibitem[Lyon \& Bergeson(2011)]{Lyon_Bergeson_2011}
{\sc \au{Lyon, M.} \& \au{Bergeson, S.~D.}} \yr{2011}  \at{The influence of
  electron screening on disorder-induced heating}.  \jt{Journal of Physics B:
  Atomic, Molecular and Optical Physics}  \bvol{44}~(18),  \pg{184014}.

\bibitem[Lyon {\em et~al.\/}(2013)Lyon, Bergeson \&
  Murillo]{Lyon_Bergeson_Murillo_2013}
{\sc \au{Lyon, M.}, \au{Bergeson, S.~D.} \& \au{Murillo, M.~S.}} \yr{2013}
  \at{Limit of strong ion coupling due to electron shielding}.  \jt{Physical
  Review E}  \bvol{87}~(3),  \pg{033101}.

\bibitem[Morawetz {\em et~al.\/}(2001)Morawetz, Bonitz, Morozov, Röpke \&
  Kremp]{Morawetz_Bonitz_Morozov_Ropke_Kremp_2001}
{\sc \au{Morawetz, K.}, \au{Bonitz, M.}, \au{Morozov, V.~G.}, \au{Röpke, G.}
  \& \au{Kremp, D.}} \yr{2001}  \at{Short-time dynamics with initial
  correlations}.  \jt{Physical Review E}  \bvol{63}~(2),  \pg{020102}.

\bibitem[More {\em et~al.\/}(1988)More, Warren, Young \&
  Zimmerman]{More_Warren_Young_Zimmerman_1988}
{\sc \au{More, R.~M.}, \au{Warren, K.~H.}, \au{Young, D.~A.} \& \au{Zimmerman,
  G.~B.}} \yr{1988}  \at{A new quotidian equation of state ({QEOS}) for hot
  dense matter}.  \jt{Physics of Fluids}  \bvol{31}~(10),  \pg{3059}.

\bibitem[Mukherjee {\em et~al.\/}(2020)Mukherjee, Jaiswal, Shukla, Hakim \&
  Thomas]{Mukherjee_Jaiswal_Shukla_Hakim_Thomas_2020}
{\sc \au{Mukherjee, R.}, \au{Jaiswal, S.}, \au{Shukla, M.~K.}, \au{Hakim, A.}
  \& \au{Thomas, E.}} \yr{2020}  \at{Measurement of temperature of a dusty
  plasma from its configuration}.  \jt{Contributions to Plasma Physics}
  \bvol{60}~(4),  \pg{e201900161}.

\bibitem[Murillo(2001)]{Murillo_2001}
{\sc \au{Murillo, M.~S.}} \yr{2001}  \at{Using fermi statistics to create
  strongly coupled ion plasmas in atom traps}.  \jt{Physical Review Letters}
  \bvol{87}~(11),  \pg{115003}.

\bibitem[O’Neil \& Rostoker(1965)]{O’Neil_Rostoker_1965}
{\sc \au{O’Neil, T.} \& \au{Rostoker, N.}} \yr{1965}  \at{Triplet correlation
  for a plasma}.  \jt{The Physics of Fluids}  \bvol{8}~(6),  \pg{1109–1114}.

\bibitem[Petrov {\em et~al.\/}(2015)Petrov, Migdal, Inogamov \&
  Zhakhovsky]{Petrov_Migdal_Inogamov_Zhakhovsky_2015}
{\sc \au{Petrov, Y.~V.}, \au{Migdal, K.~P.}, \au{Inogamov, N.~A.} \&
  \au{Zhakhovsky, V.~V.}} \yr{2015}  \at{Two-temperature equation of state for
  aluminum and gold with electrons excited by an ultrashort laser pulse}.
  \jt{Applied Physics B}  \bvol{119}~(3),  \pg{401–411}.

\bibitem[Planes \& Vives(2002)]{Planes_Vives_2002}
{\sc \au{Planes, A.} \& \au{Vives, E.}} \yr{2002}  \at{Entropic {Formulation}
  of {Statistical} {Mechanics}}.  \jt{Journal of Statistical Physics}
  \bvol{106}~(3),  \pg{827--850}.

\bibitem[Ramshaw \& Cook(2014)]{Ramshaw_Cook_2014}
{\sc \au{Ramshaw, J.~D.} \& \au{Cook, A.~W.}} \yr{2014}  \at{Approximate
  equations of state in two-temperature plasma mixtures}.  \jt{Physics of
  Plasmas}  \bvol{21}~(2),  \pg{022706}.

\bibitem[Ricotti {\em et~al.\/}(2000)Ricotti, Gnedin \&
  Shull]{Ricotti_Gnedin_Shull_2000}
{\sc \au{Ricotti, M.}, \au{Gnedin, N.~Y.} \& \au{Shull, J.~M.}} \yr{2000}
  \at{The evolution of the effective equation of state of the intergalactic
  medium}.  \jt{The Astrophysical Journal}  \bvol{534}~(1),  \pg{41–56}.

\bibitem[Rinderknecht {\em et~al.\/}(2015)Rinderknecht, Rosenberg, Li, Hoffman,
  Kagan, Zylstra, Sio, Frenje, Gatu~Johnson, Séguin, Petrasso, Amendt, Bellei,
  Wilks, Delettrez, Glebov, Stoeckl, Sangster, Meyerhofer \&
  Nikroo]{Rinderknecht_Rosenberg_Li_Hoffman_Kagan_Zylstra_Sio_Frenje_Gatu}
{\sc \au{Rinderknecht, H.~G.}, \au{Rosenberg, M.}, \au{Li, C.}, \au{Hoffman,
  N.}, \au{Kagan, G.}, \au{Zylstra, A.}, \au{Sio, H.}, \au{Frenje, J.},
  \au{Gatu~Johnson, M.}, \au{Séguin, F.}, \au{Petrasso, R.}, \au{Amendt, P.},
  \au{Bellei, C.}, \au{Wilks, S.}, \au{Delettrez, J.}, \au{Glebov, V.},
  \au{Stoeckl, C.}, \au{Sangster, T.}, \au{Meyerhofer, D.} \& \au{Nikroo, A.}}
  \yr{2015}  \at{Ion thermal decoupling and species separation in shock-driven
  implosions}.  \jt{Physical Review Letters}  \bvol{114}~(2),  \pg{025001}.

\bibitem[Salpeter(1963)]{Salpeter_1963}
{\sc \au{Salpeter, E.~E.}} \yr{1963}  \at{Density fluctuations in a
  nonequilibrium plasma}.  \jt{Journal of Geophysical Research (1896-1977)}
  \bvol{68}~(5),  \pg{1321–1333}.

\bibitem[Salpeter \& van Horn(1969)]{Salpeter_van_Horn_1969}
{\sc \au{Salpeter, E.~E.} \& \au{van Horn, H.~M.}} \yr{1969}  \at{Nuclear
  reaction rates at high densities}.  \jt{The Astrophysical Journal}
  \bvol{155},  \pg{183}.

\bibitem[Schmit {\em et~al.\/}(2010)Schmit, Dodin \&
  Fisch]{Schmit_Dodin_Fisch_2010}
{\sc \au{Schmit, P.~F.}, \au{Dodin, I.~Y.} \& \au{Fisch, N.~J.}} \yr{2010}
  \at{Controlling hot electrons by wave amplification and decay in compressing
  plasma}.  \jt{Physical Review Letters}  \bvol{105}~(17),  \pg{175003}.

\bibitem[Schmit {\em et~al.\/}(2013)Schmit, Dodin, Rocks \&
  Fisch]{Schmit_Dodin_Rocks_Fisch_2013}
{\sc \au{Schmit, P.~F.}, \au{Dodin, I.~Y.}, \au{Rocks, J.} \& \au{Fisch,
  N.~J.}} \yr{2013}  \at{Nonlinear amplification and decay of phase-mixed waves
  in compressing plasma}.  \jt{Physical Review Letters}  \bvol{110}~(5),
  \pg{055001}.

\bibitem[Scullard {\em et~al.\/}(2018)Scullard, Serna, Benedict, Ellison \&
  Graziani]{Scullard_Serna_Benedict_Ellison_Graziani_2018}
{\sc \au{Scullard, C.~R.}, \au{Serna, S.}, \au{Benedict, L.~X.}, \au{Ellison,
  C.~L.} \& \au{Graziani, F.~R.}} \yr{2018}  \at{Analytic expressions for
  electron-ion temperature equilibration rates from the {Lenard-Balescu}
  equation}.  \jt{Physical Review E}  \bvol{97}~(1),  \pg{013205}.

\bibitem[Seuferling {\em et~al.\/}(1989)Seuferling, Vogel \&
  Toepffer]{Seuferling_Vogel_Toepffer_1989}
{\sc \au{Seuferling, P.}, \au{Vogel, J.} \& \au{Toepffer, C.}} \yr{1989}
  \at{Correlations in a two-temperature plasma}.  \jt{Physical Review A}
  \bvol{40}~(1),  \pg{323–329}.

\bibitem[Shaffer {\em et~al.\/}(2017)Shaffer, Tiwari \&
  Baalrud]{Shaffer_Tiwari_Baalrud_2017}
{\sc \au{Shaffer, N.~R.}, \au{Tiwari, S.~K.} \& \au{Baalrud, S.~D.}} \yr{2017}
  \at{Pair correlation functions of strongly coupled two-temperature plasma}.
  \jt{Physics of Plasmas}  \bvol{24}~(9),  \pg{092703}.

\bibitem[Shukla {\em et~al.\/}(2017)Shukla, Avinash, Mukherjee \&
  Ganesh]{Shukla_Avinash_Mukherjee_Ganesh_2017}
{\sc \au{Shukla, M.~K.}, \au{Avinash, K.}, \au{Mukherjee, R.} \& \au{Ganesh,
  R.}} \yr{2017}  \at{Isothermal equation of state of three dimensional yukawa
  gas}.  \jt{Physics of Plasmas}  \bvol{24}~(11),  \pg{113704}.

\bibitem[Slattery {\em et~al.\/}(1980)Slattery, Doolen \&
  DeWitt]{Slattery_Doolen_DeWitt_1980}
{\sc \au{Slattery, W.~L.}, \au{Doolen, G.~D.} \& \au{DeWitt, H.~E.}} \yr{1980}
  \at{Improved equation of state for the classical one-component plasma}.
  \jt{Physical Review A}  \bvol{21}~(6),  \pg{2087–2095}.

\bibitem[Stix \& Skaley(1990)]{Stix_Skaley_1990}
{\sc \au{Stix, M.} \& \au{Skaley, D.}} \yr{1990}  \at{The equation of state and
  the frequencies of solar {P} modes.}  \jt{Astronomy and Astrophysics}
  \bvol{232},  \pg{234–238}.

\bibitem[Strachan {\em et~al.\/}(1987)Strachan, Bitter, Ramsey, Zarnstorff,
  Arunasalam, Bell, Bretz, Budny, Bush, Davis, Dylla, Efthimion, Fonck,
  Fredrickson, Furth, Goldston, Grisham, Grek, Hawryluk, Heidbrink, Hendel,
  Hill, Hsuan, Jaehnig, Jassby, Jobes, Johnson, Johnson, Kaita, Kampershroer,
  Knize, Kozub, LeBlanc, Levinton, La~Marche, Manos, Mansfield, McGuire,
  McNeill, Meade, Medley, Morris, Mueller, Nieschmidt, Owens, Park, Schivell,
  Schilling, Schmidt, Scott, Sesnic, Sinnis, Stauffer, Stratton, Tait, Taylor,
  Towner, Ulrickson, von Goeler, Wieland, Williams, Wong, Yoshikawa, Young \&
  Zweben]{Strachan_Bitter_Ramsey_Zarnstorff_Arunasalam_Bell_Bretz_Budny_Bush_Davis_et}
{\sc \au{Strachan, J.~D.}, \au{Bitter, M.}, \au{Ramsey, A.~T.}, \au{Zarnstorff,
  M.~C.}, \au{Arunasalam, V.}, \au{Bell, M.~G.}, \au{Bretz, N.~L.}, \au{Budny,
  R.}, \au{Bush, C.~E.}, \au{Davis, S.~L.}, \au{Dylla, H.~F.}, \au{Efthimion,
  P.~C.}, \au{Fonck, R.~J.}, \au{Fredrickson, E.}, \au{Furth, H.~P.},
  \au{Goldston, R.~J.}, \au{Grisham, L.~R.}, \au{Grek, B.}, \au{Hawryluk,
  R.~J.}, \au{Heidbrink, W.~W.}, \au{Hendel, H.~W.}, \au{Hill, K.~W.},
  \au{Hsuan, H.}, \au{Jaehnig, K.~P.}, \au{Jassby, D.~L.}, \au{Jobes, F.},
  \au{Johnson, D.~W.}, \au{Johnson, L.~C.}, \au{Kaita, R.}, \au{Kampershroer,
  J.}, \au{Knize, R.~J.}, \au{Kozub, T.}, \au{LeBlanc, B.}, \au{Levinton, F.},
  \au{La~Marche, P.~H.}, \au{Manos, D.~M.}, \au{Mansfield, D.~K.}, \au{McGuire,
  K.}, \au{McNeill, D.~H.}, \au{Meade, D.~M.}, \au{Medley, S.~S.}, \au{Morris,
  W.}, \au{Mueller, D.}, \au{Nieschmidt, E.~B.}, \au{Owens, D.~K.}, \au{Park,
  H.}, \au{Schivell, J.}, \au{Schilling, G.}, \au{Schmidt, G.~L.}, \au{Scott,
  S.~D.}, \au{Sesnic, S.}, \au{Sinnis, J.~C.}, \au{Stauffer, F.~J.},
  \au{Stratton, B.~C.}, \au{Tait, G.~D.}, \au{Taylor, G.}, \au{Towner, H.~H.},
  \au{Ulrickson, M.}, \au{von Goeler, S.}, \au{Wieland, R.}, \au{Williams,
  M.~D.}, \au{Wong, K.-L.}, \au{Yoshikawa, S.}, \au{Young, K.~M.} \&
  \au{Zweben, S.~J.}} \yr{1987}  \at{High-temperature plasmas in a tokamak
  fusion test reactor}.  \jt{Physical Review Letters}  \bvol{58}~(10),
  \pg{1004–1007}.

\bibitem[Tabak {\em et~al.\/}(2006)Tabak, Hinkel, Atzeni, Campbell \&
  Tanaka]{Tabak_Hinkel_Atzeni_Campbell_Tanaka_2006}
{\sc \au{Tabak, M.}, \au{Hinkel, D.}, \au{Atzeni, S.}, \au{Campbell, E.~M.} \&
  \au{Tanaka, K.}} \yr{2006}  \at{Fast ignition: Overview and background}.
  \jt{Fusion Science and Technology}  \bvol{49}~(3),  \pg{254–277}.

\bibitem[Tiwari \& Baalrud(2018)]{Tiwari_Baalrud_2018}
{\sc \au{Tiwari, S.~K.} \& \au{Baalrud, S.~D.}} \yr{2018}  \at{Reduction of
  electron heating by magnetizing ultracold neutral plasma}.  \jt{Physics of
  Plasmas}  \bvol{25}~(1),  \pg{013511}.

\bibitem[Triola(2022)]{Triola_2022}
{\sc \au{Triola, C.}} \yr{2022}  \at{Model comparisons for two-temperature
  plasma equations of state}.  \jt{Physics of Plasmas}  \bvol{29}~(11),
  \pg{112705}.

\bibitem[Tully {\em et~al.\/}(2016)Tully, Hawker \&
  Ventikos]{Tully_Hawker_Ventikos_2016}
{\sc \au{Tully, B.}, \au{Hawker, N.} \& \au{Ventikos, Y.}} \yr{2016}
  \at{Modeling asymmetric cavity collapse with plasma equations of state}.
  \jt{Physical Review E}  \bvol{93}~(5),  \pg{053105}.

\bibitem[Volkov(1999)]{Volkov_1999}
{\sc \au{Volkov, N.~B.}} \yr{1999}  \at{A model of two-temperature plasma with
  strong large-scale turbulence: dynamic equations, thermodynamics and
  transport coefficients}.  \jt{Plasma Physics and Controlled Fusion}
  \bvol{41}~(8),  \pg{1025}.

\end{thebibliography}

\end{document}